\pdfoutput=1
\RequirePackage{ifpdf}
\ifpdf 
\documentclass[pdftex]{sigma}
\else
\documentclass{sigma}
\fi

\usepackage{mathabx}
\usepackage[sans]{dsfont}
\usepackage{booktabs}
\newtheorem{condition}{Condition}[section]
\numberwithin{equation}{section}
\numberwithin{theorem}{section}
\numberwithin{lemma}{section}
\numberwithin{corollary}{section}
\numberwithin{definition}{section}
\numberwithin{example}{section}

\begin{document}

\allowdisplaybreaks

\renewcommand{\thefootnote}{$\star$}

\renewcommand{\PaperNumber}{014}

\FirstPageHeading

\ShortArticleName{Emergent Braided Matter of Quantum Geometry}

\ArticleName{Emergent Braided Matter of Quantum Geometry\footnote{This
paper is a contribution to the Special Issue ``Loop Quantum Gravity and Cosmology''. The full collection is available at \href{http://www.emis.de/journals/SIGMA/LQGC.html}{http://www.emis.de/journals/SIGMA/LQGC.html}}}

\Author{Sundance BILSON-THOMPSON~$^{\dag^1}$, Jonathan HACKETT~$^{\dag^2}$, Louis KAUFFMAN~$^{\dag^3}$\\ and Yidun WAN~$^{\dag^4}$}

\AuthorNameForHeading{S.~Bilson-Thompson, J.~Hackett, L.~Kauf\/fman and Y.~Wan}

\Address{$^{\dag^1}$~School of Chemistry and Physics, University of Adelaide, SA 5005, Australia}
\EmailDD{\href{mailto:sundance.bilson-thompson@adelaide.edu.au}{sundance.bilson-thompson@adelaide.edu.au}}

\Address{$^{\dag^2}$~Perimeter Institute for Theoretical Physics,  31 Caroline Street North,\\
\hphantom{$^{\dag^2}$}~Waterloo, ON N2L 2Y5, Canada}
\EmailDD{\href{mailto:jonathanhackett@gmail.com}{jonathanhackett@gmail.com}}

\Address{$^{\dag^3}$~Department of Mathematics, University of Illinois at Chicago,\\
\hphantom{$^{\dag^3}$}~851 South Morgan Street, Chicago, Illinois 60607-7045, USA}
\EmailDD{\href{mailto:kauffman@uic.edu}{kauffman@uic.edu}}

\Address{$^{\dag^4}$~Open Research Centre for Quantum Computing, Kinki University,\\
\hphantom{$^{\dag^4}$}~Kowakae 3-4-1, Higashi-osaka 577-0852, Japan}
\EmailDD{\href{mailto:ywan@alice.math.kindai.ac.jp}{ywan@alice.math.kindai.ac.jp}}

\ArticleDates{Received August 31, 2011, in f\/inal form March 12, 2012; Published online March 24, 2012}

\Abstract{We review and present a few new results of the program of emergent matter as braid excitations of quantum geometry that is represented by braided ribbon networks. These networks are a generalisation of the spin networks proposed by Penrose and those in models of background independent quantum gravity theories, such as Loop Quantum Gravity and Spin Foam models. This program has been developed in two parallel but complimentary schemes, namely the trivalent and tetravalent schemes. The former studies the braids on trivalent braided ribbon networks, while the latter investigates the braids on tetravalent braided ribbon networks. Both schemes have been fruitful. The trivalent scheme has been quite successful at establishing a correspondence between braids and Standard Model particles, whereas the tetravalent scheme has naturally substantiated a rich, dynamical theory of interactions and propagation of braids, which is ruled by topological conservation laws. Some recent advances in the program indicate that the two schemes may converge to yield a fundamental theory of matter in quantum spacetime.}

\Keywords{quantum gravity; loop quantum gravity; spin network; braided ribbon network; emergent matter; braid; standard model; particle physics; unif\/ication; braided tensor category; topological quantum computation}

\Classification{83C45; 83C27; 81T99; 81V25; 20F36; 18D35; 20K45; 81P68}

\vspace{-2mm}

\renewcommand{\thefootnote}{\arabic{footnote}}
\setcounter{footnote}{0}

\section{Introduction}\label{secIntro}

\subsection{An invitation to emergent matter of quantum geometry}\label{subsecInvit}
What is spacetime? What is matter? Physicists and philosophers have pondered these questions for centuries. In fact, an ultimate goal of modern physics is to f\/ind a unif\/ied answer for both questions. Recently, in order to answer these questions, a novel approach towards emergent\footnote{Here we mean coexisting quantum geometry and matter because our program indicates that a background independent quantum gravity theory may have built-in matter as topological excitations of the quantum geometry described by the theory.} matter as topological excitations of quantum geometry has been put forward and extensively developed~\cite{Bilson-Thompson2005,LouNumber,Sundance2009,Bilson-Thompson2006,Hackett2011b,Hackett2011a,Hackett2007,WanHackett2008,WanHe2008b,WanHe2008a,Isabeau2008,WanLee2007,Wan2008,Wan2007}. Provided with the results of two recent papers~\cite{Hackett2011b,Hackett2011a} along this course, an article that of\/fers a precise review and outlook of this research line seems timely.

A brief historical account is as follows. In 2005, Bilson-Thompson
proposed a topological matter model, the helon model~\cite{Bilson-Thompson2005}, which is based on the preon models of Harari and Shupe~\mbox{\cite{preon1979Harari, preon1979Shupe}} and is more elementary than the Standard Model (SM) of particles by interpreting the elementary particles as braids of three ribbons. At the time it was proposed, the helon model took the form of a combinatoric game rather than a rigorous theory. In this model, the integral twists of ribbons of braids are interpreted as the quantized electric charges of particles. The permutations of twists on certain braids naturally account for the color charges of quarks and gluons. This model incorporates a simple scheme of the color interaction and the electro-weak interaction with lepton and baryon number manifestly conserved. It may also be able to account for the three generations of elementary fermions.

In 2006, Bilson-Thompson, Markopoulou and Smolin~\cite{Bilson-Thompson2006} coded the helon model in certain background independent Quantum Gravity models such as Loop Quantum Gravity (LQG) and Spin Foam (SF) models, by identifying helons with emergent topological excitations of embedded trivalent spin networks that label the states in LQG. Hereafter this will be called the trivalent scheme. Developments of the trivalent scheme~\cite{LouNumber,Sundance2009,Hackett2007,Isabeau2008} allowed the helon model to be used as a dictionary between the 3-strand braids of embedded trivalent spin networks and the SM particles. The trivalent scheme led to a new perspective; instead of treating the helon model as yet another model of elementary particles, one can encode it in LQG and SF models to make a~theory of both spacetime and matter. The dynamics governing particle interactions would then be a consequence of the dynamics of the discrete building blocks of quantum spacetime. In this setting, matter is emergent from quantum spacetime, and the corresponding low energy ef\/fective theories may give rise to general relativity coupled with quantum f\/ields.

Unfortunately, results on the stability of braided states~\cite{Bilson-Thompson2006} in the trivalent scheme strongly suggested that the dynamics of spacetime would allow particle propagation, but not interactions. In ef\/fect, braids in the trivalent scheme are ``too stable''. To address this issue, and because of the geometrical correspondence between framed 4-valent spin networks and 3-space, a 4-valent scheme was developed~\cite{WanHackett2008,WanHe2008b,WanHe2008a,WanLee2007,Wan2008,Wan2009,Wan2007}. In the 4-valent scheme, the topological structures that can potentially be identif\/ied with particles are also 3-strand braids, each of which is formed by the three common edges of two adjacent 4-valent nodes of embedded, framed 4-valent spin networks. The 4-valent scheme gives rise to forms of braid propagation and interaction that are analogous to the dynamics of particles. Nevertheless, the lack of suf\/f\/icient super-selection rules over an enormous zoo of 3-strand braids in the 4-valent scheme withholds a Rosetta Stone that maps the braids to the SM particles. On the other hand, the 4-valent braids may be more elementary, high-energy entities whose low energy limit produces the SM particles~\cite{Wan2008,Wan2009}.

Very recently, two papers by Hackett~\cite{Hackett2011b, Hackett2011a} and work by Bilson-Thompson reported here in Section~\ref{sec:3v4vcorrespondence}, provide a framework that may encode both the trivalent and 4-valent schemes. This would allow the economical reproduction of SM particle states that occurs in the trivalent scheme, and the propogation and interactions that occur in the 4-valent scheme to be combined into a single theory.

As a historical remark, the idea that matter is topological defects of spacetime is an old dream that dates back to 1867 when Lord Kelvin proposed that atoms were knots in the ether~\cite{Kelvin}. Kelvin's idea failed largely due to the limited knowledge of atomic and subatomic structure at the time. Nevertheless, this dream has persisted in physicists thereafter. Various proposals of topological matter have arisen as physicists deepen and broaden their recognition of nature. An example is the topological Geon model due to Wheeler and others~\cite{Geon3Hartle1964,Geon5Sorkin1996,Geon2Misner1959, Geon4Perry1999,Geon1Wheeler1955} but the geons therein were unstable and classical. To make stable geons~\cite{Geon2Misner1959}, Finkelstein invented the notion of topological conservation laws that also led to advances in condensed matter physics, e.g., topologically conserved excitations in the sine-Gordon theory. Finkelstein's idea had not been compatible with quantum gravity until the recent work by Markopoulou et al.~\cite{fotini2005kribs, fotini2007bi, fotini2003} that motivated our work. Analogously, certain condensed matter systems have quasi-particles as collective modes, e.g., phonons and rotons in superf\/luid~He$^4$. A recent example in condensed matter physics is a unif\/ication scheme due to Wen et al.~\cite{Wen2006gravity, Wen2004light}, where gauge theories and linearised gravity appear to be low energy ef\/fective descriptions of a new phase, the string-net condensate of lattice spin systems.

In the rest of the Introduction, we brief\/ly introduce concepts and ideas that underpin our approach to emergent matter. We leave main discussions on the trivalent and 4-valent schemes to other sections.

\subsection{Noiseless subsystems}\label{subsecNS}

To appreciate the ideas of emergent matter of embedded, framed spin networks, one needs to understand two notions, namely noiseless subsystems and spin networks. Let us address the former f\/irst. Noiseless subsystems, put forward in quantum information and computation for quantum error correction~\cite{noise2004Holbrook, noise2000Kempe, noise2004Kribs, noise1997Zanardi}, are subsets of states of a quantum system that are preserved under the evolution algebra of the system, and hence protected from any error. Markopoulou et al.\ adopted the idea of noiseless subsystems to solve the problem of the low energy limit in background independent quantum gravity theories~\cite{Dreyer2006,fotini2005kribs,fotiniUnpub}.

Background independence brings in dif\/f\/iculties that make taking the low energy limit of a~background independent theory of quantum gravity~\cite{fotini2007bi} a big open issue, although attempts have been made in various approaches of quantum gravity. In LQG, e.g.,~\cite{cohere2005Ashtekar, cohere2007Thiemann,cohere2002Thiemann, ThiemannBook} use the method of coherent states, which are, according to quantum physics, the quantum states closest to classical ones. For another example, in SF models, \cite{lqgProp2008Alesci,lqgProp2007Rovelli, lqgProp2007Alesci, Christensen2009npoint, lqgProp2008Perini} utilize $n$-point correlations functions, which is reasonable because all semiclassical observables are either correlation functions or their derivatives. In view of this, Markopoulou and Kribs~\cite{fotini2005kribs, fotini2007bi, fotiniUnpub} proposed a new way to resolve this issue by looking for conserved quantities in background independent theories of gravity. Since the aforementioned noiseless subsystems are conserved states under the evolution algebra of a~quantum system, there should be conserved quantities associated with them. This is why one can adapt the method of noiseless subsystems to f\/ind conserved quantities of quantum geometry in a large class of background independent theories.

The f\/irst application of noiseless subsystems in LQG~\cite{Dreyer2006} gives a possible explanation of black hole entropy and how symmetries can emerge from a dif\/feomorphism invariant formulation of quantum gravity. The noiseless substructures of braided ribbon networks, which appear to encode the particle states in models of quantum gravity such as LQG, are braids whose associated conserved quantities are called reduced link invariants~\cite{LouNumber,Bilson-Thompson2006,Hackett2011b, Hackett2011a,Hackett2007}.

\subsection{Spin networks}\label{subsecSnet}
Penrose invented spin networks as a fundamental discrete description of spacetime~\cite{Penrose1971, Penrose1972}; later, Rovelli and Smolin found a more generalized version of spin networks to label the states in LQG Hilbert space~\cite{Rovelli1995spinnet}. Although the context of spin networks in this article is mainly LQG and its path integral formulation, SF models, it will be clear that our results do not really depend on these models but f\/ind their natural home in a generalisation of Penrose's version. Spin networks also arise in lattice gauge theories~\cite{ Cherrington2008lgt, Cherrington2009lgt, lgauge1987Kowala, lgauge1975Kogut} and topological f\/ield theories~\cite{tqft1994Crane,tqft1994Foxon, tqft1992Ooguri, tqft1992Turaev}, which are not discussed here.

\subsubsection{Penrose's spin networks}

Penrose noticed the fundamental incompatibility between General Relativity and Quantum Physics, the problem of the concept of continuum, and the divergences in quantum f\/ield theories. He thought that resolving this incompatibility demands a discrete notion of spacetime at the Planck scale, where the classical notion of spacetime is no longer valid. Consequently, the concept of time and space gives way to a more fundamental notion, the microscopic causal relation between quantum events\footnote{The underlying philosophy is relationalism, as opposed to reductionism, reviewed in~\cite{Rovelli2008rev, RovelliBook, Smolin2005bi}.}. Knowing that spin (or angular momentum) is intrinsic and characteristic to both quantum systems and classical spacetime, Penrose used combinatoric graphs, consisting of lines intersecting at vertices, to represent the fundamental states of spacetime. Each line in a graph is labeled by a spin, an integer or half integer. Hence, such a graph is called a spin network. Later on, \cite{MoussourisThesis} showed that the classical 3-dimensional angles of space can be recovered from trivalent spin networks. Note that these spin networks are unembedded and, opposed to those in LQG, are a direct construction of fundamental quantum spacetime, rather than obtained from quantizing spacetime or General Relativity by any means.

\subsubsection{Spin networks from LQG}\label{subsubSecSNLQG}
LQG is a non-perturbative, canonical quantization of General Relativity\footnote{Other non-perturbative approaches to quantum gravity also exist; however, here we focus on LQG.}. The background independence of General Relativity actually does not leave any room for perturbative quantization~\cite{AshtekarBook, BaezBook,RovelliBook, Smolin2005bi}. LQG assumes a $(d+1)$-dimensional dif\/ferential manifold $M$ with a foliation $M=\Sigma\times\mathbb{R}$, without metric but merely a dif\/ferential structure with a Lie algebra\footnote{Usually $\mathfrak{su}(2)$ or $\mathfrak{so}(3)$.} valued connection 1-form f\/ield. We take $d=3$.

Our approach does not directly depend on any specif\/ic techniques and results of LQG but is only inspired by them, so we do not review the technical settings and quantization procedure in LQG.

\begin{figure}[h]
\centering
\includegraphics[scale=0.8]{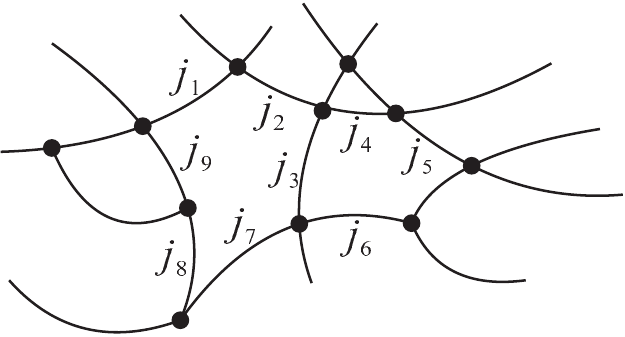}
\caption{A portion of a
generic spin network. Some labels are omitted.}\label{spinnet}
\end{figure}

In LQG, the states are spin networks, which are graphs embedded in $\Sigma$ (Fig.~\ref{spinnet}). An edge~$e$ is a f\/lux line, labelled by an irreducible representation $j_e$ of a Lie group (usually $SU(2)$ or $SO(3)$). A vertex is labelled by an intertwiner that is the invariant tensor of the labels on the edges meeting at the vertex.

LQG has produced many physical results, which are extensively reviewed in~\cite{Rovelli2008rev,Smolin2006, Smolin2004inv}. The result most relevant to this review is the evidence that space at the Planck scale is discrete because spin network states are eigenstates of operators corresponding to geometric measurements such as area and volume. For example the area operator $\hat{A}$ acting on a 2-dimensional surface $S$ has the spectrum
\begin{gather*}
\hat{A}[S]\left|\Gamma\right>\propto l^2_p\sum\limits_{i\in\{\Gamma\cup
S\}}\sqrt{j_i(j_i+1)}\left|\Gamma\right>,
\end{gather*}
where $l_p$ is the Planck length, and $\Gamma$ is a spin network that has no vertices but only edges in~$S$.\footnote{If $\Gamma$ has edges on $S$, a degeneracy arises and calls for regularization methods to obtain the correct spectrum~\cite{RovelliBook}.} Likewise, the intertwiners on the nodes of a spin network in a region determine the 3-volume of the region. This fundamental discreteness resolves the singularity problem and also eliminates ultraviolet divergences, as it provides a natural cutof\/f at the Planck scale to the physical spectrum of the theory.

In general, a vertex of a spin network can have any valence greater than two, the number of edges meeting at the vertex, as seen in Fig.~\ref{spinnet}. We may consider a basis of spin networks with def\/inite valences, i.e., spin networks respectively with three, four, and higher valences, such that a generic LQG state is a linear combination of these basis states. one may also think that trivalent spin networks may be suf\/f\/icient to provide a complete basis that spans all spin networks. This is plausible and is suggested by Rovelli~\cite{RovelliBook, Rovelli1995spinnet} for the case of $SU(2)$ and $SO(3)$. The trivalent spin networks represent a basis of LQG state space. That is, associated with a~spin network $\Gamma$ is a state $\left|\Gamma\right>$, and for two such states $\left|\Gamma\right>$ and $\left|\Gamma'\right>$, $\left<\Gamma|\Gamma'\right>=\delta_{\Gamma\Gamma'}$. Spin network labels on edges and nodes are representations of the group elements, labeling the graphs in the classical conf\/iguration space, and the corresponding intertwiners.

Trivalent spin networks have dif\/f\/iculty in representing 3-space because their nodes have zero 3-volume. But each 4-valent node yields a 3-volume\footnote{This correspondence is at the Planck level. Whether it holds in a continuum limit is still open.}~\cite{Rovelli1996vol,Loll1995vol}. \cite{fotini1997} also suggests one may carry this correspondence to any higher valence. In this review, we shall study both trivalent and 4-valent spin networks.

\begin{figure}[h]
\centering
\includegraphics[scale=0.7]{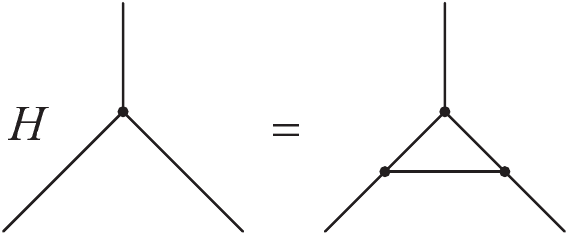}
\caption[Action of the Hamiltonian constraint operator.]{Sketch of
of the Hamiltonian constraint acting on a vertex. Irrelevant details are ignored.}\label{hConstraint}
\end{figure}

Trivalent spin networks acquire dynamics by evolving under the action of the Hamiltonian constraint operator of LQG, which also helps to realize the 4D dif\/feomorphism invariance of the theory. The well-accepted form of the Hamiltonian constraint acts only on vertices (Fig.~\ref{hConstraint}) Thiemann~\cite{Thiemann1996hc, Thiemann1996hc1,Thiemann1996hc2}; hence, it behaves as a local move that evolves a spin network state to another. In fact, LQG has a path integral formulation, SF models, casting the evolution of spin networks in a systematic, covariant way~\cite{RovelliBook}.

In many studies of SF models~\cite{BarrettCrane1998,Freidel2007sf}, spin networks and their histories are unembedded, combinatoric graphs. The key dif\/ference between embedded and unembedded spin networks is that the edges in the former can knot, braid, and link. The role of these knots, braids, and links has been a big open issue. Nevertheless, in this article we will show the correspondence between some of these topological structures of embedded spin networks and matter. \cite{WanHackett2008Inf}~of\/fers another perspective.

\begin{figure}[h]
\centering
\includegraphics[scale=1]{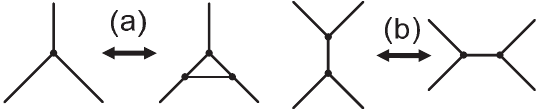}
\caption[Evolution moves of trivalent spin networks.]{(a): expansion
and contraction move. (b):~exchange move.}\label{evolMove}
\end{figure}

According to SF models~\cite{Baez1999,Baez1997sf, Rovelli1998rev,
RovelliBook}, trivalent spin networks evolve under two more
moves, shown in Fig.~\ref{evolMove}(a) and~(b), including the one in
Fig.~\ref{hConstraint}.

\begin{figure}[h]
\centering
\includegraphics[scale=1]{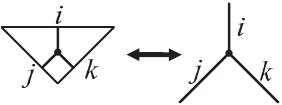}
\caption{Duality between a 2-simplex and a trivalent vertex.}\label{dual2d}
\end{figure}
\begin{figure}[h]
\centering
\includegraphics[scale=0.5]{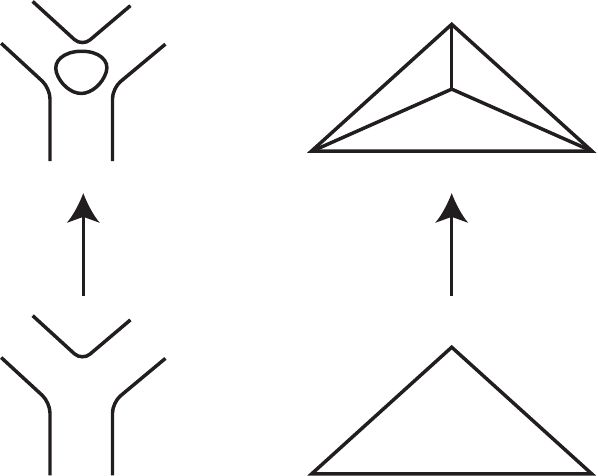} \quad \includegraphics[scale=0.5]{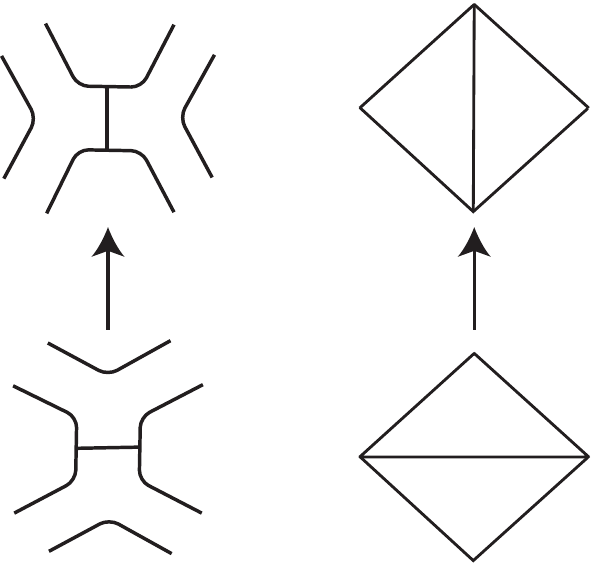} \quad \includegraphics[scale=0.5]{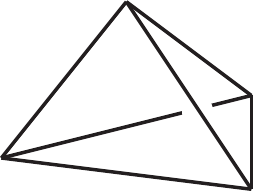}

{\small \hspace*{5mm} (a) \hspace{31mm} (b) \hspace{19mm} (c)}
  \caption{Pachner moves. (a) $1\rightarrow 3$. (b) $2\rightarrow 2$.}
  \label{pachner}
\end{figure}

Trivalent spin networks are the dual skeletons of triangulations of 2D surfaces (Fig.~\ref{dual2d}), in which a node (an edge) is dual to a 2-simplex, i.e., a triangle (a side of the triangle). This is consistent with 2D because the vertices of trivalent spin networks have zero 3-volume. Hence, the evolution moves of trivalent spin networks are dual to the Pachner moves~\cite{Pachner} that relate triangulations of the same surface (Fig.~\ref{pachner}). This topological interpretation indicates that summing over histories of the evolution of certain fundamental building blocks produces a quantum spacetime. That is, one can build $(n+1)$D spacetime from the evolution of $n$-valent spin networks. This picture is partly implemented in SF models and fully implemented in another formulation of quantum gravity, Group Field Theories (GFT)~\cite{Oriti2005,OritiThesis,Oriti2007}.

A subtlety exists, however. A spin network with structureless edges and nodes contains less information than an exact dual of a 2D simplicial triangulation, in which two triangles can be glued along a side in two opposite ways. To remedy this, trivalent vertices and their edges should be framed to disks and ribbons respectively (Fig.~\ref{pachner}(a)). We refer to these (embedded) framed spin networks as {\bf (braided) ribbon networks}\footnote{For embedded spin networks, the duality is in general only local, i.e., restricted to a single node. This restriction is unnecessary in 2D because a braided ribbon network is dual to a topological manifold globally~\cite{depietri}.}. We also name the evolution moves on (braided) ribbon networks the {\bf adapted Pachner moves} (Fig.~\ref{pachner}\footnote{This f\/igure is adopted
from \cite{Isabeau2008} with the author's permission.}).
The criteria for a legal 2D Pachner move is that the triangles before and after the move bound a 3D tetrahedron (Fig.~\ref{pachner}). Interestingly, Major and Smolin~\cite{Major1995a, MajorThesis,Major1995, Smolin2002cc} suggested that when LQG contains a~non-zero cosmological constant, the
corresponding spin networks become framed: edges and vertices become
ribbons and disks in the trivalent case, and become tubes and
spheres in the 4-valent case. We shall loosely refer to the (embedded) framed 4-valent spin networks as 4-valent (braided) ribbon networks. Likewise, 4-valent ribbon networks evolve under a set of 4-valent adapted dual Pachner moves (Section~\ref{subSecEvol}). Moreover, labels of framed spin networks turn out to be representations of the quantum group, e.g., $SU_q(2)$. In GFTs, the spin networks arise as framed but unembedded.

Our third ansatz requires the embedded 4-valent spin networks to evolve under adapted dual Pachner moves. These moves are adapted and are thus dif\/ferent
from those in SF models. In many studies of spin foam models,
the spin networks and spin foam histories are taken to be abstract,
or unembedded. Our results do not directly apply to these models, as
the topological structures our results concern arise from the
embedding of the graphs in a topological three manifold. But neither
do such models give dynamics for states of loop quantum gravity,
which are embedded. A question therefore is how to build spin-foam like dynamics for embedded spin networks. The adapted dual Pachner moves to be def\/ined in Section~\ref{subSecEvol} for the 4-valent braided ribbon networks will give an answer.

\subsection{Three Ansatzes}\label{subsec3A}

All of the above encourages a unif\/ication scheme in which matter emerges as topological excitations of the braided ribbon networks. We thus posit that braided ribbon networks are the most fundamental entities of nature\footnote{Since all the information due to embedding can be characterized purely combinatorially, to be pointed out later, embedded spin networks (or combinatorial spin networks with embedding data) are more general than the unembedded ones~\cite{Kauffman1992,Kauffman2007}.}, beyond LQG and SF Models and actually in accordance with Penrose's original proposal of spin networks, with, however, a~great deal of generalization. More precisely, the unif\/ication scheme and its results we have obtained are based on three ansatzes:
\begin{enumerate}\itemsep=0pt
\item Spacetime is pre-geometric and discrete at the fundamental scale.
\item The discrete space is a superposition of basis states represented by braided ribbon networks.
\item The braided ribbon networks evolve under a set of local moves.
\end{enumerate}
These ansatzes are independent of the spacetime dimension. Here, we consider only $(3+1)$-dimensional spacetime, consistent with the observable universe\footnote{Why our spacetime is (3+1)-dimensional is a big open issue in physics and philosophy. The earliest reasonable argument, due to Ehrenfest in 1917~\cite{Ehrenfest1917}, was that atoms are unstable unless in 3D. Recently, anthropic arguments also arose. Nevertheless, all these arguments sound {\it a posteriori}, and a theory that naturally gives rise to our $(3+1)$ spacetime is still missing.}. The braided ribbon networks are in general graced with spin network labels, which are otherwise removed in this article because our results obtained so far do not depend on them. The set of local moves include only the adapted dual Pachner moves; however, it may extend to incorporate other moves in future. We hope that classical spacetime would exist as certain limit of the pre-geometric history of the evolution of these graphs.

\section{The trivalent scheme}\label{sec3V}

The trivalent scheme is the most natural case in which to embed the helon model~\cite{Bilson-Thompson2005}, which provides a mapping between braided network states and the fermions and bosons of the SM. Here we will brief\/ly discuss the helon model in terms of abstract braided structures, and how those structures may be mapped to particle states and quantities such as hypercharge, baryon number and lepton number. We will then discuss how these braids may be characterised by appropriately chosen topological invariants. At the end of this section we will discuss how such braided structures may be embedded in framed spin networks. The possibility of a unif\/ied treatment of trivalent and tetravalent networks is discussed in Section~\ref{sec:3v4vcorrespondence}.

\subsection{The helon model}
\label{sec:helon_model}
In the helon model, the subcomponents of SM particles occurring in certain preon models are replaced by a framed braid on three strands. The strands are joined to two surfaces of non-zero size (which we may think of as one disk at each end), and we will suppose there is a way of distinguishing these end surfaces so that the braids have a ``top'' and ``bottom'' (as shown in the left of Fig.~\ref{fig:generators}).
For brevity, let us refer to these end surfaces as ``caps''. Such a braid (all three strands and the two caps) constitutes a two-dimensional surface 
and we will immediately restrict our attention to 
orientable surfaces. The three strands between the caps can in general be distinguished by their relative crossings, and it becomes meaningful to speak of the f\/irst, second, and third strand. An entire braid on three strands will represent a single type of fermion or boson.

The individual strands in a braid can carry twists, as mentioned above, and we identify right-handed and left-handed twists as positive and negative electric charges. This is the basis of the name `helon' (evoking the image of a helix) for a single strand. The requirement that we consider only orientable surfaces restricts us to twists that are multiples of $2\pi$, so let us interpret a twist through $\pm 2\pi$ on any helon (that is, any strand) as an electric charge of $\pm e/3$. Twists through $\pm 4\pi$, $\pm 6\pi$, and so on will not be considered, as we shall see below that extra twists can be regarded as equivalent to crossings. Besides the helons carrying twist through $-2\pi$ and $+2\pi$, there is a third type to consider, carrying no twist. We will denote the three types of helons as~$H_-$,~$H_+$, and~$H_0$ respectively.

Adapting a scheme originally devised by Harari~\cite{preon1979Harari} and Shupe~\cite{preon1979Shupe}, we construct braids composed of three $H_+$s (corresponding in electric charge to positrons), three $H_-$s (corresponding to electrons), a single $H_+$ and two $H_0$s (corresponding to anti-down quarks), a single $H_-$ and two $H_0$s (corresponding to down quarks), a single $H_0$ and two $H_+$s (corresponding to up quarks), a single $H_0$ and two $H_-$s (corresponding to anti-up quarks), and three $H_0$s (corresponding to neutrinos). This scheme reproduces the fermions of the f\/irst generation of the SM, with no extra particles. Braids consisting of a mix of $H_+$ and $H_-$ helons are not allowed when constructing fermions (but are in fact used to construct the gluons). We identify the permutations of braids containing two helons of one type, and one of another (e.g.~$H_+ H_+ H_0$) with the three colour charges of QCD, and write the helons in ordered triplets for convenience (this is, of course, simply notation). The quarks are then as follows (subscripts denote colour):
\begin{center}
\begin{tabular}{llllll}
  $H_+H_+H_0$ & \hspace{-0.4mm}($u_B$)\hspace{1.2mm}
     & $H_+H_0H_+$ & \hspace{-0.4mm}($u_G$) \hspace{1.2mm} & $H_0H_+H_+$ & \hspace{-0.4mm}($u_R$)\\
  $H_0H_0H_+$ & ($\overline{d_B}$)
     & $H_0H_+H_0$ & ($\overline{d_G}$)
                       & $H_+H_0H_0$ & ($\overline{d_R}$)\\
  $H_-H_-H_0$ & ($\overline{u_B}$)
     & $H_-H_0H_-$ & ($\overline{u_G}$)
                       & $H_0H_-H_-$ & ($\overline{u_R}$)\\
  $H_0H_0H_-$ & ($d_B$)
     & $H_0H_-H_0$ & ($d_G$) & $H_-H_0H_0$ & ($d_R$)\\
\end{tabular}
\end{center}
while the leptons are
\begin{center}
\begin{tabular}{llllll}
$H_+H_+H_+$ & \hspace{-0.4mm}$(e^+)$ \hspace{1.2mm}
   & $H_0H_0H_0$ & ($\nu_e$) \hspace{1.2mm}
     & $H_-H_-H_-$ & ($e^-$).
\end{tabular}
\end{center}

Note that in this scheme we have identif\/ied neutrinos, but not anti-neutrinos.
This has occurred because while the $H_-$ may be regarded as the anti-partner to the~$H_+$, there is no anti-partner to the~$H_0$ helon. This apparent problem will be turned to our advantage in Section~\ref{sec:quantum_numbers}.

\subsection{Topological invariants of trivalent braids}\label{sec:top_inv}

\begin{figure}[h]
\centering
  \includegraphics[scale=0.4]{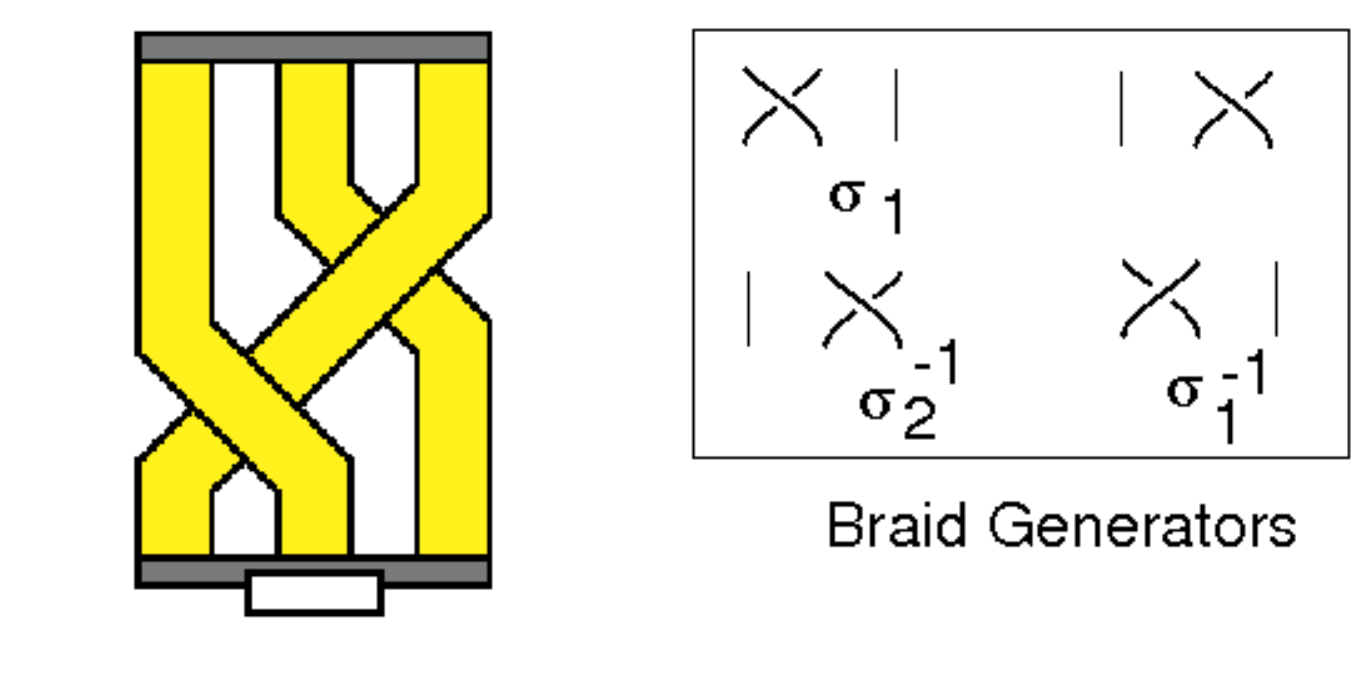}
  \caption{A braid, with the bottom cap denoted by a rectangular block (left) and the generators of the braid group on three strands (right).}
  \label{fig:generators}
\end{figure}

The braids on three strands may be characterised by an invariant called the {\em pure twist number}, f\/irst described in~\cite{Sundance2009}. This is a triple of real numbers which count the twist remaining on each strand when the braid is deformed such that all crossings are removed. This is possible because any braid on $n$ strands can be written as a product of the generators $\sigma_1,\ldots,\sigma_{n-1}$, and their inverses, where $\sigma_i$ crosses the $i^\mathrm{th}$ strand in front of the $(i+1)^\mathrm{th}$ strand, and $\sigma^{-1}_i$ crosses the $i^\mathrm{th}$ strand behind the $(i+1)^\mathrm{th}$ strand. The sequence of $\sigma$ factors def\/ining a braid is called its {\em braid word}. Clearly, in the case of braids on three strands we are only concerned with~$\sigma_1$,~$\sigma_2$ and their inverses. The generators induce permutations of the strand ordering. The generator $\sigma_1$ induces the permutation $P_{1,2}$ (that is, it swaps the $1^\mathrm{st}$ and $2^\mathrm{nd}$ strands), while the generator $\sigma_2$ induces the permutation $P_{2,3}$. Notice also that the same permutation is induced by a generator or its inverse, $\sigma^{-1}_i$. Therefore the generators contain more information than the permutations~-- in particular the direction of the crossing is specif\/ied by the generators (as shown in Fig.~\ref{fig:generators}).
\begin{figure}[h]
\centering
  \includegraphics[scale=0.2]{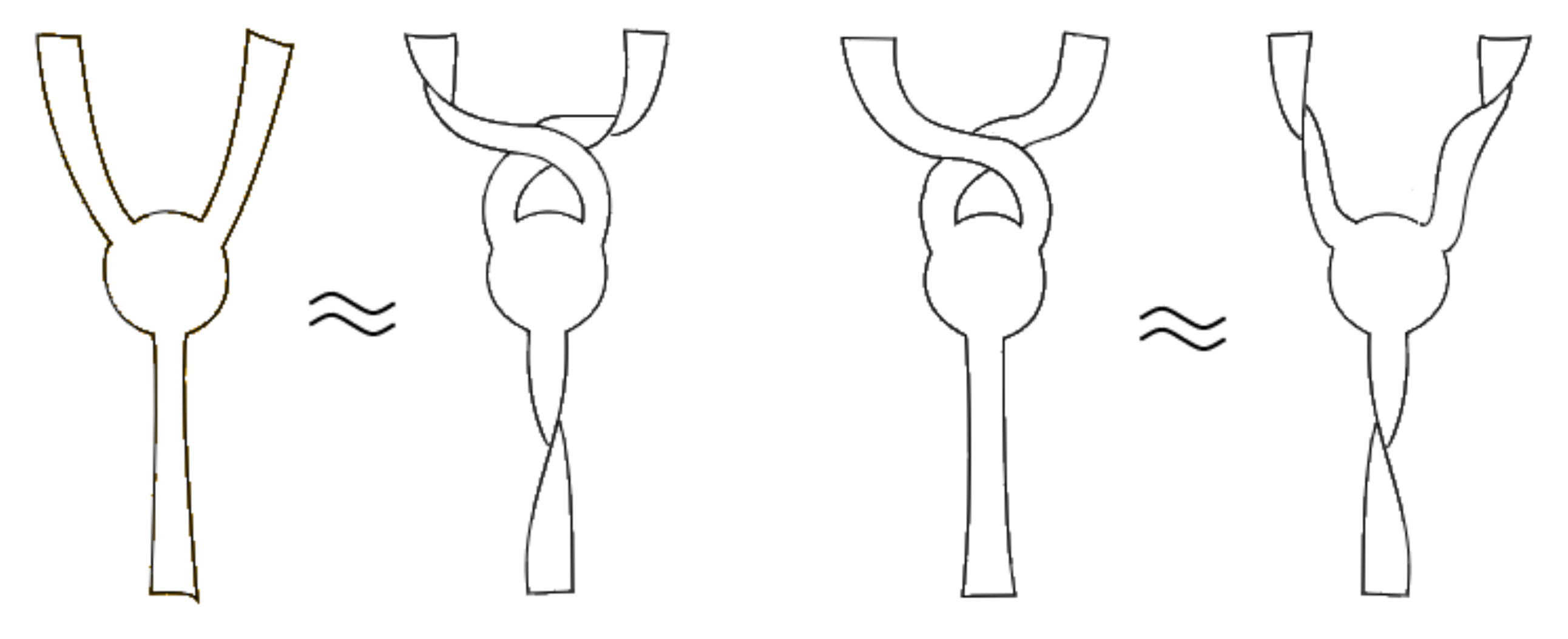}
  \caption{Swapping crossings for twist.}
  \label{fig:trinion_flip}
\end{figure}
\begin{figure}[h!]
\centering
  \includegraphics[scale=0.2]{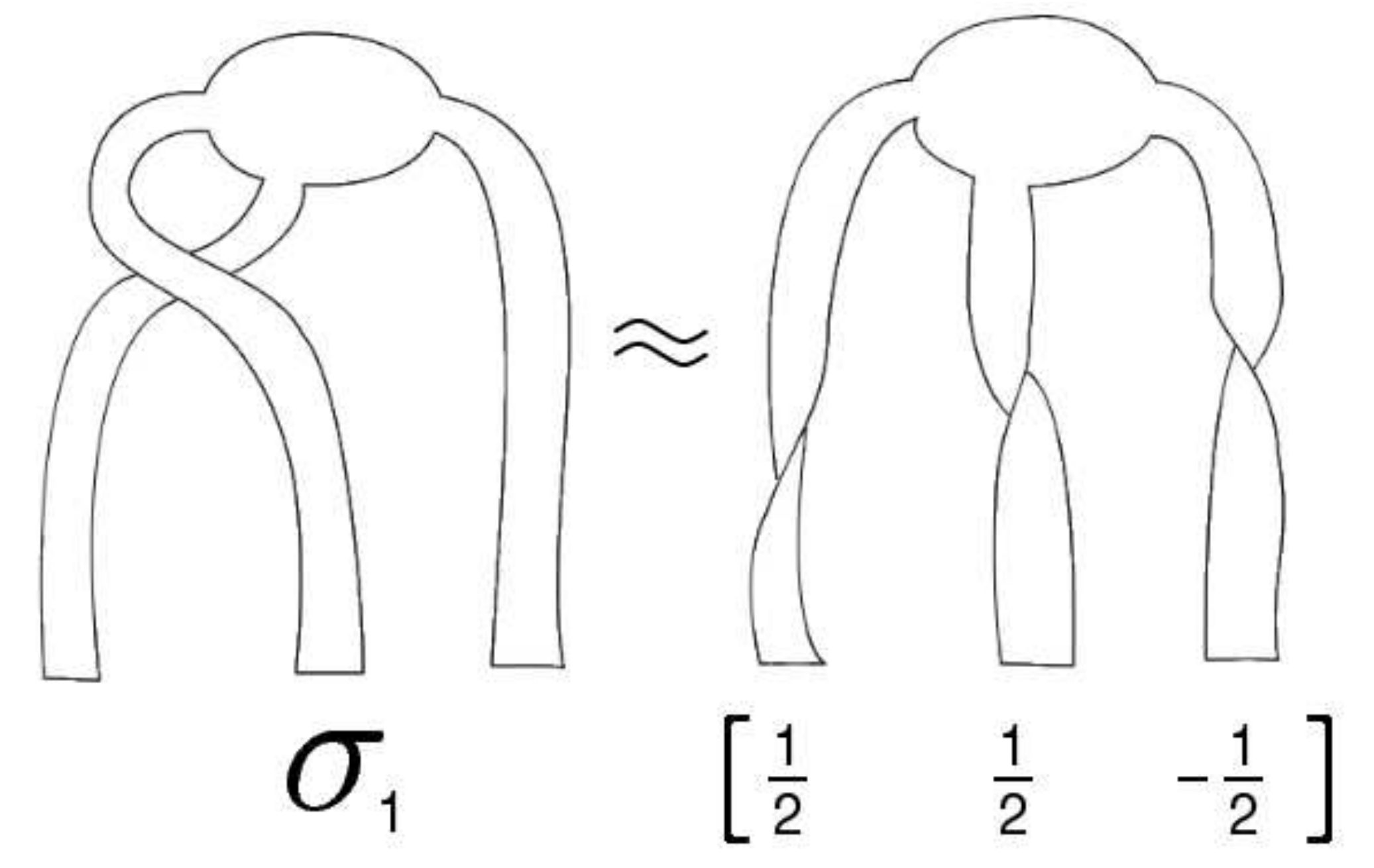}
  \caption{Converting a braid generator into twists.}
  \label{fig:GenFlip}
\end{figure}

It is convenient to def\/ine a standardised form for framed braids in which all the twist is isotoped to the top of the braid. Then we can write $[r,s,t]{\cal B}$ where ${\cal B}$ is an ordinary braid word and $[r,s,t]$ is a triple of multiples of half-integers which catalogue the twists in the ribbons. We shall call this triple of numbers the {\em twist-word}. Thus a framed braid on three strands is completely specif\/ied by the twist word and the braid word. Since these braids are taken to exist on networks embedded in a manifold of three dimensions, we also allow the braids to be deformed such that the top end of the braid is f\/lipped over (ef\/fectively feeding it through the strands). This allows us to undo crossings and hence simplify the braid structure, as illustrated in Fig.~\ref{fig:trinion_flip} where we show how a disk with three untwisted ribbons emerging from it (but with two ribbons crossed), can be converted into a disk with uncrossed ribbons and oppositely-directed half-twists in the upper and lower ribbons by f\/lipping the disk over (in the illustration, a negative half-twist in the lower ribbon and positive half-twists in the upper ribbons). In Fig.~\ref{fig:GenFlip}, we show the same process performed on a disk whose (crossed) upper ribbons have been bent downwards to lie besides and to the left of the (initially) lower ribbon. This conf\/iguration is nothing other than a framed braid on three strands corresponding to the generator~$\sigma_1$ (with cap at the bottom omitted). Keeping the ends of the ribbons f\/ixed as before and f\/lipping over the cap so as to remove the crossings now results in three unbraided (i.e.\ trivially braided) strands, with a~positive half-twist on the leftmost strand, a  positive half-twist on the middle strand, and a~negative half-twist on the rightmost strand. Hence the associated twist-word is $[\frac{1}{2},\frac{1}{2},-\frac{1}{2}]$. This illustrates that in the case of braids on three strands, each of the crossing generators can be isotoped to uncrossed strands bearing half-integer twists. By variously bending the top two ribbons down to the right of the bottom ribbon, and/or taking mirror images, and f\/lipping the cap at the top of the braid appropriately we can determine that the generators may be exchanged for twists according to the pattern:
\begin{gather}
\sigma_1      \rightarrow \left[\frac{1}{2},  \frac{1}{2}, -\frac{1}{2}\right], \qquad
\sigma_1^{-1}\rightarrow\left[-\frac{1}{2},-\frac{1}{2},\frac{1}{2}\right],\nonumber\\
\sigma_2\rightarrow\left[-\frac{1}{2},\frac{1}{2}, \frac{1}{2}\right], \qquad
\sigma_2^{-1} \rightarrow \left[\frac{1}{2},-\frac{1}{2},-\frac{1}{2}\right].
\label{eq:sigmaToTwist}
\end{gather}

All braids on three strands can be built up as products of these generators. It should therefore be clear to the reader that we may entirely eliminate the crossings from a braid on three strands. When we do so we uncross the strands one generator at a time (hence permuting them by the permutation~$P_{\sigma_i}$ associated with the crossing $\sigma_i$ being eliminated) and introduce the twists indicated in equation~(\ref{eq:sigmaToTwist}). In general, this means that we iterate the process
\begin{gather}
[a_1, a_2, a_3][b_1, b_2, b_3]\sigma_i\sigma_j\ldots\sigma_m
\rightarrow [a_1+b_1, a_2+b_2, a_3+b_3]\sigma_i\sigma_j\cdots\sigma_m \nonumber  \\
\qquad{} \rightarrow P_{\sigma_i}([a_1+b_1, a_2+b_2, a_3+b_3])[x,y,z]\sigma_j\cdots\sigma_m,
\label{eq:sigmaTwistIteration}
\end{gather}
where $[x,y,z]$ is the twist-word associated to $\sigma_i$ (as listed in equation~(\ref{eq:sigmaToTwist}), when $i$ is specif\/ied), and $P_{\sigma_i}([a, b, c])$ is the permutation of $[a,b,c]$ arising from $\sigma_i$. The process given in equation~(\ref{eq:sigmaTwistIteration}) is iterated until the braid word becomes the identity.

We shall refer to the form of a braid in which all the crossings have been exchanged for twists as the {\em pure twist form}. The list of three numbers which characterise the twists on the strands in the pure twist form will be referred to as the {\em pure twist-word}. The pure twist-word is of interest because it is a topological invariant (since it is obtained when a braid is reduced to a particular simple form i.e.\ all crossings removed). The iteration of equation~(\ref{eq:sigmaTwistIteration}) therefore describes an algorithm for calculating the pure twist-word of any three-strand braid. This algorithm was f\/irst described in~\cite{Sundance2009}, where further details and examples can be found.

One criticism that has been levelled at this research program is that braids appear somewhat ad-hoc and unnatural, however it follows from the discussion above that the use of braids in the helon model is a convenience, and that each such braid can be related to a topological invariant which is independent of the way the structure is drawn. When we speak of a particular braid corresponding to a type of fermion, we are therefore simply referring to an equivalence class of topological structures and using one distinctive member of that equivalence class to refer to the entire class. We will see shortly how such general topological structures may be embedded in trivalent spin networks.

It should be noted in the interests of clarity, however, that it is too early to assign specif\/ic choices of braiding structure (and hence specif\/ic equivalence classes) to any SM particles. Although the electric and colour charges are determined by the twist assignments listed above as permutations of~$H_+$s,~$H_-$s, and~$H_0$s, the task of matching topological characteristics to proper\-ties such as spin, generation, hypercharge, etc.\ is still a work-in-progress. This is discussed below. Furthermore, a way of modelling transitions between generations using braids (and hence describing Cabbibo-mixing and neutrino oscillations) has not yet been found. In light of this, it seems premature to assign def\/inite topological structures to the gauge bosons, and to the fermions of each generation~-- although such an assignment is a def\/inite goal of this research program. The braids shown in Figs.~\ref{fig:generators} and~\ref{fig:all_fermion} represent our best attempts at the present time of matching particle properties to the braid multiplication. Their role in the wider models discussed here is yet to be fully determined.

\subsection{Quantum numbers of particle states}\label{sec:quantum_numbers}

\begin{figure}[t]
\centering
\includegraphics[height=35mm]{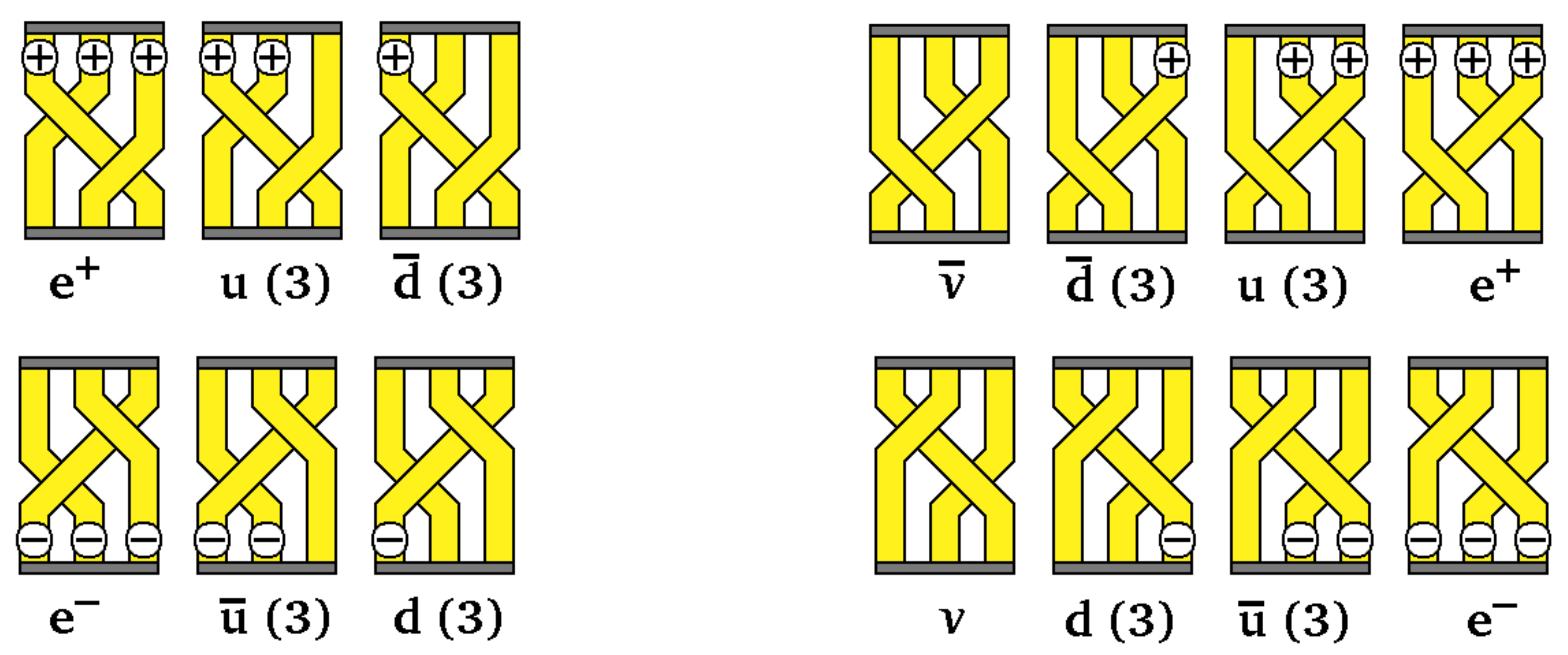}
\caption{The fermions formed by adding zero, one, two or three charges to a neutral
braid. Charged fermions come in two handedness states each, while $\nu$ and
$\overline{\nu}$ come in only one each. (3)~denotes that there are three possible
permutations, identif\/ied as the quark colours. The bands at top and bottom
represent the nodes and connections to the external network (``the rest of the Universe'').}
\label{fig:all_fermion}
\end{figure}

Given the discussion above, it is clear that the node at the ``top'' of a braid def\/ines an internal direction which is unaf\/fected by rotating the diagram of a braid. The pure twist form is found by working through each crossing in turn, starting with the crossing closest to the node at the ``top''. Therefore rotating a diagram of a braid on a page does not change which equivalence class that braid belongs to (and hence we may always perform such a rotation without compromising the validity or usefulness of our model). Let us then pick a certain braid with untwisted strands (i.e.\ composed entirely of $H_0$s) to act as a basic structure for one generation of fermions. Unlike rotation, taking the mirror image of such a braid will produce a member of a dif\/ferent topological equivalence class, in general. Such a braid and its mirror image can be regarded as left-handed and right-handed fermions. By adding one, two, or three twisted strands ($H_+$s or $H_-$s, but not both at the same time), we construct left-handed and right-handed fermions with overall positive and negative charge. This is illustrated in Fig.~\ref{fig:all_fermion}. Notice that for any given non-zero magnitude of charge, there are four states (left-handed and right-handed particle, and left-handed and right-handed antiparticle, where positive and negative electric charge distinguish between particles and antiparticles), but for the case of zero charge there exist only two states (left-handed and right-handed). We identify these with the neutral left-handed fermion (neutrino) and right-handed antifermion (anti-neutrino).

The helon model allows us to describe electric charge and colour charge in terms of the topological structure of braids, but we can similarly describe hypercharge and isospin. We begin by assigning a scalar quantity $\beta$ to braids, such that $\beta=+1$ for the braids on the top row of Fig.~\ref{fig:all_fermion}, and $\beta=-1$ to the braids on the bottom row. This ef\/fectively distinguishes between fermions with a net positive and net negative charge (and establishes a def\/inition of the equivalent quantity for the neutrinos). Of course $\beta$ provides only a very crude distinction.  To rectify this shortcoming we will def\/ine a new quantity, given by one-third the number of ``more positive'' helons, minus one-third the number of ``less positive'' helons. We shall denote this quantity by the symbol $\Omega$. To clarify, $H_+$ helons are considered ``more positive'' than $H_0$ helons, which are ``more positive'' than $H_-$ helons. If $N(H_+)$ is the number of $H_+$ helons, $N(H_0)$ the number of~$H_0$s and~$N(H_-)$ the number of~$H_-$s within a triplet, and remembering that~$H_+$ and~$H_-$ helons never occur within the same braided triplet, we may write
\begin{gather*}
\Omega=\beta\left(\frac{1}{3}N(H_+)+\frac{1}{3}N(H_-)-\frac{1}{3}N(H_0)\right).
\end{gather*}
Hence we have $\Omega=+1$ for the $e^+$, $\Omega=+1/3$ for the $u$, $\Omega=-1/3$
for the $\overline{d}$, and $\Omega=-1$ for the anti-neutrino. For the electron,
anti-up, down, and neutrino the signs are reversed. With this def\/inition,
noting that $N(H_0)=3-(N(H_+)+N(H_-))$ and the total electric charge of a~fermion
is given by
\begin{gather*}
Q = \beta\left(\frac{1}{3}N(H_+)+\frac{1}{3}N(H_-)\right),
\end{gather*}
it is easy to show that
\begin{gather}
Q = \frac{1}{2}\left(\beta + \Omega\right).
\label{eq:QbetaOmega}
\end{gather}
For the quarks and anti-quarks $\Omega$ reproduces the SM values of strong hypercharge, while for
the leptons~$\beta$ reproduces the SM values of weak hypercharge. We also observe
that for quarks~$\beta/2$ reproduces the values of the third component of strong
isospin, while for leptons $\Omega/2$ reproduces the values of the third component
of weak isospin (in short, the roles of~$\beta$ and~$\Omega$ as isospin and
hypercharge are reversed for quarks and leptons). With these correspondences the
Gell-Mann--Nishijima relation $Q=I_3+Y/2$ for quarks may trivially be derived from
equation~(\ref{eq:QbetaOmega}). This construction does not distinguish between left-handed and right-handed fermions, and so it does not match all values of weak isospin and hypercharge in the SM. The reader should remember that the values of isospin and hypercharge in the SM are assigned in an entirely {\em ad hoc} manner, to ref\/lect the observed asymmetry of the weak interaction, while in the helon model these values are constructed. It is possible that with further work another set of topologically-def\/ined quantities may be found which will enable the helon model to describe the asymmetry of the weak interaction.

\subsection{Interactions and embedding in trivalent networks}\label{sec:embed_trivalent}

\begin{figure}[h]
\centering
  \includegraphics[scale=0.25]{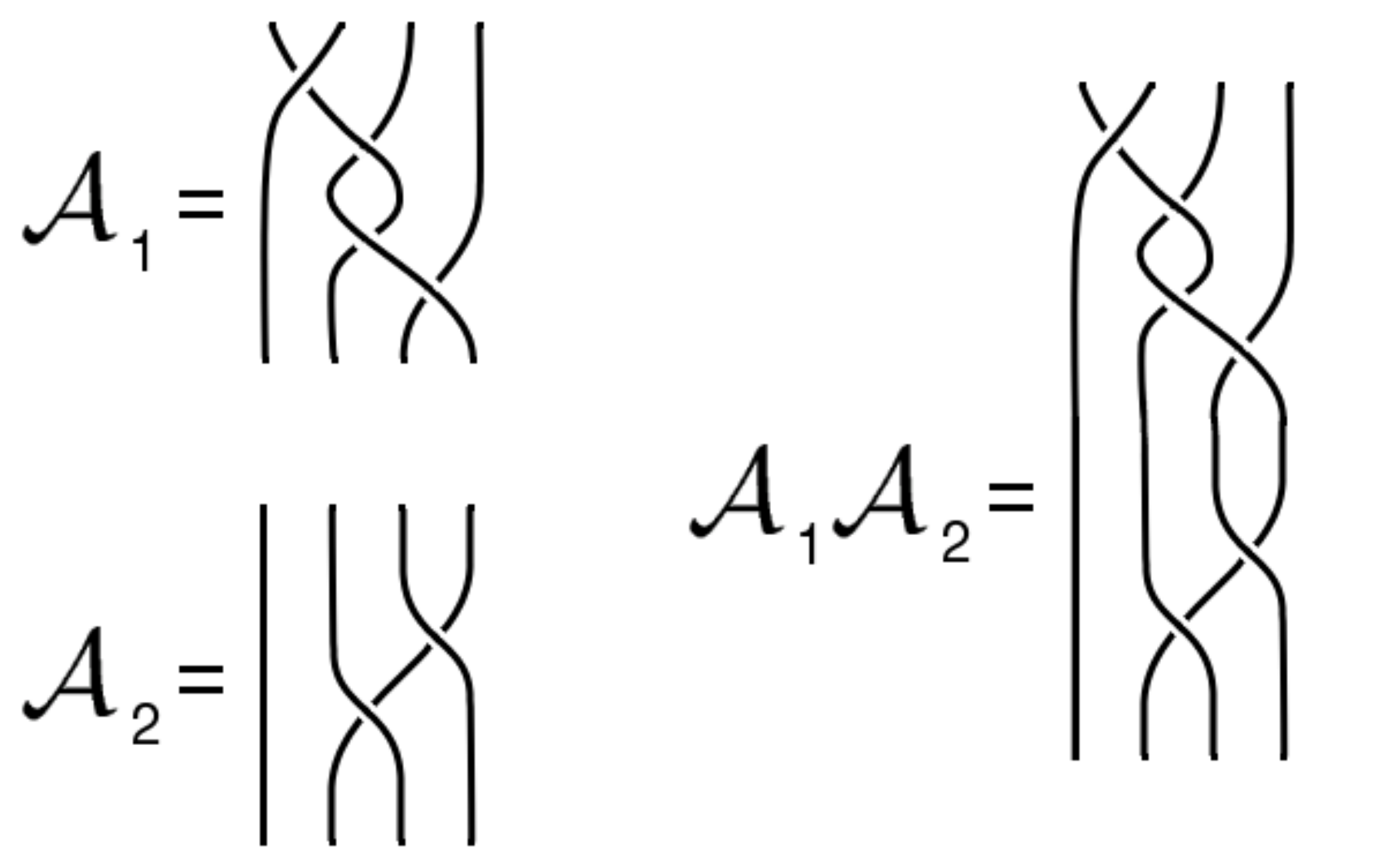}
  \caption{The product of two braids.}
  \label{fig:braid_product}
\end{figure}

Although the helon model does not provide a dynamical framework, it is possible to represent the electroweak interactions in a simple manner, by forming the braid product of several braids. The product of two braid diagrams is accomplished by adjoining the strands of the second braid to the corresponding strands of the f\/irst braid, as in Fig.~\ref{fig:braid_product}. The product of two braids, ${\cal A}_1$ and~${\cal A}_2$, is therefore a single braid, the braid word of which is the concatenation of the generators in the braid words of~${\cal A}_1$ and~${\cal A}_2$. It therefore follows that if~${\cal A}_2$ is the top-to-bottom mirror image of ${\cal A}_1$ the product of ${\cal A}_1$ and ${\cal A}_2$ will contain no crossings. The reader can easily check that this is true for any braid in the top row of Fig.~\ref{fig:all_fermion}, and the corresponding braid in the bottom row. Such a process suggests particle-antiparticle annihilation. More generally, when the product of two braids is formed, and then decomposed into several braids, the twists on the strands may be shuf\/f\/led so that the outgoing braids are dif\/ferent from the incoming braids. In this way an interaction such as $u + e^- \rightarrow d + \nu_e$ may be modelled, with the structure of the intermediate braid product suggesting the structure of a boson. A more detailed discussion of how particle interactions can be modelled using braid products is given in~\cite{Bilson-Thompson2005}.

\begin{figure}[h]
\centering
  \includegraphics[scale=0.25]{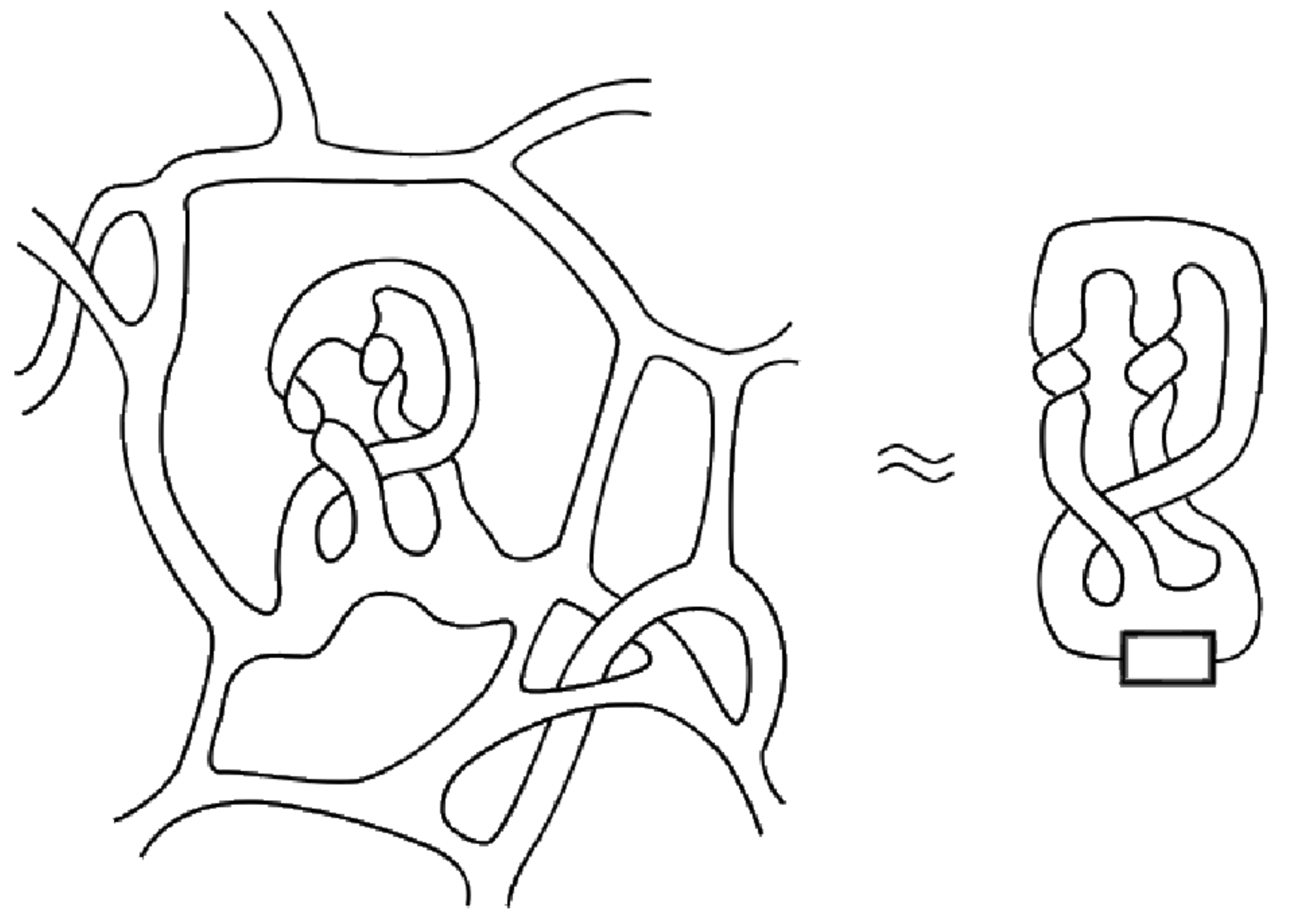}
  \caption{A framed braid on three strands (right), equivalent to a braided substructure in a network (left).}
  \label{fig:braid_in_network}
\end{figure}
\begin{figure}[h]
\centering
  \includegraphics[scale=0.6]{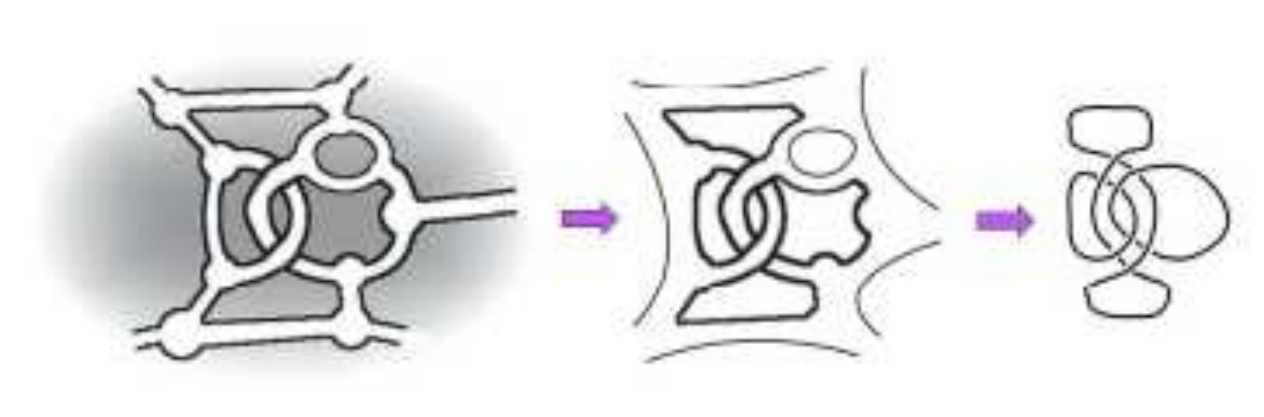}
  \caption{Finding the reduced link of a topological structure within a ribbon network.}
  \label{fig:Reducedlink}
\end{figure}

Having established the possibility of a mapping between braids and SM particles, we now turn to the matter of embedding the helon model in trivalent spin networks. In this scheme, a braid occurs as a single node from which three strands (helons) emerge. That is, the cap at the top of a braid is identif\/ied as a trivalent node. The strands braid around each other, and then join to the rest of the network at three other nodes (see Fig.~\ref{fig:braid_in_network}). It was shown in~\cite{Bilson-Thompson2006} that such embedded braids can be characterised by the linking of the edges of the braid. If we take a diagram of a topological substructure in a trivalent network, and trace along the left and right edges of each strand~-- discarding any unlinked closed loops~-- we obtain a diagram of a link corresponding to that structure. This construction is illustrated in Fig.~\ref{fig:Reducedlink}. The link obtained may in general be simplif\/ied by applying a series of Reidemeister moves to the diagram, to obtain a ``reduced link''. The reduced link is an invariant of a braid, so that it does not change no matter how much the network of nodes and strands is deformed within the manifold in which it is embedded. Likewise, the standard trivalent evolution moves (1-3 expansion move and 2-2 exchange move) do not change the reduced link. The def\/inition of the reduced link provides another way of seeing that, as claimed above, rotating a diagram of a braid on a page does not change its equivalence class, since a rotated reduced link is equivalent to an unrotated reduced link, under the Reidemeister moves.

It is here that the dif\/f\/iculty regarding interactions occurs. In order for any process analogous to the braid product to take place, two braids must be able to combine and then split apart. If the outgoing particle states are to be dif\/ferent from the incoming particle states, then the associated reduced links should also be dif\/ferent. But as noted above, it was shown that this is not possible under the 1-3 and 2-2 evolution moves. While this ensures that individual leptons and quarks will not decay (since the associated reduced link cannot be changed), it also appears to prohibit any non-trivial particle interactions. It is possible that this issue could be addressed by postulating a new evolution move. However in Section~\ref{sec:3v4vcorrespondence} we wish to explore an alternative approach, in which we construct a correspondence between braided networks in the trivalent and tetravalent cases.

\section{The 4-valent scheme}\label{sec4V}
Having seen the results and limitations of the trivalent scheme, we shall move on to the 4-valent scheme, which, although begun as an extension to the trivalent one, turned out to be a rich, fully dynamical theory of braids of 4-valent braided ribbon networks. The 4-valent scheme has been studied and cast in two parallel but complementary formalisms, namely the graphic and the algebraic formalisms. While the former of\/fers a more intuitive picture, the latter provides a more convenient playground for theorem-proving and investigating the properties of braids and their dynamics. Consequently, in this section, we shall adopt the graphic formalism for illustrative purposes only but the algebraic formalism extensively.

\subsection{4-valent braided ribbon networks}\label{subSecGraphNotation}

\begin{figure}[h]
\centering
\includegraphics[scale=0.8]{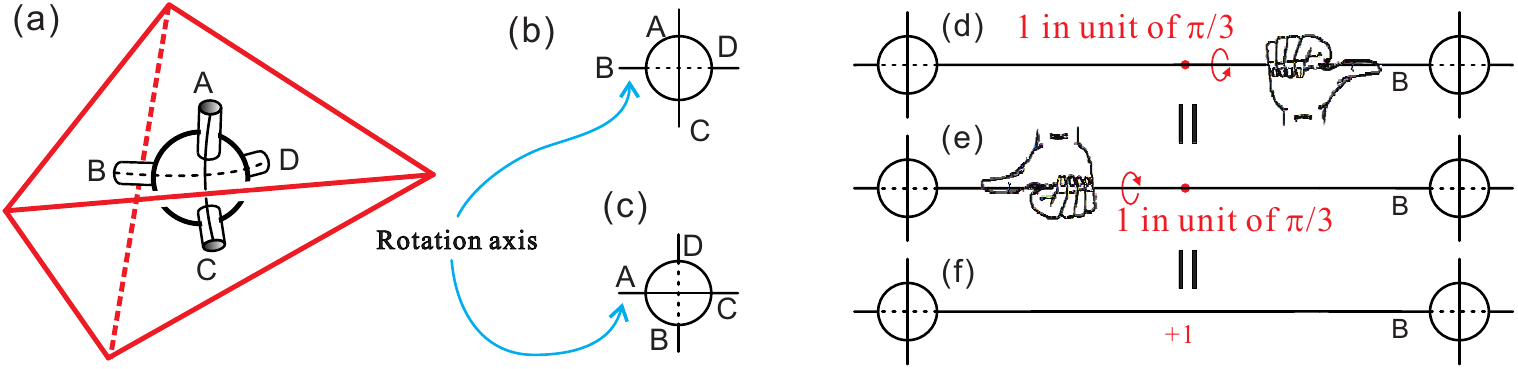}
\caption{(a) is a tetrahedron and its dual node; a tube has three racing stripes. In a diagram, a node
can have two states, (b)~$\oplus$ and~(c)~$\ominus$. (f)~shows an unambiguous label of the two right-handed $\pi/3$ twists in~(d) and~(e).} \label{notation}
\end{figure}

4-valent braided ribbon networks are framed 4-valent spin networks embedded in~$\mathbb{R}^3$. The local duality between a node of a 4-valent braided ribbon network and a tetrahedron allows us to represent a node by a 2-sphere with four circular punctures and its edges by tubes that are welded at the punctures. This is depicted in Fig.~\ref{notation}. We assume that each node is rigid and non-degenerate, such that it can only be translated and rotated, its punctures where its edges are attached are f\/ixed, and no more than two edges of a node are co-planar\footnote{The reader may wonder why our discussion of braided ribbon graphs relies so much on the classical structure of 3D space. As the model is not assumed to take place at some large scale where a 3D tangent space has emerged, this is an important epistemological point and can be answered in the following way: One should regard these networks as holding combinatorial conditions that happen to be expressed in the language of three-dimensional space. It would make the statement of the model overly complex to state such combinatorial conditions abstractly entirely in terms of graphs and orderings, but this can be done in principle. We realize that this situation is not entirely satisfactory.}. Because a tubular edge is dual to a triangular face of a tetrahedron, a tube implicitly carries three racing stripes that record the twist of a tube, which dictates how two tetrahedra are glued on a common face. Section~\ref{secJon} gives a further discussion of the racing stripes.

In a projection, we simplify the tube-sphere notation in Fig.~\ref{notation}(a) to a circle-line notation ((b) or (c)), in which solid lines piercing through the circle represent tubes that are above in the 3D notation, while a~dashed line
represents two lines as the tubes that are under. There is no information loss in doing so because one can always arrange a node in the either the state in Fig.~\ref{notation}(b) or (c) by isotopy before projecting it. Owing to the local duality between a node and a tetrahedron and the symmetry on the latter, if we grab an edge of a node, the other three edges are still on an equal footing, inducing a rotation symmetry w.r.t.\ the edge being grabbed (Fig.~\ref{notation}(b) \& (c)), which will be explained shortly. Therefore, in a projection one can assign states to a node w.r.t.\ its rotation axis. If the rotation axis is an edge in the back (front), the node is in state $\oplus$ ($\ominus$) and is called a $\oplus$-node ($\ominus$-node), as in Fig.~\ref{notation}(b) ((c)).

Nonetheless, if the three edges other than the rotation axis of a node must be distinguished, e.g., to be seen shortly in the case of the nodes of a braid, a degeneracy of each state of the node arises. Consider Fig.~\ref{notation}(b), the node is in $\oplus$ w.r.t.\ edge~$B$; rotating the node about $B$ produces permutations of the other edges and puts them in dif\/ferent conf\/igurations. Ignoring the twists and crossings of the edges that are created by rotations, which will be our next topic, it takes a full rotation for the node to roll back to its original conf\/iguration. In conf\/igurations $(A,\ D,\ C)$\footnote{This notation records the order of the three edges from top to bottom w.r.t.\ the rotation axis (see Fig.~\ref{notation}(b)).}, $(D,\ C,\ A)$, and $(C,\ A,\ D)$, the node is in $\oplus$ w.r.t.~$B$, whereas in $(A,\ C,\ D)$, $(C,\ D,\ A)$, and $(D,\ A,\ C)$, the node f\/lips to $\ominus$. That is, each state has a 3-fold degeneracy; or in other words, each state is a triplet. The six sub-states in total record the full conf\/iguration of the node w.r.t.\ the rotation axis.

If we denote a full rotation by $2\pi$, then the amount of rotation keeping a node within a state triplet is $2\pi/3$; however, a $\pi/3$ rotation causes to jump back and forth between two state triplets. Note again that these types of rotations are not the ones with a rigid metric but rather discrete and purely topological. Details of these rotations will be studied in Section~\ref{subsubSecRot}.

\begin{figure}[h]
\centering
\includegraphics[scale=0.7]{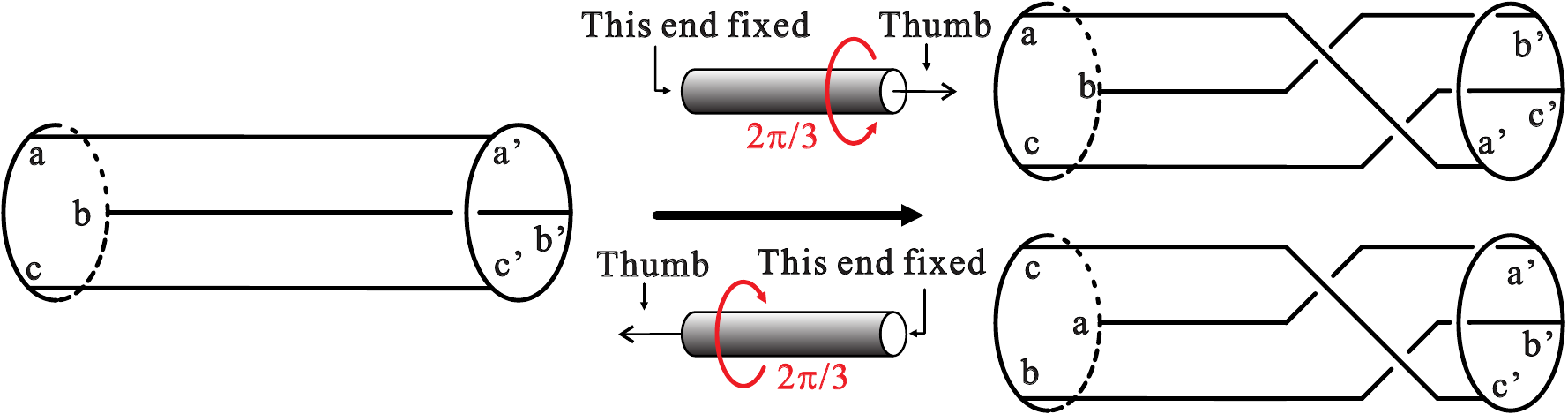}
\caption{Unambiguity of the handedness of twists. A tube is represented by two circles as ends connected by three lines as racing stripes. Twists are depicted by the crossings of the racing stripes.} \label{racingstrip}
\end{figure}

Naturally, an edge can be twisted discretely. The discussion above of rotations shows that the smallest distinguishable twist is $\pi/3$; higher distinguishable twists in the projection must be integral multiples of $\pi/3$. Fig.~\ref{notation}(d)-(f) shows how the handedness and hence the sign of a twist is unambiguously def\/ined. This unambiguity appears to be more lucid with the help of racing stripes, as in Fig.~\ref{racingstrip}, which purports that one can def\/ine the handedness of twists on a tube without referring to either end of the tube. Bearing this in mind, we will stick to our simpler notation without racing stripes throughout Section~\ref{sec4V}.

\subsection{Braids}\label{subSecBraids}

\begin{figure}[h]
\centering
\includegraphics[scale=0.8]{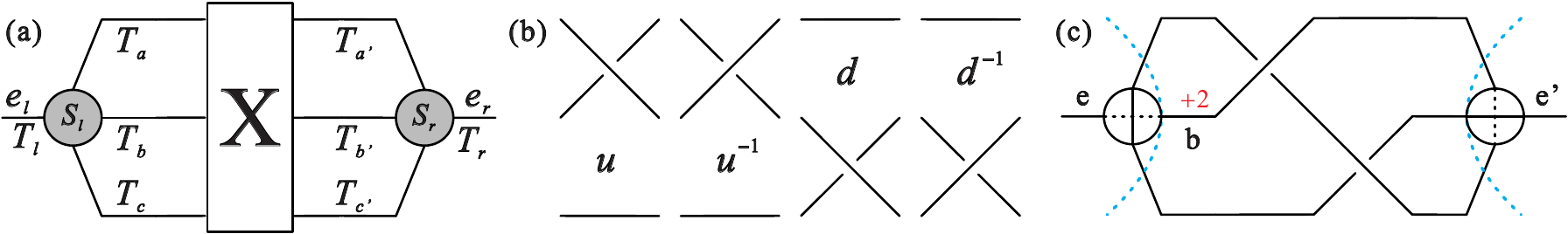}
\caption{(a)~A generic {\bf 3-strand braid}. (b)~The four
generators of $X$. (c)~An example.}%
\label{braidGen}
\end{figure}

The graphic notation enables us to f\/ind an interesting type of topological excitations of 4-valent braided ribbon networks, namely {\bf 3-strand braids} or {\bf 4-valent braids}, def\/ined in Fig.~\ref{braidGen}(a).
A~3-strand braid is made of two {\bf end-nodes} that share three {\bf strands}, which are generically braided and twisted. Each end-node has an {\bf external edge} attached elsewhere in the network. This def\/inition is unambiguous because a braid can always be arranged horizontally as in Fig.~\ref{braidGen}(a). We disallow the strands of a braid to tangle with any other edges in the network, including the braid's external edges.

A 3-strand braid is characterized by an 8-tuple $\{T_l,S_l,T_a,T_b,T_c,X,S_r,T_r\}$ that consists of a~pair of end-node states $(S_l, S_r)$, a pair of {\bf external twists} $(T_l,T_r)$, a crossing sequence $X$, and a~triple of internal twists $(T_a, T_b,T_c)$. $(S_l, S_r)$ is valued in $\{+,-\}$. $S$'s inverse is~$-S$ or~$\bar{S}$. $X$ codes the braiding (from left to right) of the three strands; it must be an element in~$B_3$, the group of ordinary braids of three strands, and is thus generated by the four crossings in Fig.~\ref{braidGen}(b). All twists are valued in $\mathbb{Z}$ in units of~$\pi/3$.

An $X$ of {\bf order} $n=|X|$, the number of crossings, reads $X=x_1x_2\cdots x_i\cdots x_n$, where $x_i\in\{u,u^{-1},d,d^{-1}\}$ is the $i$-th crossing. We also assign $+1$ or $-1$, the crossing number, to each generator according to its handedness, i.e., $u=d=1$ and $u^{-1}=d^{-1}=-1$. Thus, an $x_i$ is an abstract crossing in a multiplication but $+1$ or $-1$ in a summation. Suppose $X'$ is a segment of~$X$, then $|X|\geq |X'|$ and three cases exist: Assume $X'=x_1x_2\cdots x_n$, 1) if $X=X'x_{n+1}\cdots$, we write $X'\preccurlyeq X$, 2) if $X=x_{i_1}x_{i_2}\cdots x_{i_m}X'$, we write $X'\curlyeqprec X$, and 3) otherwise, we write $X'\prec X$. Note that $\preccurlyeq\neq \curlyeqprec$ unless $X=X'$. $X$ clearly induces a permutation $\sigma_X$ of the three strands of a~braid, which takes value in the permutation group $S_3=\{\mathds{1},(1\ 2), (1\ 3), (2\ 3), (1\ 2\ 3), (1\ 3\ 2)\}$.
A braid can take either of these two algebraic forms,
\begin{gather}
  \left._{T_l}^{S_l}\hspace{-0.5mm}[(T_a,T_b,T_c)_X]^{S_r}_{T_r},\right.\label{eqAlgNotation}\\
  \left._{T_l}^{S_l}\hspace{-0.5mm}[_X(T_{a'},T_{b'},T_{c'})]^{S_r}_{T_r},\right.\label{eqAlgNotationEquiv}
\end{gather}
where the triple $(T_a,T_b,T_c)$ on the left of the $X$ and the triple $(T_{a'},T_{b'},T_{c'})$ on the right are related by $(T_a,T_b,T_c)\sigma_X=(T_{a'},T_{b'},T_{c'})$, indicating that $\sigma_X$ is \emph{left-acting} on the triple of internal twists. Equivalently, we have $(T_a,T_b,T_c)=\sigma^{-1}_X(T_{a'},T_{b'},T_{c'})$, where $\sigma^{-1}_X$ is the inverse of $\sigma_X$ and is \emph{right-acting}. For instance, the braid in
Fig.~\ref{braidGen}(c) can be written as
$\left._{\hspace{0.5mm}0}^+\hspace{-0.2mm}[(0,2,0)_{u^{-1}d}]^-_0\right.$
or
$\left._{\hspace{0.5mm}0}^+\hspace{-0.2mm}[_{u^{-1}d}(2,0,0)]^-_0\right.$,
with induced permutations $\sigma_{u^{-1}d}=(3\ 1\ 2)$ and $\sigma^{-1}_{u^{-1}d}=(2\ 3\
1)$. Besides, $(T_a, T_b, T_c)=(T_d, T_e, T_f)$ means $T_a=T_d$, $T_b=T_e$, and $T_c=T_f$, and we also have $(T_a, T_b, T_c)\pm(T_d, T_e, T_f)=(T_a\pm T_d, T_b\pm T_e, T_c\pm T_f)$.

We shall see that dif\/ferent types of braids pattern the corresponding 8-tuples dif\/ferently. For a trivial braid, its $X$ is trivial, and $\sigma_X=\mathds{1}$; hence, the generic notation uniquely boils down to $\left._{T_l}^{S_l}\hspace{-0.5mm}[T_a,T_b,T_c]^{S_r}_{T_r}\right.$.

Four-valent braids are noiseless topological excitations of braided ribbon networks~\cite{ Bilson-Thompson2006, fotini2007bi, fotini2003}. We emphasize that 4-valent braids are 3D structures that are better studied in their 2D projections, which are called {\bf braid diagrams} hereafter. In fact, we are dealing with equivalence classes of braids because each braid is equivalent to an inf\/inite number of braids under a set of isotopy moves (to be introduced soon), whose 2D projections relate equivalent braid diagrams. Equivalence classes of braids are in one-to-one correspondence with those of braid diagrams~\cite{WanLee2007,Wan2009, Wan2007}; hence, we need not distinguish a braid from its diagrams. That is, by a braid we mean all its isotopic braids, and we study this braid by its equivalence class of braid diagrams. In what follows, we may use braids and braid diagrams interchangeably.

The generators of the $X$ of a 4-valent braid obey the well-known braid relations of $B_3$, which are
\begin{gather}
udu^{-1}=d^{-1}ud,\qquad u^{-1}du=dud^{-1},\qquad udu=dud.
\label{eqBraidRel}
\end{gather}
We assume in any $X$, these relations have been applied, such that, e.g., $udu^{-1}d^{-1}$ should have been written as $udu^{-1}d^{-1}=d^{-1}udd^{-1}=d^{-1}u$ by the f\/irst relation above. This assumption ensures that each 4-valent braid we study has the least number of crossings among all the braids related to it by equation~(\ref{eqBraidRel}). As such, we consider braid diagrams with the same number of crossings and equivalent crossing patterns as the same!

\subsection{Equivalence moves}\label{subSecEquivMove}

As aforementioned, equivalent braids and hence equivalent braided ribbon networks are related by a set of local equivalence moves that act on the nodes and edges of a network without altering the dif\/feomorphism class of the embedding of the network. One type of these moves are the well-known Reidermeister moves~\cite{Reidemeister}, which are translations of nodes and continuous deformation of ribbons~\cite{Wan2009,Wan2007}. We shall focus on the other type, discrete rotations of nodes, which are peculiar to braided ribbon networks.

\subsubsection[$\pi/3$-rotations: generators of rotations]{$\boldsymbol{\pi/3}$-rotations: generators of rotations}
\label{subsubSecRot}

With respect to any of its four edges, a node admits discrete, purely topological rotation symmetries that are not those with a rigid metric and do not af\/fect the dif\/feomorphism class of the embedding of the node.
\begin{figure}[ht]
\centering
\includegraphics[scale=0.8]{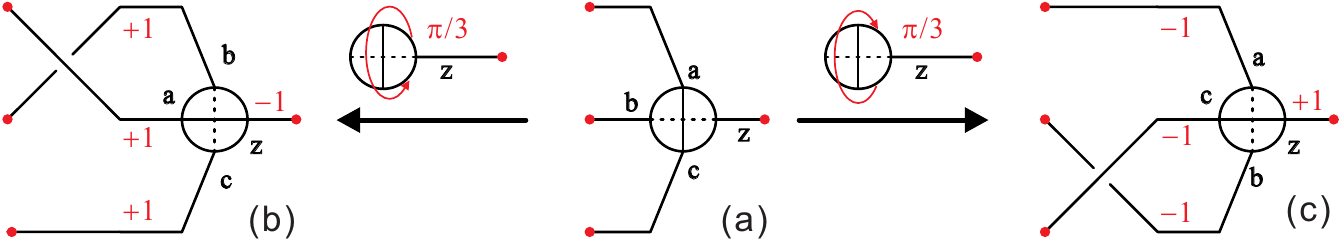}
\caption{(b) \& (c) are results of (a) by
rotating the $\oplus$-node in (a) w.r.t.\ edge $z$ in two directions
respectively. Red dots are assumed connected elsewhere and f\/ixed.}
\label{pi3rot+}
\end{figure}
Section \ref{subSecGraphNotation} points out that each end-node state is a triplet preserved by a $2\pi/3$ rotation but mapped to the other state by a $\pi/3$ rotation. Because two consecutive $\pi/3$ rotations comprise a $2\pi/3$ rotation, $\pi/3$ rotations are generators of all possible discrete rotations of a node w.r.t.\ an edge of the node. We now show how $\pi/3$ rotations af\/fect a subgraph, in which a node is rotated. In either a $\oplus$ or a $\ominus$ state, a node can be rotated in two opposite directions. Fig.~\ref{pi3rot+} shows the case where a node is in a $\oplus$-state w.r.t.\ its rotation axis before the $\pi/3$ rotation is done. The ef\/fect of a~$\pi/3$ rotation on a $\ominus$-node is the mirror image of Fig.~\ref{pi3rot+}. A~$\pi/3$ rotation of a node always f\/lips the state of the node, creates a crossing of two edges of the node, and generates a~$\pm 1$ twist on the rotation axis and a $\mp 1$ twist on each of the other three edges.

\begin{figure}[h]
\centering
\includegraphics[scale=0.8]{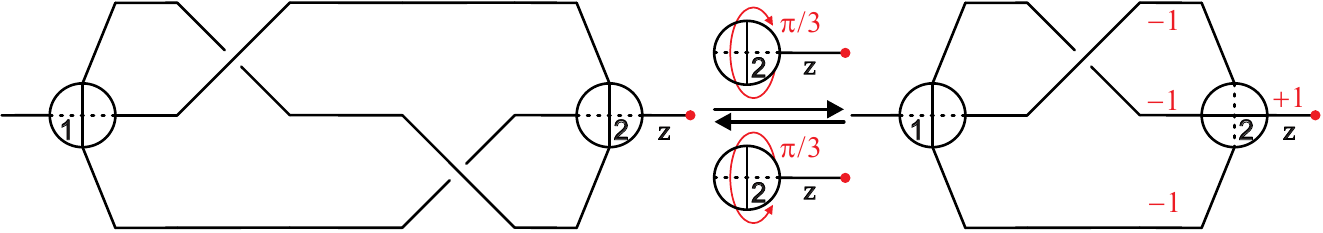}
\caption{Two braids are equivalent under a $\pi/3$-rotation of node 2.} \label{equibraids}
\end{figure}

Since our key topological structures are braids, we may ask how the rotations act on a braid. In this case, e.g.\ in Fig.~\ref{equibraids}, we only allow an external edge of a braid to be a rotation axis~\cite{Wan2007}. In Fig.~\ref{equibraids} the left braid is equivalent to the right one with one less crossing. This observation motivates a classif\/ication of braids as if they are isolated structures. We name a few important def\/initions, whose details are referred to~\cite{WanLee2007, Wan2007}. A braid is \textbf{reducible} if it is equivalent to a~braid with fewer crossings and otherwise \textbf{irreducible}. A braid is {\bf left-}, {\bf right-}, or \textbf{two-way-reducible} if it can be reduced by rotations on either its left, right, or both end-nodes. A braid equivalent to a trivial braid, i.e., a braid without crossings, is \textbf{completely reducible}. The algebraic form of a rotation on a braid as a whole can be found in~\cite{WanHe2008a}.

As equivalence moves, the rotations and translations should have associated invariants. Although such an invariant exist on arbitrary sub-graphs of braided ribbon networks, in which equivalence moves act~\cite{WanLee2007,Wan2007}, we are more interested in restricting the sub-graphs to be 4-valent braids only. In this case, a braid bears two invariants of discrete rotations, namely its {\bf ef\/fective twist} and {\bf ef\/fective state}~\cite{WanHe2008a},
\begin{gather}
\Theta =T_l+T_r+\sum\limits_{i=a}^{c}T_i-2\sum\limits_{i=1}^{|X|}x_i,\qquad
\chi  =(-)^{|X|}S_lS_r.\label{eqETES}
\end{gather}
Moreover, we shall see that both quantities in equation~(\ref{eqETES}) are also conserved quantities of braid interactions.

\subsubsection{Braid representations}\label{subsubSecBraidRep}

As an equivalence class, a braid should be studied in terms of certain convenient representatives of the class. Each braid bears a {\bf unique representation}, in which the braid has twist-free external edges\footnote{This uniqueness is def\/ined modulo the ordinary braid relations, as explained in Section~\ref{subSecBraids}.}. Here is why. Were there two braids with twist-free external edges in one equivalence class, they must be related by rotations, which is contradictory to the condition that any rotation creates twists on an external edge. This representation is convenient for the study of braid dynamics in general.

All irreducible braids in a class must have the same number of crossings; otherwise, the longer ones should either be reducible or belong to another class. Therefore, irreducible braids in a~class must have the smallest number of crossings. We henceforth call an irreducible braid an {\bf extremum} of a class. A braid is in an {\bf extremal representation} if it is represented by its extremum. The extremal representation is not unique because a braid has an inf\/inite number of extrema~\cite{WanHe2008a,Wan2009}. An extremal representation is called a {\bf trivial representation} if the associated extremum has no crossings.

\subsection{Dynamics: evolution moves}\label{subSecEvol}

To def\/ine the models our results apply to, we have to choose a set of dynamical moves. In SF and other models of spin networks, dual Pachner moves are a common choice~\cite{Baez1999, Rovelli1998rev, RovelliBook}, as seen in Section~\ref{subsubSecSNLQG} for the trivalent scheme. We now discuss the dual Pachner moves on 4-valent braided ribbon networks.

Let us f\/ix a non-singular topological manifold $\mathcal{M}$ and choose a triangulation of it in terms of tetrahedra embedded in $\mathcal{M}$ whose union is homeomorphic to $\mathcal{M}$. Any such simplicial triangulation of~$\mathcal{M}$ has a natural dual that is a framed 4-valent graph embedded in $\mathcal{M}$. The framing determines how the tetrahedra are glued on their faces. Thus, a Pachner move on the triangulation should result in a local move in the framed graph, i.e., the dual Pachner move.

Nevertheless, not every embedding of a framed four valent graph in $\mathcal{M}$ is dual to a triangulation of $\mathcal{M}$. Examples of obstructions to f\/inding the dual include the case of braids (e.g., Fig.~\ref{braidGen}). This is an embedding of a graph that could not have arisen from taking the dual of a regular simplicial triangulation of $\mathcal{M}$. We note that these obstructions are local, in the sense that a sub-graph of the embedded graph could be cut out and replaced by another sub-graph that would allow the duality to a triangulation of~$\mathcal{M}$.

The question then is how to def\/ine the dual Pachner moves on
sub-graphs that are not dual to any part of a triangulation
of $\mathcal{M}$. The answer is that we do not. We thus have the
basic rule:

\textit{Basic rule\footnote{This rule and other possible rules are discussed in detail in~\cite{Isabeau2008}.}:} The evolution moves on 4-valent braided ribbon networks are the dual Pachner moves that are allowed only on subgraphs which are dual to a 3-ball.

\begin{figure}[h]
\centering
\includegraphics[scale=0.9]{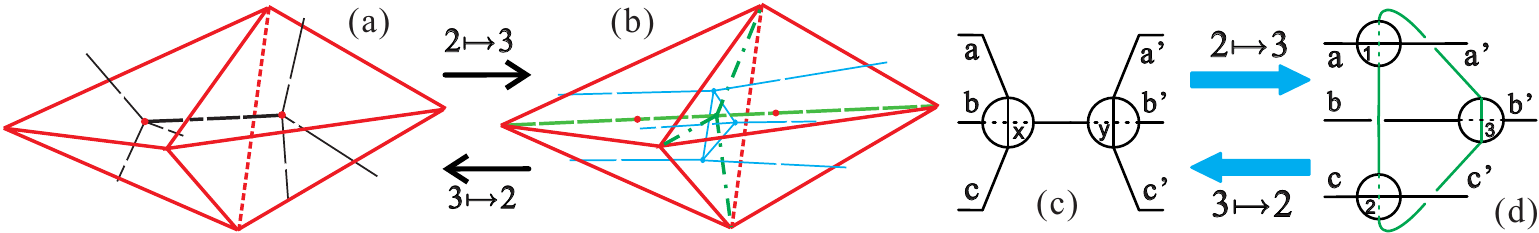}
\caption{The $2\leftrightarrow 3$ Pachner moves between (a) two and (b) three tetrahedra. The dual $2\leftrightarrow 3$ move between~(c) two nodes in the $\oplus$-state and~(d). (c) and~(d) are dual to~(a) and~(b). The red and dashed green lines in (b) outline the three tetrahedra. Green edges in~(d) are generated by the dual $2\rightarrow 3$ move. If the two nodes in~(c) are in the $\ominus$-state, the result of a $2\rightarrow 3$ move is the left-right mirror image of (d).}\label{2to3++}
\end{figure}

\begin{figure}[h]
\centering
\includegraphics[scale=0.8]{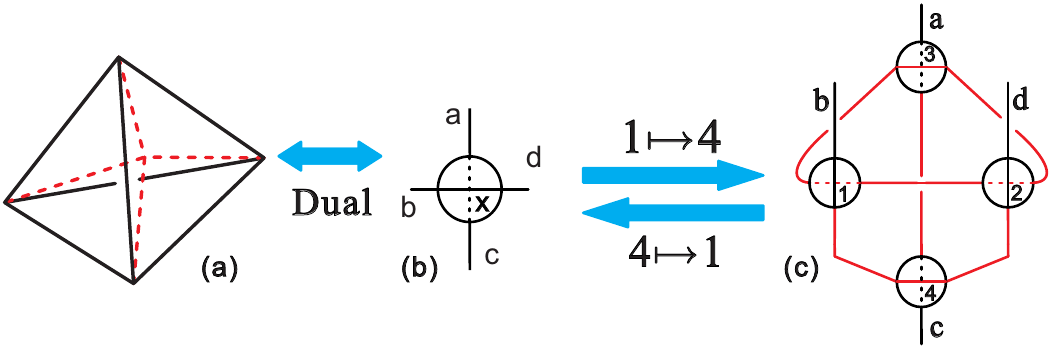}
\caption{(a) The $1\leftrightarrow4$ Pachner moves between one and four tetraheda. The dual $1\leftrightarrow4$ moves are between~(b) and~(c). Red edges in (c) are the generated by the dual $1\rightarrow4$ move.}%
\label{1to4+}%
\end{figure}

The Pachner moves and the dual evolution moves that obey the basic rule on the 4-valent braided ribbon networks, namely the $2\rightarrow 3$ ($3\rightarrow 2$) and $1\rightarrow 4$ ($4\rightarrow 1$) moves, are respectively explained by Figs.~\ref{2to3++} and~\ref{1to4+} and their captions.
The result of a dual Pachner move is unique up to equivalence moves, i.e., the discrete rotations and adapted Reidemeister moves. The basic rule we posit, which dictates the legitimacy of a dual Pachner move, actually boils down to the following conditions.

\begin{condition}
\label{conEvolMove} A legal dual Pachner move falls into the following three cases.
\begin{enumerate}\itemsep=0pt
\item Two nodes allow a $2\rightarrow 3$ move iff they share only one edge and can be put by equivalence moves in the same state with the common edge twist-free $($Fig.~$\ref{2to3++}(b)$ or its mirror image$)$.
\item Three nodes allow a $3\rightarrow 2$ move iff they and their edges can be set by equivalence moves as either Fig.~$\ref{2to3++}(c)$ or its mirror image, where there is a contractible loop made of three twist-free edges.
\item A $1\rightarrow 4$ move is always doable; however, a $4\rightarrow 1$ on four nodes is doable iff the four nodes together with their edges can be arranged in the form of either Fig.~$\ref{1to4+}(c)$ or its mirror image, in which there is a contractible loop and all common edges are twist-free.
\end{enumerate}
\end{condition}

In view of Condition~\ref{conEvolMove}, a $2 \rightarrow3$ move is illegal on the two end-nodes of the braid in Fig.~\ref{braidGen}(c), so is a $3\rightarrow 2$ move on any three nodes that contain this pair, which makes this braid stable under single evolution moves. This generalises to the statement that any nontrivial braid with internal twists is stable under single evolution moves. The {\bf stable braids} are thus considered noiseless subsystems~\cite{WanLee2007} and local excitations with conserved quantities~\cite{WanHackett2008,WanHe2008a}. Section~\ref{subSecDisc} has more on the stability and locality of 4-valent braids.

The dual Pachner moves in SF models contain only the permutation relations of edges, which are suf\/f\/icient only for triangulations. In contrast, our dual Pachner moves are adapted to the embedded case by recording the spatial relation (under or above) of the edges and nodes in a~projection of the 3D graph; they are thus called the {\bf adapted dual Pachner moves} and are able to endow embedded 4-valent spin networks with a spin-foam like dynamics. This answers the question raised at the end of Section~\ref{subsubSecSNLQG}.

\subsection{Dynamics: propagation, direct and exchange interaction of braids}\label{subSecBrainInt}

In order that the stable braids, as topological excitations of the 4-valent braided ribbon networks, can be candidates for particles or pre-particles, they must be dynamical. Indeed, the evolution moves endow the stable braids with a rich dynamics: they can propagate and interact. We brief\/ly address braid propagation f\/irst.

\looseness=-1
Since a braid can be considered an insertion in an edge, it makes sense to speak of them propagating to the left or to the right along that edge. To help visualize this in the diagrams we will always arrange a braid so that the edge of the graph it interrupts runs horizontally on the page. There are two types of propagation of braids, namely induced propagation and active propagation.

Under the evolution moves, especially the $1\leftrightarrow 4$ moves, the ambient network of a braid may expand on one side of a braid but contract on the other side, such that the braid ef\/fectively moves towards the side of contraction. We call this phenomenon {\bf induced propagation} because the braid is not directly involved in the evolution and remains the same. Any braid can propagate in this induced way.

\looseness=-1
Opposed to induced propagation is {\bf active propagation}, which occurs only to specif\/ic network conf\/igurations and braids; it is called active because the braid's structure undergoes intermediate changes (and probably permanent change of its internal twists) during the propagation. Braids that can propagate in this way are called {\bf actively propagating} and otherwise {\bf stationary} or {\bf non-actively propagating}. Nevertheless, active braid propagation needs some special care that may modify the overall settings of the 4-valent scheme, we thus refer to~\cite{WanLee2007,Wan2009} for details.

Two braids may propagate and meet each other in a situation, such that they can interact. Two interaction types exist: direct interaction and exchange interaction. We shall illustrate these striking behaviours of the braids by f\/igures and present some key results of the dynamics in the algebraic notation.

\looseness=-1
We f\/irst elucidate the direct interaction. In a {\bf direct interaction} of two adjacent braids, one can merge with the other, through a sequence of evolution moves. We shall describe this by a complete example and then by the generalised def\/inition.
\begin{figure}[t]
\centering
\includegraphics[scale=0.67]{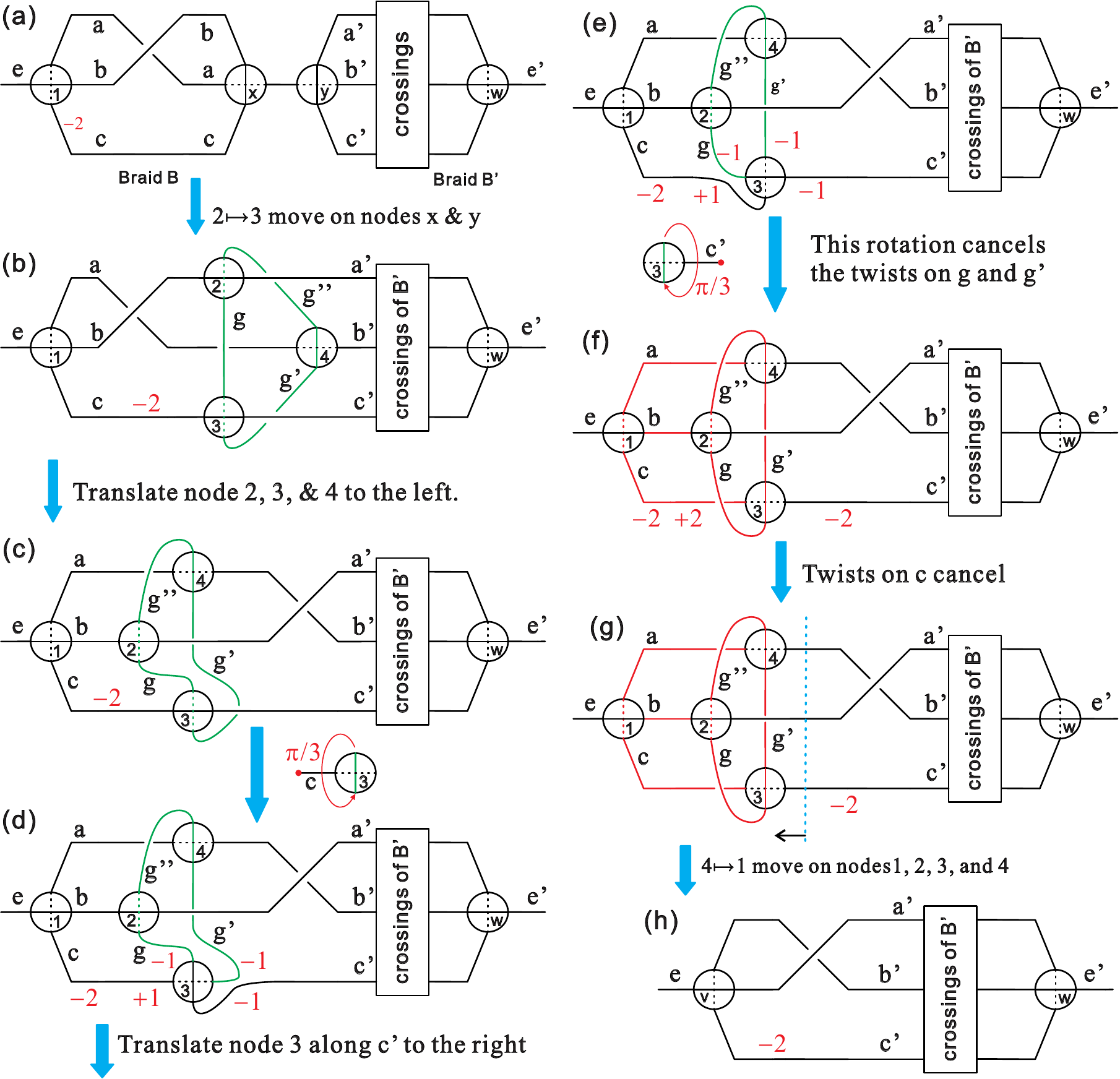}
\caption{An example of direct right-interaction.}
\label{rightInt1xLongAll}
\end{figure}
\begin{figure}[h]
\centering
\includegraphics[scale=0.7]{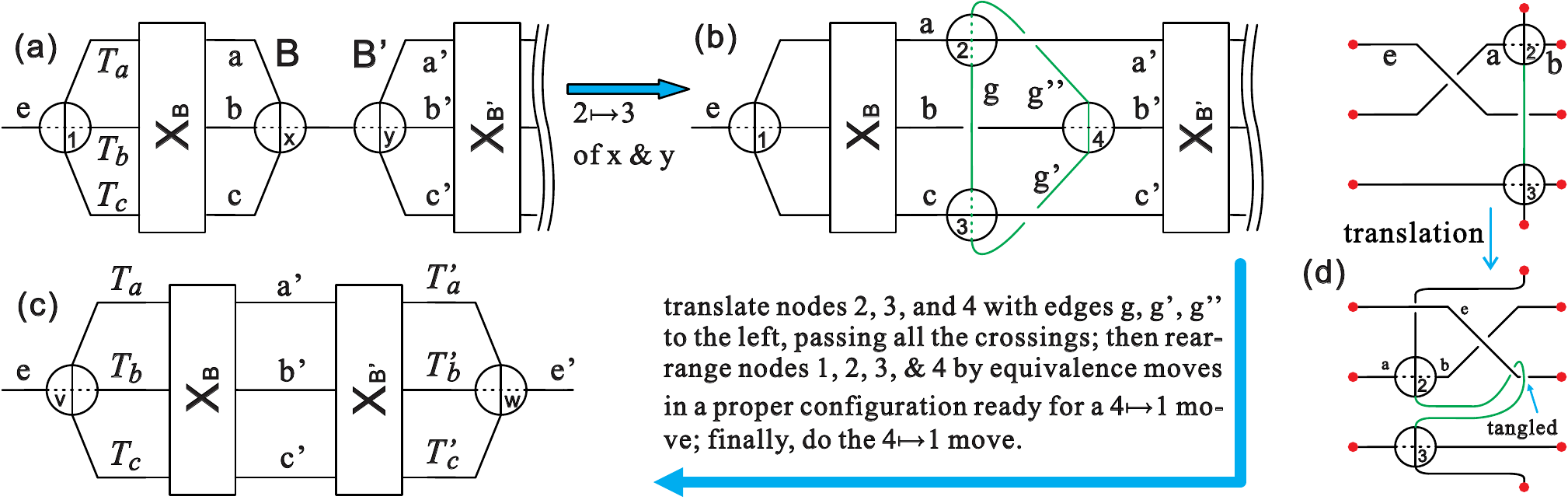}
\caption{(a)-(c) Def\/inition of the direct right-interaction. (d)~An obstructed translation.} \label{intdef}
\end{figure}
Fig.~\ref{rightInt1xLongAll} depicts all the steps that the two braids in (a) take to become the single braid in (h) to complete an interaction. To supplement this f\/igure, a few remarks are in order. In this example, the equivalence moves~-- the rotations and translations~-- are nondynamical but are there only to set the nodes and edges in a proper conf\/iguration that manifests the legitimacy of an evolution move, e.g.\ the f\/inal $4\rightarrow 1$ move (compare it to Fig.~\ref{1to4+}).

Braid $B$ in Fig.~\ref{rightInt1xLongAll}(a) is an example of what we call {\bf actively interacting} braids because all the moves are done basically on $B$'s nodes and edges; $B'$, except its left end-node, plays no role. Moreover Fig.~\ref{rightInt1xLongAll} is an example of {\bf direct right-interaction}, in the sense that the actively interacting braid merges with the braid on its right. In fact, a direct interaction always involves at least one actively interacting braid.

For simplicity braid $B'$ is assumed twist-free. One may notice that the twist of $-2$ on strand $c$ of braid $B$ appears again in the braid in Fig.~\ref{rightInt1xLongAll}(h). This is not a coincidence but an instance of certain conservation laws that braid interactions follow.

Bearing this example in mind, Fig.~\ref{intdef} serves as a transparent graphical def\/inition of the {\bf direct right-interaction} of two braids. Note that translating the nodes 2, 3, and 4 together with their common edge~$g$, $g^{\prime}$, and $g^{\prime\prime}$ to the left, passing through all of $B$'s crossings~$X_B$ in Fig.~\ref{intdef} is certainly not always possible because $X_B$ may obstruct the translation by creating a~tangle like that in Fig.~\ref{intdef}(d). This translation is guaranteed viable only when braid~$B$ is at least completely right-reducible (see Fig.~\ref{rightInt1xLongAll}). Note also that nodes 1, 2, 3, and 4 with their edges cannot always be arranged by equivalence moves to meet Condition~\ref{conEvolMove}(3), such that the f\/inal $4\rightarrow1$ move can be done to complete the interaction. If all steps in Fig.~\ref{intdef} are possible, braid~$B$ must be actively interacting.

\begin{figure}[ht]
\centering
\includegraphics[scale=0.7]{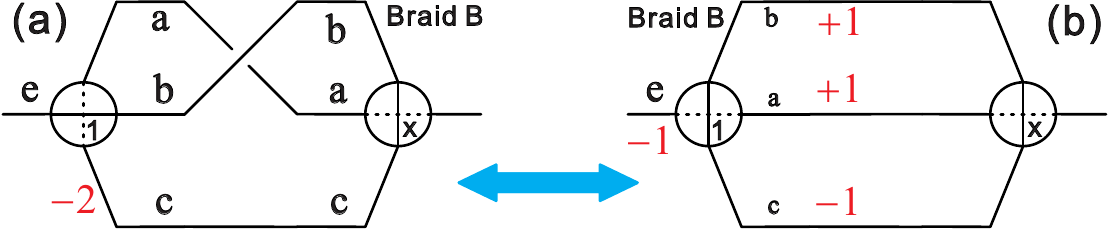}
\caption{The braid $B$ in~Fig.~\ref{rightInt1xLongAll} is equivalent to a trivial braid.}%
\label{rightint1xlongequiv}%
\end{figure}

It is a theorem~\cite{WanLee2007,WanHackett2008} that if a braid is actively interacting, this constrains the braid to be equivalent to a trivial braid with both end-nodes in the same state (e.g.\ Fig.~\ref{rightint1xlongequiv}). Recall that an actively interacting braid is studied most conveniently in its trivial representation, which should read $\left._{T_l}^{S}\hspace{-0.5mm}[T_a,T_b,T_c]^{S}_{T_r}\right.$.

The discussion above is all about direct right-interaction. {\bf Direct left-interaction} is def\/ined analogously and can be visualized as the left-right mirror image of right-interaction. Thus, in what follows, we assume direct left-interaction is understood. We denote a direct interaction of two braids $B_1$ and $B_2$ by $B_1+_{\mathrm{d}}B_2=B$; whether it is a left- or right-interaction is manifest contextually.

A formal division of braids in now natural. We temporarily denote the set of stable braids by~$\mathfrak{B}^S_0$. Nevertheless, for reasons to become transparent we enlarge $\mathfrak{B}^S_0$ by adding two more braids:
\begin{gather}
B_0^{\pm}=\left._0^{\pm}\hspace{-0.5mm}[0,0,0]^{\pm}_0,\right.\label{eqB0pm}
\end{gather}
which are completely trivial\footnote{Note again that at this point spin network labels are not in play. Including spin network labels has two immediate consequences. Firstly, like any other braid in $\mathfrak{B}^S$, $B_0^{\pm}$ are not just two braids but an
inf\/inite set of braids coloured by dif\/ferent sets of spin network labels. Secondly, $B_0^{\pm}$ are only trivial topologically but neither algebraically nor physically.}. $B_0^{\pm}$ are actually unstable because they are dual to a 3-ball. If we tolerate their instability they are obviously actively interacting. As such, let us still call the enlarged set the set of stable braids but denote it by $\mathfrak{B}^S$. $\mathfrak{B}^S$ admits a disjoint union of three subsets:
\begin{gather*}
\mathfrak{B}^S=\mathfrak{B}^b\sqcup\mathfrak{B}^f\sqcup\mathfrak{B}^s,
\end{gather*}
where $\mathfrak{B}^b$, $\mathfrak{B}^f$, and $\mathfrak{B}^s$ are the sets respectively of actively interacting braids (including $B_0^{\pm}$), all actively propagating braids that do not actively interact, and stationary braids. The meaning of the superscripts will be made clear later. Although actively interacting braids are also actively propagating~\cite{WanLee2007,Wan2008}, they are excluded from $\mathfrak{B}^f$. It is a theorem that $\forall\, B\in\mathfrak{B}^b$, the ef\/fective state of~$B$, $\chi_B\equiv 1$~\cite{WanHe2008a}.

We now dwell on the algebra of direct interactions. For simplicity, let us consider $B,B'\in\mathfrak{B}^b$, $B=\left. _{T_l}^S\hspace{-0.5mm}[ T_a,T_b,T_c] _{T_r}^S\right.$ on the left of $B'=\left. _{\ T_l'}^{\hspace{1.2 mm}S'}\hspace{-0.5 mm}[ T'_a,T'_b,T'_c] _{T'_r}^{S'}\right.$. If $B$ and $B'$ satisfy the interaction Condition \ref{conEvolMove}(1) trivially, i.e., if $T_r+T'_l=0$ and $S=S'$, then the direction interaction of $B$ and $B'$ is simply
\begin{gather}
B+_{\mathrm{d}} B'\overset{T_r+T'_l=0}{=}
\left._{T_l}^S\hspace{-0.5 mm}[ (T_a,T_b,T_c) + (T'_a,T'_b,T'_c)]
_{T'_r}^S =\hspace{1 mm}_{T_l}^S\hspace{-0.5 mm}[
T_a+T'_a,T_b+T'_b,T_c+T'_c] _{T'_r}^S.\right.
\label{eqConnectedsumAlg}
\end{gather}
If, however, $B$ and $B'$ satisfy the interaction condition nontrivially, i.e.\ if either $B$'s right end-node or $B'$'s left end-node must be rotated to make them meet Condition~\ref{conEvolMove}(1), some algebra is needed. We would refer to~\cite{WanHackett2008,WanHe2008a} for the details; rather, we present the Lemma~\ref{lemmIntGen} as the general result. Note that we put an actively interacting braid in its trivial representation and an arbitrary braid in its unique representation.

\begin{lemma}\label{lemmIntGen}
Given a braid
$B=\left._{T_l}^{\hspace{0.5mm}S}\hspace{-0.5mm}[T_a,T_b,T_c]^{S}_{T_r}\right.\in\mathfrak{B}^b$
on the left of a non-actively interacting braid,
$B'=\left._{\hspace{1mm}0}^{S_l}\hspace{-0.5mm}[(T'_a,T'_b,T'_c)_X]^{S_r}_{0}\right.$,
with the interaction condition satisfied by $(-)^{T_r}S=S_l$, the
direct interaction of $B$ and $B'$ results in a braid
\begin{gather*}
B''=\!\left._{\hspace{6mm}0}^{(-)^{T_l}S}\hspace{-0.8mm}
[((P^S_{-T_l}(T_a,T_b,T_c))\!+\!(P^{(-)^{T_r}S}_{-T_l-T_r}(T'_a,T'_b,T'_c))\!+\!(T_l+T_r,\cdot,\cdot))_{X_l((-)^{T_r}S,-T_l-T_r)X}]^{S_r}_{0},\right.
\end{gather*}
where $(T_l+T_r,\cdot,\cdot)$ is the short for
$(T_l+T_r,T_l+T_r,T_l+T_r)$.
\end{lemma}

The $P^{S_l}_m$ in the $B''$ in Lemma \ref{lemmIntGen} is a permutation on the triple, determined by $S_l$ and $m$ and valued in the group $S_3$. Fig.~\ref{pi3rot+} and its mirror images readily show $P^+_{+1}=(1\ 2)$, $P^+_{-1}=(2\ 3)$, $P^-_{+1}=(3\ 2)$, and $P^-_{-1}=(1\ 2)$. More general equalities can be derived recursively and found in~\cite{WanHackett2008}. The functions $X_l(S_l,m)$ and $X_r(S_r,n)$ return crossing sequences generated by the rotations needed to set $B$'s right end-node and $B'$'s left end-node ready for a $2\rightarrow 3$ move. We quote its def\/inition in the following equation but refer their detailed properties to~\cite{WanHe2008a}:
\begin{gather}
X_l(+,m)=
\begin{cases}  (ud)^{-m/2}& \text{if $m$ is even,}
\\
d(ud)^{(-1-m)/2} &\text{if $m$ is odd,}
\end{cases}\nonumber\\
X_l(-,m)=
\begin{cases}  (du)^{-m/2}& \text{if $m$ is even,}
\\
u(du)^{(-1-m)/2} &\text{if $m$ is odd,}
\end{cases}\nonumber\\
X_r(+,n) =
\begin{cases}  (ud)^{-n/2}& \text{if $n$ is even,}
\\
(ud)^{(1-n)/2}d^{-1} &\text{if $n$ is odd,}
\end{cases}\nonumber\\
X_r(-,n)=
\begin{cases}  (du)^{-n/2}& \text{if $n$ is even,}
\\
(du)^{(1-n)/2}u^{-1} &\text{if $n$ is odd,}
\end{cases}\label{eqCrossing}
\end{gather}
where $n,m\in\mathbb{Z}$. In equation~(\ref{eqCrossing}), a positive exponent of a crossing sequence means, for example, $(ud)^2=udud$, while a negative one means, for instance, $(ud)^{-2}=d^{-1}u^{-1}d^{-1}u^{-1}$.  Lemma~\ref{lemmIntGen} is independent of the trivial representation chosen for~$B$~\cite{WanHe2008a}. Equation~(\ref{eqConnectedsumAlg}) is merely a special case of Lemma~\ref{lemmIntGen}. We are now ready for a main result.

\begin{theorem}\label{theoConserve}
Given $B=\left._{T_l}^{\hspace{0.5mm}S}\hspace{-0.5mm}[T_a,T_b,T_c]^{S}_{T_r}\right.\in\mathfrak{B}^b$,
and any braid,
$B'=\left._{\hspace{1mm}0}^{S_l}\hspace{-0.5mm}[(T'_a,T'_b,T'_c)_X]^{S_r}_{0}\right.\in\mathfrak{B}^S$, such that $B+B'=B''\in\mathfrak{B}^S$, the effective twist number $\Theta$ is an additive conserved quantity, while the effective state $\chi$ is a multiplicative conserved quantity, namely%
\begin{gather*}
\Theta_{B''} =\Theta_B+\Theta_{B'},\qquad
\chi_{B''}  =\chi_B\chi_{B'}.
\end{gather*}
\end{theorem}

This theorem, proved in~\cite{WanHe2008a}, demonstrates that the two invariants of equivalence moves of braids, $\Theta$ and $\chi$, are also conserved charges of direct interactions. Conservation laws play a pivotal role in revealing the underlying structure of a physical theory. By invariants and conserved quantities we are able to determine how the content of our theory may relate to particle physics and if extra inputs are required. In Section~\ref{subSecCPT}, we try to identify our conserved quantities with particle quantum numbers.

Since $\chi_B\equiv 1 \ \forall\, B\in\mathfrak{B}^b$, Theorem~\ref{theoConserve} shows that if $B$ directly interacts with a braid with $\chi=-1$, the result must be a braid with $\chi=-1$ too and is thus not in $\mathfrak{B}^b$. Moreover, $\mathfrak{B}^b+_{\mathrm{d}}\mathfrak{B}^S\setminus\mathfrak{B}^b\subseteq\mathfrak{B}^S\setminus\mathfrak{B}^b$.

Because fermions usually do not directly interact with each other but can interact with (gauge) bosons, the fact that a direct interaction always involves at least one actively interacting braid implies an analogy between actively (non-actively) interacting braids and bosons (fermions). The evidence of this analogy will become stronger soon, after we study exchange interaction. This analogy manifests the meaning of the superscript ``$b$'', as for bosons, of $\mathfrak{B}^b$, the set of actively interacting braids. As to the set of non-actively interacting braids, we divided it into $\mathfrak{B}^f$ and $\mathfrak{B}^s$, and we are more inclined to consider the former analogous to the set of fermions because the braids in $\mathfrak{B}^f$ are chiral~\cite{WanLee2007}, analogous to the chirality of the SM elementary fermions, which manifest the superscript ``$f$'' in $\mathfrak{B}^f$.

The result of a direct interaction of two braids is unique! Nonetheless, this uniqueness may cease to hold if the braided ribbon networks are graced with spin network labels, such that the result of an evolution move becomes a superposition of outcomes with the same topology but dif\/ferent sets of labels.

\begin{figure}[h]
\centering
  \includegraphics[scale=0.7]{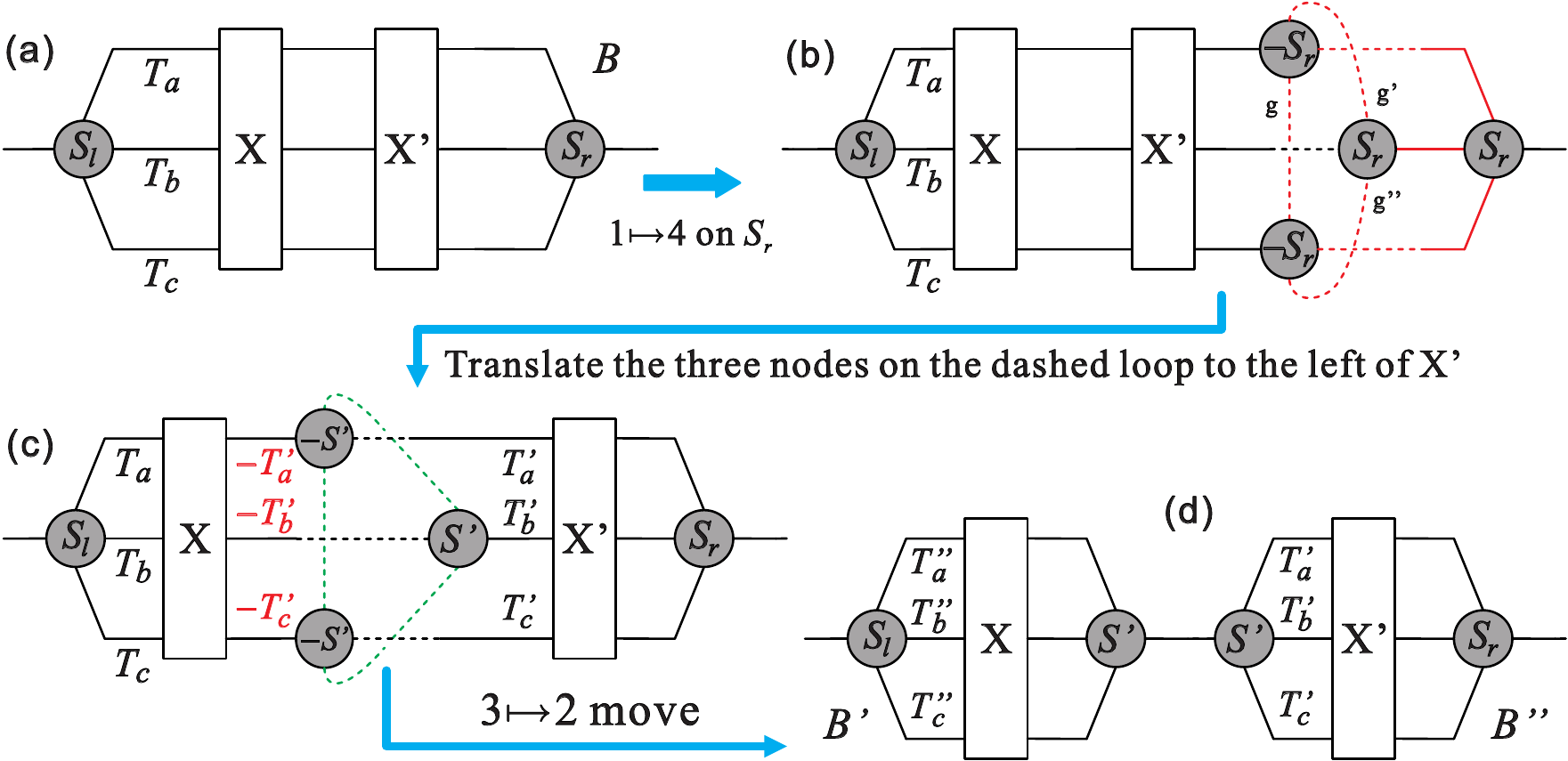}
  \caption{Def\/inition of braid decay:
  $B\overset{\rightsquigarrow}{\rightarrow} B'+B''$,
  $B,B'\in\mathfrak{B}^S$, $B''\in\mathfrak{B}^b$. The dashed lines
  emphasize the dependence of their relative positions
  on the state~$S_r$.}  \label{DecayDef}
\end{figure}
It is natural to ask if a braid can decay, like a particle. The answer is Yes. A braid decay is a~reversed process of a direct interaction. Hence, through decay a braid always radiates an actively interacting braid, to its left or its right. We thus dif\/ferentiate a left-decay from a~right-decay, which are symbolized respectively by $B\overset{\leftsquigarrow}{\rightarrow}B'+B''$, indicating that $B'\in\mathfrak{B}^b$, and $B\overset{\rightsquigarrow}{\rightarrow} B'+B''$ because $B''\in\mathfrak{B}^b$.

Fig.~\ref{DecayDef} def\/ines right-decay. Fig.~\ref{DecayDef}(a) shows a right-reducible braid, $B\!=\!\left.^{S_{l}}\hspace{-0.5mm}[(T_{a},T_{b},T_{c})_{XX'}]^{S_r},\right.$ with a chosen reducible crossing segment $X'$, $X'\curlyeqprec X_{\rm maxred}\curlyeqprec X$, where $X_{\rm maxred}$ is the maximal reducible segment of $X$. This indicates that, in contrast to direct interaction, a braid may decay in multiple ways, each corresponding to a choice $X'$ that must be specif\/ied in a decay process. As the reverse of direct left-interaction, Fig.~\ref{DecayDef} should be easily understood. Its algebraic form~is
\begin{gather*}
B=\left.^{S_{l}}\hspace{-0.5mm}[(T_{a},T_{b},T_{c})_{XX'}]^{S_r}\right.
 \overset{\rightsquigarrow}{\rightarrow}
\left.^{S_{l}}\hspace{-0.5mm}[((T_a,T_b,T_c)-\sigma^{-1}_X(T'_a,T'_b,T'_c))_X]^{S'}\right.+
\left.^{S'}\hspace{-0.5mm}[(T'_a,T'_b,T'_c)_{X'}]^{S_r}\right.\nonumber\\
\hphantom{B=\left.^{S_{l}}\hspace{-0.5mm}[(T_{a},T_{b},T_{c})_{XX'}]^{S_r}\right.}{}  =B'+B'',
\end{gather*}
where $S'=(-)^{|X'|}S_r$, $B''\in\mathfrak{B}^b$, and the ``$+$'' denotes the adjacency of $B'$ and $B''$. Left-decay is def\/ined similarly. The relation between decay and direct interaction ensures that ef\/fective twist~$\Theta$ and ef\/fective state $\chi$ are also additively and multiplicatively conserved in braid decay.

Not only a reducible braid but also an irreducible~-- in fact any~-- braid can radiate. What an irreducible braid emits is either of $B_0^{\pm}$ in equation~(\ref{eqB0pm}). In fact, because a $1\rightarrow 4$ can always take place on either end-node of any braid, a subsequent $3\rightarrow 2$ on three of the four nodes generated by the $1\rightarrow 4$ move results in either a $B_0^+$ or a $B_0^-$, depending on the state of the end-node. This is why $B_0^{\pm}$ are included in $\mathfrak{B}^S$ although they are unstable. It is therefore plausible that $B_0^{\pm}$ are analogous to gravitons.

We now proceed to the exchange interaction of braids. Unlike a direct interaction, an exchange interaction can be def\/ined on the whole $\mathfrak{B}^S$ as a map, $+_{\mathrm{e}}: \mathfrak{B}^S\times\mathfrak{B}^S\rightarrow\mathfrak{B}^S\times\mathfrak{B}^S$. Exchange interactions get this name because each such process always involves an exchange of a virtual actively interacting braid. It is useful to keep track of the direction of the f\/low of the actively interacting braid during an exchange interaction. Therefore, we dif\/ferentiate a \textit{left} and a \textit{right exchange interaction}, respectively denoted by $\overset{\leftarrow}{+}_{\mathrm{e}}$ and $\overset{\rightarrow}{+}_{\mathrm{e}}$. The arrow indicates the ``f\/low'' of the virtual actively interacting braid.
\begin{figure}[h]
\begin{center}
  \includegraphics[scale=0.7]{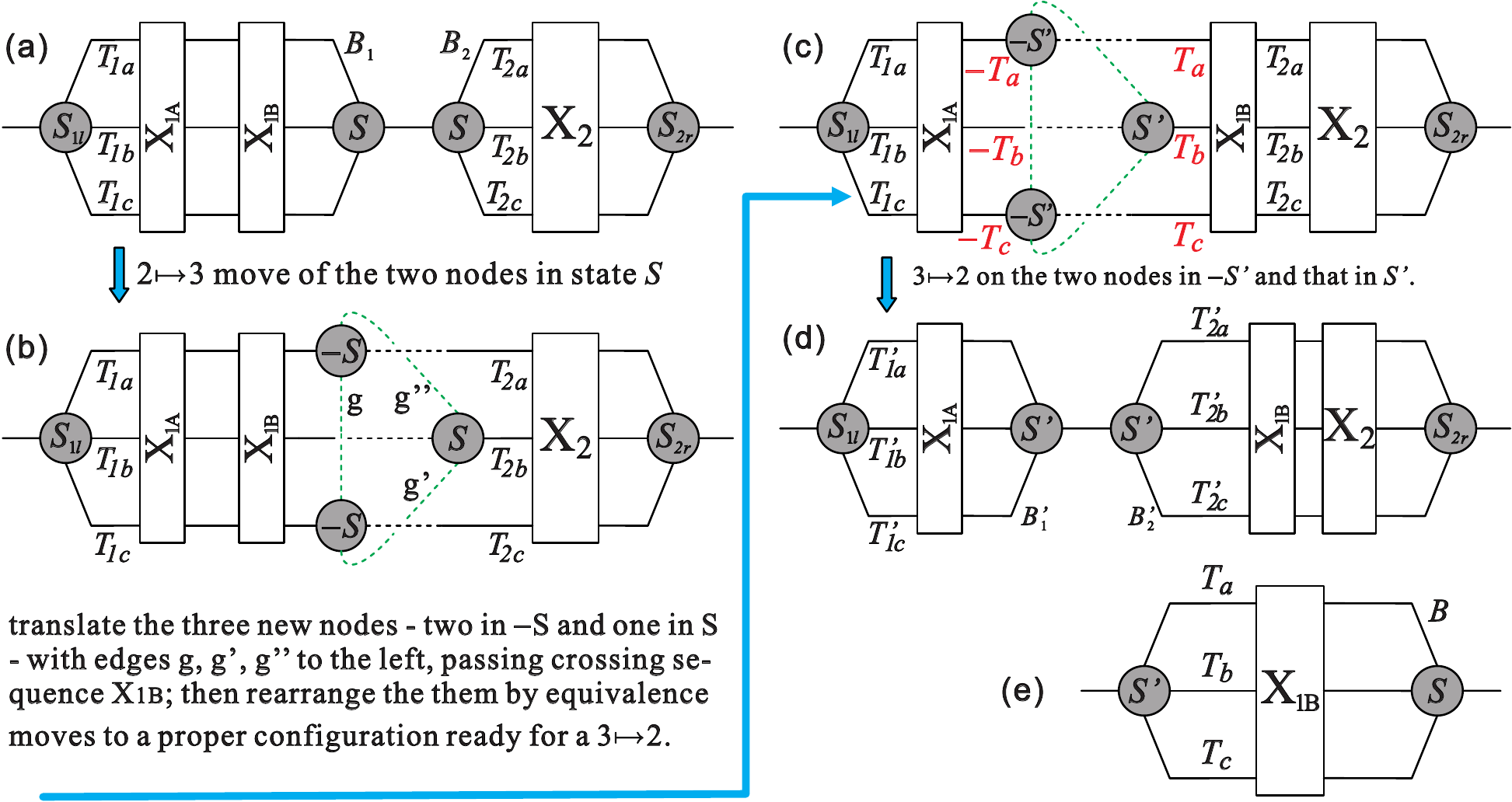}
  \caption{(a)-(d): Def\/inition of right exchange interaction:
  $B_1 \overset{\rightarrow}{+}_{\mathrm{e}}B_2\rightarrow B'_1+B'_2$,
  $B_1,B_2,B'_1,B'_2\in\mathfrak{B}^S$. (e) The virtual actively interacting braid exchanged from $B_1$ to $B_2$.}
  \label{type2IntDef}
\end{center}
\end{figure}

The graphical def\/inition of a {\bf right exchange interaction} is illustrated in Fig.~\ref{type2IntDef}. The left exchange interaction is def\/ined likewise. We now make a few remarks.

Firstly, we begin with the $B_1=\left.^{S_{1l}}\hspace{-0.5mm}[(T_{1a},T_{1b},T_{1c})_{X_{1A}X_{1B}}]^{S}\right.$ and $B_2=\left.^{S}\hspace{-0.5mm}[(T_{2a},T_{2b},T_{2c})_{X_2}]^{S_{2r}}\right.$, which are two stable braids, in Fig.~\ref{type2IntDef}(a). The zero external twists are omitted as~$B_1$ and~$B_2$ are in their unique representations. $B_1$'s right end-node and~$B_2$'s left end-node are set in the same state, $S$, to satisfy the interaction condition. We also assume that $B_1$ has a reducible crossing segment, say $X_{1B}$.

Secondly, if $X_{1B}$ is not trivial, i.e.\ if $B_1$ is right-reducible, translating the three nodes on the loop in (b) and rearranging them by equivalence moves into a proper conf\/iguration ready for a $3\rightarrow 2$ move generates a pair of triples of internal twists, $-(T_a,T_b,T_c)$ and $(T_a,T_b,T_c)$, and causes $S'=(-)^{|X_{1B}|}S$, as in~(c). Taking the permutations induced by $X_{1A}$ and by $X_{1B}$ into account, in~(d) we have the relations:{\samepage
\begin{gather}
(T'_{1a},T'_{1b},T'_{1c}) =(T_{1a},T_{1b},T_{1c})-\sigma^{-1}_{X_{1A}}(T_a,T_b,T_c),\label{eqR1}\\
(T'_{2a},T'_{2b},T'_{2c}) =(T_a,T_b,T_c)+\sigma^{-1}_{X_{1B}}(T_{2a},T_{2b},T_{2c}),\label{eqR2}
\end{gather}
which produce the two adjacent braids, $B'_1$ and $B'_2$, related to $B_1$ and $B_2$.}

Thirdly, according to direct interaction, the only possible triple $(T_a,T_b,T_c)$ in Fig.~\ref{type2IntDef}(c) is exactly the same as the triple of internal twists of an actively interacting braid~-- $B=\left.^{S'}\hspace{-0.5mm}[(T_a,T_b,T_c)_{X_{1B}}]^{S}\right.$~-- in Fig.~\ref{type2IntDef}(e). For an actively interacting braid of the form in Fig.~\ref{type2IntDef}(e), $S'=(-)^{|X_{1B}|}S$; hence, $B$'s left and right end-nodes are respectively in the same states as that of the left end-node of $B'_2$ in Fig.~\ref{type2IntDef}(d) and that of the right end-node of $B_1$ in Fig.~\ref{type2IntDef}(a). Thus, the form of braid~$B'_2$ in Fig.~\ref{type2IntDef}(d) must be precisely the result of the direct interaction of~$B$ and~$B_2$, which by Lemma~\ref{lemmIntGen} is
\begin{gather*}
B +_{\mathrm{d}}\; B_2 =
\left.^{S'}\hspace{-0.5mm}[((T_a,T_b,T_c)+\sigma^{-1}_{X_{1B}}(T_{2a},T_{2b},T_{2c}))_{X_2}]^{S_{2r}}\right.=B'_2,
\end{gather*}
which validates the relation in equation~(\ref{eqR2}).

Therefore, the process of the right exchange interaction def\/ined in Fig.~\ref{type2IntDef} is as if $B_1$ and $B_2$ interact with each other via exchanging a virtual actively interacting braid $B$, then become~$B'_1$ and~$B'_2$. Or one may say that an exchange interaction is mediated by an actively interacting braid. This reinforces the analogy between actively interacting braids and bosons\footnote{More generally, this should imply the analogy between actively interacting braids and particles that mediate interactions, which should potentially include super partners of gauge bosons.}. Note that in an exchange interaction, there does not exist an intermediate state in which only the virtual actively interacting braid is present because our def\/inition of a braid requires the presence of its two end-nodes. The following theorem summarizes the above as another main result. (The case of left exchange interaction is similar.)

\begin{theorem}\label{theoType2Int}
Given two adjacent braids, $B_1,B_2\in\mathfrak{B}^S$.
$B_1=\left.^{S_{1l}}\hspace{-0.5mm}[(T_{1a},T_{1b},T_{1c})_{X_{1A}X_{1B}}]^{S}\right.$
is on the left and has a reducible
crossing segment $X_{1B}$, and
$B_2=\left.^{S}\hspace{-0.5mm}[(T_{2a},T_{2b},T_{2c})_{X_2}]^{S_{2r}}\right.$,
there exists a braid $B\in\mathfrak{B}^b$,
$B=\left.^{S'}\hspace{-0.5mm}[(T_a,T_b,T_c)_{X_{1B}}]^{S}\right.$
with $S'=(-)^{|X_{1B}|}S$, such that it mediates the exchange
interaction of $B_1$ and $B_2$ to create $B'_1,
B'_2\in\mathfrak{B}^S$, i.e.,
\begin{gather*}
  B_1\overset{\rightarrow}{+}_{\mathrm{e}}B_2\rightarrow B'_1+B'_2
 =\left.^{S_{1l}}\hspace{-0.5mm}[((T_{1a},T_{1b},T_{1c})-\sigma^{-1}_{X_{1A}}(T_a,T_b,T_c))_{X_{1A}}]^{S'}\right.\nonumber\\
\hphantom{B_1\overset{\rightarrow}{+}_{\mathrm{e}}B_2\rightarrow B'_1+B'_2
 =}{}
+
\left.^{S'}\hspace{-0.5mm}[((T_a,T_b,T_c)+\sigma^{-1}_{X_{1B}}(T_{2a},T_{2b},T_{2c}))_{X_2}]^{S_{2r}}.\right.
\end{gather*}
\end{theorem}

The fourth remark is, if $X_{1B}=\mathbb{I}$, however, $(T_a,T_b,T_c)=(0,0,0)$, and hence $(T'_{1a},T'_{1b},T'_{1c})=(T_{1a},T_{1b},T_{1c})$, $(T'_{2a},T'_{2b},T'_{2c})=(T_a,T_b,T_c)$, and $S'=S$. That is, $B'_1=B_1$ and $B'_2=B_2$. Thus, the virtual actively interacting braid exchanged in the interaction is either $B_0^+$ or $B_0^-$, which were aforementioned to be analogous to gravitons.

Here is the f\/inal remark. For the same reason that a braid can decay in multiple ways, two braids can also have exchange interactions in more than one way, as opposed to direct interaction. This non-uniqueness of exchange interaction has an analogy in particle physics. For instance, quarks have both electric and color charges, both photons and gluons can mediate forces on quarks. The relation between actively interacting braids and bosons is not yet an actual identif\/ication, however. In fact, if each actively interacting braid corresponded to a boson, there would be too many ``bosons''. The underlining physics of how two braids can have exchange interactions in dif\/ferent ways deserves further study.

It should be emphasized that each individual exchange interaction is a process that yields a~unique result\footnote{With spin network labels the result is not unique any more because two topologically equal braids can be decorated by dif\/ferent sets of labels, and an interaction should result in a superposition of braids labelled dif\/ferently.}. An expression like $B_1\overset{\rightarrow}{+}_{\mathrm{e}}B_2$ is only formal. Only when the exact forms of~$B_1$ and~$B_2$ with their reducible segments are explicitly given does $B_1\overset{\rightarrow}{+}_{\mathrm{e}}B_2$ acquire a precise and unique meaning. In computing an exchange interaction, we have to specify our choice of the reducible crossing segment of the braid that gives out the virtual actively interacting braid. For any such choice Theorem~\ref{theoType2Int} holds.

The following Theorem shows that the additive and multiplicative conserved quantities of direct interaction in Theorem~\ref{theoConserve} are also conserved in the same manner under exchange interaction.
\begin{theorem}
Given two neighbouring stable braids, $B_1,B_2\in\mathfrak{B}^S$,
such that an exchange interaction $($left or right or both$)$ on them is
doable, i.e., $B_1+_{\mathrm{e}}B_2\rightarrow B'_1+B'_2$,
$B'_1,B'_2\in\mathfrak{B}^S$, the effective twist $\Theta$ is an
additive conserved quantity, while the effective state $\chi$ is a
multiplicative conserved quantity, namely
\begin{gather*}
\Theta_{B'_1}+\Theta_{B'_2}  \overset{+_{\mathrm{e}}}{=} \Theta_{B_1}+\Theta_{B_2},\qquad
\chi_{B'_1}\chi_{B'_2}  \overset{+_{\mathrm{e}}}{=}
\chi_{B_1}\chi_{B_2},
\end{gather*}
independent of the virtual actively
interacting braid being exchanged during the exchange interaction.
\end{theorem}

This Theorem is proved in \cite{Wan2008}. Consequently, exchanges of actively interacting braids give rise to
interactions between braids that are charged under the topological
conservation rules. The conservation of $\Theta$ is analogous to the
charge conservation in particle physics.

\subsection{Dynamics: CPT and braid Feynman diagrams}\label{subSecCPT}

In this section we discuss the charge conjugation, parity, and time reversal symmetries of stable braids, and the braid Feynman diagrams. We shall present some key results with a few remarks.

\subsubsection{C, P, and T}\label{subsubSecCPT}

Though not separately, as a theorem the combined action CPT is a symmetry in any Lorentz invariant, local f\/ield theory. Being a physical model of QFT, the SM respects CPT-symmetry too. This motivates the search for the possible discrete, non-equivalent, transformations of 4-valent braids and the testing of their correspondence with C, P, and T transformations. Whether our braids would eventually be mapped to or more fundamental than the SM particles, they should possess quantum numbers that are transformed by~C,~P, and~T. Conversely, investigating the action of discrete transformations on our braid excitations can help us to construct quantum numbers of a braid.

A large number of possible discrete transformations of the 4-valent braids exist, however, for example the permutation group $S_3$ on the triple of internal twists, several copies of $\mathbb{Z}_2$ that f\/lip a twist, an end-node state, and a crossing respectively, etc. The challenge then is to f\/ilter out the unwanted discrete transformations. Surprisingly, the braid dynamics introduced above turns out to  select exactly seven legal discrete transformations of the braids, as C, P, T, and their products add up to seven in total (eight including identity)~\cite{WanHe2008b}. Now we brief\/ly show how this ``super-selection'' works and direct the reader to~\cite{WanHe2008b,Wan2009} for details. Since QFT does not bear a~transformation that magically changes a particle to something else, the dynamics of the braids of embedded 4-valent spin networks, namely propagation and interaction, naturally constrains what discrete transformations are allowed on braids, which is summarized as a guideline in the following condition.

\begin{condition}\label{condLegalDT}A legal discrete transformation $\mathcal{D}$ on an arbitrary braid $B$ must be an automorphism on $\mathfrak{B}^b$, $\mathfrak{B}^f$, and $\mathfrak{B}^s$ separately.
\end{condition}

We expect the discrete transformations to be representation independent; therefore, the study of their ef\/fects should be made on braids in their generic forms (Fig.~\ref{braidGen}(a) and equation~(\ref{eqAlgNotation})). Table~\ref{tabDT} displays the result, in which explicit identif\/ication of the legal discrete transformations with C, P, and T is made.
\begin{table}[h]
\caption{The group of discrete transformations on a generic braid diagram. In column-3, a $-$ ($+$) means the propagation direction of the braid is f\/lipped (unaf\/fected). For comparison, column-4 is the action
of the group on a one-particle state, with 3-momentum $\mathbf{p}$, 3rd component $\sigma$ of spin~$J$, and charge~$n$. The primed and unprimed twists are related by $(T_{a^{\prime}%
},T_{b^{\prime}},T_{c^{\prime}})=(T_{a},T_{b},T_{c})\sigma_{X}$, as in Fig.~\ref{braidGen}(a).} \label{tabDT}

\vspace{1mm}

\centering
\begin{tabular}
[c]{c c c | c}
\toprule
Discrete  & Action on  & Prop- & Action on\\
Transformation & $B=\left.  _{T_{l}}^{S_{l}}\hspace{-0.5mm}[(T_{a}%
,T_{b},T_{c})_{X}]_{T_{r}}^{S_{r}}\right.  $ &
Direction & $\left\vert \mathbf{p},\sigma,n\right\rangle $\\[0.75ex]\midrule
$\mathds{1}$ & $\left.  _{T_{l}}^{S_{l}}\hspace{-0.5mm}%
[(T_{a},T_{b},T_{c})_{X}]_{T_{r}}^{S_{r}}\right.  $ &
$+$ & $\left\vert \mathbf{p},\sigma,n\right\rangle $\\[0.75ex]\midrule
$\mathcal{C}$ & $\left.  _{-T_{l}%
}^{\ \ \bar{S_{l}}}\hspace{-0.5mm}[-(T_{a},T_{b},T_{c})_{\mathcal{I}_{X}(X)}]_{-T_{r}%
}^{\bar{S_{r}}}\right.  $ & $+$ & $\propto\left\vert \mathbf{p},\sigma,n^{c}\right\rangle $\\[0.75ex]\midrule
$\mathcal{P}$ & $\left.  _{T_{r}}^{\bar{S_{r}}}\hspace{-0.5mm}[(T_{a^{\prime}%
},T_{b^{\prime}},T_{c^{\prime}})_{\mathcal{R}(X)}]_{T_{l}}^{\bar{S_{l}}}\right.
$ & $-$ & $\propto\left\vert -\mathbf{p},\sigma,n\right\rangle $\\[0.75ex]\midrule
$\mathcal{T}$ & $\left. _{T_{r}}^{S_{r}}\hspace{-0.5mm}[(T_{c^{\prime}%
},T_{b^{\prime}},T_{a^{\prime}})_{\mathcal{S}_{c}\mathcal{R}(X)}]_{T_{l}}^{S_{l}%
}\right.  $ & $-$ & $\propto(-)^{J-\sigma}\left\vert
-\mathbf{p},-\sigma,n\right\rangle$\\[0.75ex]\midrule
$\mathcal{CP}$ & $\left.  _{-T_{r}}^{\ \ S_{r}}\hspace{-0.5mm}[-(T_{a^{\prime}%
},T_{b^{\prime}},T_{c^{\prime}})_{X^{-1}}]_{-T_{l}}^{S_{l}}\right.
$ & $-$ & $\propto\left\vert -\mathbf{p},\sigma,n^{c}\right\rangle$\\[0.75ex]\midrule
$\mathcal{CT}$ & $\left.  _{-T_{r}}^{\ \ \bar{S_{r}}%
}\hspace{-0.5mm}[-(T_{c^{\prime}},T_{b^{\prime}},T_{a^{\prime}})
_{\mathcal{I}_{X}\mathcal{S}_{c}\mathcal{R}(X)}]_{-T_{l}}^{\bar{S_{l}}}\right.  $ &
$-$ & $\propto(-)^{J-\sigma}\left\vert -\mathbf{p},-\sigma
,n^{c}\right\rangle $\\[0.75ex]\midrule
$\mathcal{PT}$ & $\left.  _{T_{l}}^{\bar{S_{l}}%
}\hspace{-0.5mm}[(T_{c},T_{b},T_{a})_{\mathcal{S}_{c}(X)}]_{T_{r}}^{\bar{S_{r}}%
}\right.  $ & $+$ & $\propto(-)^{J-\sigma}\left\vert
\mathbf{p},-\sigma,n\right\rangle$\\[0.75ex]\midrule
$\mathcal{CPT}$ & $\left. _{-T_{l}}^{\ \
S_{l}}\hspace{-0.5mm}[-(T_{c},T_{b},T_{a})
_{\mathcal{I}_{X}\mathcal{S}_{c}(X)}]_{-T_{r}}^{S_{r}}\right.  $ &
$+$ & $\propto(-)^{J-\sigma}\left\vert \mathbf{p},-\sigma
,n^{c}\right\rangle $\\\bottomrule
\end{tabular}
\end{table}

In column~2 of Table~\ref{tabDT}, $\mathcal{R}$, $\mathcal{I}_X$, and $\mathcal{S}_c$ are discrete operations on the crossing sequence $X$ of a braid, respectively def\/ined by $\mathcal{R}:\ X=x_1x_2\cdots x_n\mapsto x_n x_{n-1}\cdots x_1$, $\mathcal{I}_X:\ X=x_1x_2\cdots x_n\mapsto x^{-1}_1x^{-1}_2\cdots x^{-1}_n$, and $\mathcal{S}_c$: $\forall\, x_i\in X$, $x_i\mapsto d$ if $x_i=u$ and $x_i\mapsto u$ if $x_i=d$. Hence, $X^{-1}=\mathcal{I}_X\mathcal{R}(X)$.  These operations are commutative and are elaborated in~\cite{WanHe2008b,Wan2009}.

In Table~\ref{tabDT}, we chose to denote C, P, and T transformations in the Hilbert space by calligraphic letters $\mathcal{C}$, $\mathcal{P}$, and $\mathcal{T}$ because braids are topological excitations of embedded spin networks that are the states in the Hilbert space describing the fundamental spacetime. One can easily check that the eight discrete transformations including identity in the f\/irst column of Table~\ref{tabDT} indeed form a group, which is actually the largest group of legal discrete transformations of 3-strand braids~\cite{WanHe2008b,Wan2009}. This is the f\/irst reason why they can be identif\/ied with C, P, T, and their products.

We emphasize again that these discrete transformations of braids are not equivalence moves; they take a braid to inequivalent ones, as seen in Table \ref{tabDT}. Staring at the 3rd column of the table, one readily f\/inds that some characterizing quantities of a braid, e.g., the ef\/fective twist and ef\/fective state, are invariant under some transformations but not under others. Table~\ref{tabDT} is obtained by utilizing these topological characterizing quantities, without involving spin network labels. We do not count in the phase and sign factors in the 4th column of Table~\ref{tabDT} either. All the transformations are restricted to local braid states, rather than a full evolution picture. Given these, surprisingly, the map between the legal discrete transformations of braids and those on single particle states appears to be unique.

According to column-4 in Table~\ref{tabDT}, the transformations $\mathcal{P}$, $\mathcal{T}$, $\mathcal{CP}$, and $\mathcal{CT}$ reverse the three momentum of a one-particle state. But then, what is the momentum of a braid? Fortunately, we need not to explicitly def\/ine the 3-momentum of a braid for the moment to tell the discrete transformations that can f\/lip the momentum because it should always agree with the braid's local propagation direction, however it is def\/ined~\cite{WanHe2008b}. Therefore, the discrete transformations reversing the 3-momentum of a braid are exactly those f\/lipping the propagation direction of the braid. This also helps to pin down Table~\ref{tabDT}.

\looseness=1
The ef\/fective twist $\Theta$ and ef\/fective state $\chi$ are representation-independent invariants and dynamically conserved quantities of a braid, while charges, e.g.\ electric and color charges, are quantum numbers of a particle. Hence, only~$\Theta$,~$\chi$, and functions of them can be candidates for certain charges of a braid. As we know, the electric charge of a particle is an integral multiple of~$1/3$, an additively conserved quantity, and a result of $U(1)$ gauge symmetry. The ef\/fective twist of a braid has three similar traits. We have seen the f\/irst two and now talk about the third  property. The framing that inf\/lates a spin network edge to a tube is in fact a $U(1)$ framing. That is, a tube is essentially an isomorphism from $U(1)$ to $U(1)$, which is characterized by its twists. A twist-free tube is an identity map, whereas a twisted tube represents a non-trivial isomorphism. These suggest we interpret $\Theta$ or an appropriate function of it as the ``electric charge'' of a braid, which may in turn explain the origin and quantization of electric charge.

Here is the f\/inal remark on Table~\ref{tabDT}. On a single particle state, a $\mathcal{CPT}$ has one more ef\/fect than a $\mathcal{C}$ because it also turns $\sigma$, the $z$-component spin, to $-\sigma$. We notice that in Table~\ref{tabDT}, the last transformation does more to the braid than the second transformation: It swaps the f\/irst  and the third elements in the triple of internal twists of the braid.  Although we do not know yet what features of a braid correspond to $\sigma$, we can now consider the last transformation in Table~\ref{tabDT} as $\mathcal{CPT}$. \cite{WanHe2008b,Wan2009} argue that spin network labels should play the role that determines the ``spin'' of a braid state.

C, P, and T group stable braids into CPT-multiplets. The braids in a multiplet are not equivalent but may share some traits. It is instructive to f\/ind how a CPT-multiplet of braids is characterized. Theorem~\ref{theoCPTmultiplet} shows that only CPT-multiplets of actively interacting braids have a topological character.

\begin{theorem}
\label{theoCPTmultiplet}
All actively interacting braids in a CPT-multiplet have the same number of crossings if each of them is in its unique representation. This number uniquely characterizes the CPT-multiplet.
\end{theorem}

The proof of the theorem can be found in \cite{WanHe2008b,Wan2009}. Theorem~\ref{theoCPTmultiplet} does not apply to non-actively interacting braids. In fact, we can always f\/ind two non-actively interacting braids with $m$ crossings ($m>1$), in their unique representations, which are not related to each other by any discrete transformation. For example, the 2-crossing braids $^{\hspace{0.75mm}S_l}\hspace{-0.5mm}[(T_a,T_b,T_c)_{ud^{-1}}]^{S_r}$ and $^{\hspace{0.75mm}S'_l}\hspace{-0.5mm}[(T'_a,T'_b,T'_c)_{uu}]^{S'_r}$ can never belong to the same CPT-multiplet, regardless of their internal twists and end-node states. Nevertheless, it is still true that {\it all the non-actively interacting braids in a CPT-multiplet have the same number of crossings if they are in the same type of representation}~\cite{WanHe2008b}. This is simply because the discrete transformations do not alter the representation type and the number of crossings of a braid.

Having seen the ef\/fects of C, P, and T on single braid excitations, we now discuss the action of these discrete transformations on braid interactions. {\it Braid interactions turn out to be invariant under CPT, and more precisely, under C, P, and T separately}~\cite{WanHe2008b,Wan2008}. By this we mean, say, for a direct interaction under a C, $\mathcal{C}(B)+_{\mathrm{d}} \mathcal{C}(B')=\mathcal{C}(B+_{\mathrm{d}}B')$, while under a P, it means $\mathcal{P}(B')+_{\mathrm{d}}\mathcal{P}(B) =\mathcal{P}(B+_{\mathrm{d}}B')$. Note that the P-transformation of a direct interaction swaps its direction. A subtlety arises in the case of T, however. An interaction involves the time evolution of a spin network. To apply our T-transformation to an interaction, one should reverse all the dynamical moves. Hence, a T-transformation turns a direct interaction into a~decay. That is, to show the invariance of $B+_{\mathrm{d}}B'\longrightarrow B''$ under time reversal, it suf\/f\/ices to show that $\mathcal{T}(B'')\overset{\rightsquigarrow}{\rightarrow} \mathcal{T}(B')+\mathcal{T}(B)$. Analogously, the invariance of an exchange interaction, $B_1\overset{\rightarrow}{+}_{\mathrm{e}}B_2\rightarrow B'_1+B'_2$, under time reversal reads $\mathcal{T}(B'_2)\overset{\leftarrow}{+}_{\mathrm{e}}\mathcal{T}(B'_1)\rightarrow\mathcal{T}(B_2)+\mathcal{T}(B_1)$. The case of braid decay follows similarly~\cite{Wan2008}.

This observation of the absence of CP-violation in our theory does not comply with the SM of particles, which seems to infer an issue that the interactions of braids are deterministic, in the sense that an interaction of two braids produces a def\/inite new braid. Nevertheless, this may not be a problem at all because we have only worked with def\/inite vertices of interactions. In terms of vertices we have a~def\/inite result for an interaction as to the case in QFT; this is similar to what has been done in SF models or GFTs. Besides, one can certainly argue that if our braids are more fundamental entities, the CP-violation in particle physics need not hold at this level. Putting this CP-violation problem aside, however, a fully quantum mechanical picture should be probabilistic\footnote{One should note that a few theoretical physicists may not agree on this.}.

If the absence of CP-violation was truly an issue, we would consider braids with the same topological structure but a~dif\/ferent sets of spin network labels as physically dif\/ferent. One may adapt some SF methods to assign amplitudes to the adapted dual Pachner moves of the braided ribbon networks. An evolution move may then yield outcomes with the same topological conf\/iguration but dif\/ferent spin network labels; each outcome has a certain probability amplitude. As a result, an interaction of two braids may give rise to superposed braids, each of which has a certain probability to be observed, with the same topological content but dif\/ferent set of spin network labels. With this, CP-violating interactions may arise.

Note that the current study of discrete transformations of braids would not be impacted by just adding spin network labels in a straightforward way in to our scheme. One reason is that the discrete transformations of the braids do not change the spin network label of each existing edge of the network. One may try to construct discrete transformations that change the spin network labels on braids, but one does not have {\it a priori} a reason to make a special choice among the many arbitrary ways of doing this.

\subsubsection{Asymmetry of braid interaction}

Both direct and exchange interactions are asymmetric. A brief description is as follows. The statement that direct interaction is asymmetric means given an actively interacting braid $B$ and an arbitrary braid $B'$, in general either of the direct right interaction $B+_{\mathrm{d}} B'$ or the left interaction $B'+_{\mathrm{d}} B$ cannot occur because of the violation of the corresponding interaction condition. Even if both interactions are feasible, $B+_{\mathrm{d}} B'$ and $B'+_{\mathrm{d}} B$ are two inequivalent braids in general, which reads $B+_{\mathrm{d}} B'\ncong B'+_{\mathrm{d}} B$. Two exceptions exist. In the cases where $B$ and $B'$ meet certain constraints, $B+_{\mathrm{d}} B'$ and $B'+_{\mathrm{d}} B$ can simply be equal~\cite{WanHe2008a}.

On the other hand, interestingly, $B+_{\mathrm{d}}B'$ and $B'+_{\mathrm{d}}B$ can be related by discrete transformations. Because P-transformation swaps the two braids undergoing a direct interaction, i.e., $\mathcal{P}(B+_{\mathrm{d}}B')=\mathcal{P}(B')+_{\mathrm{d}}\mathcal{P}(B)$, we immediately have
\begin{gather*}
B'+_{\mathrm{d}}B=
\mathcal{P}(B+_{\mathrm{d}}B'),\qquad \mathrm{if}\quad B=\mathcal{P}(B),\qquad B'=\mathcal{P}(B'),
\end{gather*}
where the $\mathcal{P}$ can be replaced by $\mathcal{CP}$ by the same token. Note that, however, time reversal cannot relate $B+_{\mathrm{d}}B'$ and $B'+_{\mathrm{d}}B$ because it turns a direct interaction into a decay.

As aforementioned, the set of actively interacting braids is closed under direct interaction. This and the asymmetry of direct interaction then give rise to the following theorem~\cite{WanHe2008a,WanLee2007}.
\begin{theorem}
The set of actively interacting braids $\mathfrak{B}^b$ is an algebra under direct interaction, namely $\mathfrak{B}^b+_{\mathrm{d}}\mathfrak{B}^b=\mathfrak{B}^b$. This algebra is associative and non-commutative.
\end{theorem}

It follows that braid decay is also asymmetric but we shall skip this and move on to the asymmetry of the exchange interaction, which is subtler.

The asymmetry of the exchange interaction is three-fold. Firstly, for $B_1,B_2\in\mathfrak{B}^S$, in general $B_1\overset{\rightarrow}{+}_{\mathrm{e}} B_2\ncong B_2\overset{\rightarrow}{+}_{\mathrm{e}} B_1$ ($B_1\overset{\leftarrow}{+}_{\mathrm{e}} B_2\ncong B_2\overset{\leftarrow}{+}_{\mathrm{e}} B_1$), which is called the {\bf asymmetry of the f\/irst kind}. Secondly, in general $B_1\overset{\rightarrow}{+}_{\mathrm{e}} B_2\ncong B_1\overset{\leftarrow}{+}_{\mathrm{e}} B_2$, which is termed the {\bf asymmetry of the second kind}. The {\bf asymmetry of the third kind} states that generically $B_1\overset{\rightarrow}{+}_{\mathrm{e}} B_2\ncong B_2\overset{\leftarrow}{+}_{\mathrm{e}} B_1$. As in the case of direct interaction, when $B_1$ and $B_2$ satisfy certain constraints, symmetric exchange interactions arise; however, a subtlety should be noted. Since two braids may have dif\/ferent exchange interactions, $B_1\overset{\rightarrow}{+}_{\mathrm{e}} B_2$ and $B_2\overset{\rightarrow}{+}_{\mathrm{e}} B_1$ cannot be equal for all possible interactions of~$B_1$ and~$B_2$ may interact. The right question to ask is, taking the right exchange interaction as an example: For any $B_1$ and $B_2$, does there exist an instance of $B_1\overset{\rightarrow}{+}_{\mathrm{e}} B_2$ and one of $B_2\overset{\rightarrow}{+}_{\mathrm{e}} B_1$ among all possible ways of these two interactions, such that $B_1\overset{\rightarrow}{+}_{\mathrm{e}} B_2= B_2\overset{\rightarrow}{+}_{\mathrm{e}} B_1$? \cite{Wan2008}~answers this question for the f\/irst two kinds of asymmetry. This asymmetry of the third kind is new and not studied in \cite{Wan2008} but it would not be hard by following the derivations in~\cite{Wan2008}.

Like direct interactions, asymmetric exchange interactions may be related by discrete transformations, but only for the asymmetry of the third kind. Since in the asymmetry of the third kind, the positions of the braids and the interaction direction are both swapped, according to Table~\ref{tabDT}, we immediately see that only P and CP are able to do this because they are the only ones that can f\/lip a braid horizontally without changing the vertical order of the braid's internal twists. Therefore, we obtain equation~(\ref{eqAsymmEIntCPT}), in which $\mathcal{P}$ can be replaced by $\mathcal{CP}$,
\begin{gather}
B_2\overset{\leftarrow}{+}_{\mathrm{e}} B_1=
\mathcal{P}(B_1\overset{\rightarrow}{+}_{\mathrm{e}} B_2),\qquad \mathrm{if}\quad B_1=\mathcal{P}(B_1),\qquad B_2=\mathcal{P}(B_2).
\label{eqAsymmEIntCPT}
\end{gather}

\subsubsection{Braid Feynman diagrams}
An ef\/fective theory of the dynamics of 4-valent braids based on Feynman diagrams is possible, which are called {\bf braid Feynman diagrams}. Unlike the usual QFT Feynman diagrams having no internal structure, each braid Feynman diagram is an ef\/fective description of the whole dynamical process of a braid interaction and has internal structures that record the evolution of the braid and its surroundings.

We use {\includegraphics[height=0.1219in,
width=0.6417in]{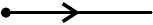}} and {\includegraphics[height=0.1219in,
width=0.6417in]{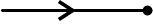}} for respectively outgoing and ingoing
braids in $\mathfrak{B}^f$,
{\includegraphics[height=0.1219in, width=0.6417in]{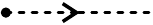}} and
{\includegraphics[height=0.1219in, width=0.6417in]{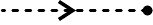}} for
respectively outgoing and ingoing non-actively propagating
braids\footnote{These braids can still propagate in an induced way.}
in $\mathfrak{B}^s$. Outgoing and ingoing braids in
$\mathfrak{B}^b$ are better represented by
{\includegraphics[height=0.1643in, width=0.5025in]{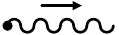}} and
{\includegraphics[height=0.1643in, width=0.5025in]{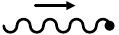}}
respectively.

\begin{figure}[h]
\centering
\includegraphics[scale=0.7]{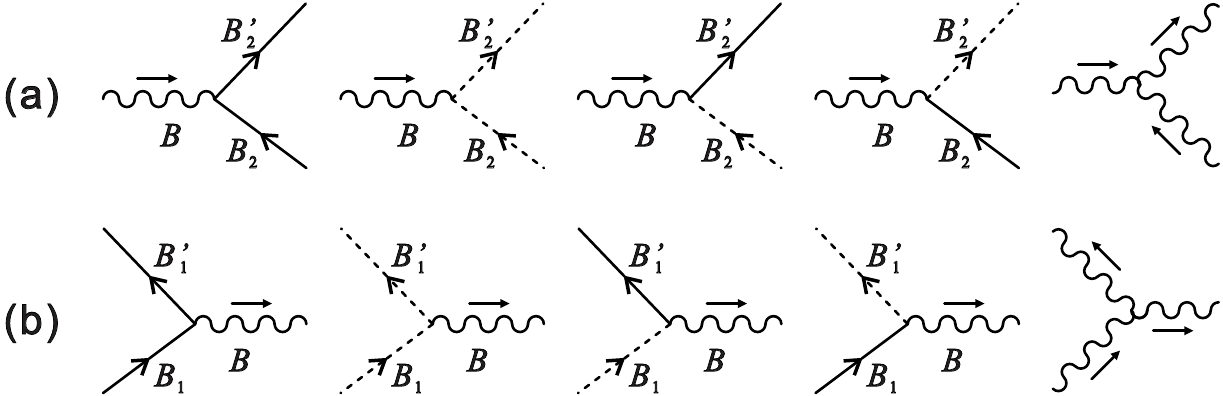}
\caption{(a) All vertices of direct right-interaction. (b) All vertices of right-decay. Time f\/lows up.}
\label{rType1Feynman}
\end{figure}

In accordance with left and right-decay, we will henceforth denote
left and right direct interactions by
$\overset{\leftarrow}{+}_{\mathrm{d}}$ and
$\overset{\rightarrow}{+}_{\mathrm{d}}$ respectively. Note that if the two interacting braids are both in~$\mathfrak{B}^b$, the direction of the direct interaction is irrelevant because the result is independent of which of the two braids plays the active role in the interaction. Since
$\mathfrak{B}^b\overset{\rightarrow}{+}_{\mathrm{d}}\mathfrak{B}^b\subseteq\mathfrak{B}^b$
and
$\mathfrak{B}^b\overset{\rightarrow}{+}_{\mathrm{d}}(\mathfrak{B}^f\sqcup\mathfrak{B}^s)\subseteq\mathfrak{B}^f\sqcup\mathfrak{B}^s$,
the only possible single vertices of right direct interaction and of right decay are respectively listed in Fig.~\ref{rType1Feynman}(a) and (b), whose left-right mirror images are vertices of direct left-interaction and left-decay.
The arrows over the wavy lines in Fig.~\ref{rType1Feynman} are important for they dif\/ferentiate a direct interaction from a decay.

\begin{figure}[h]
\centering
\includegraphics[scale=0.7]{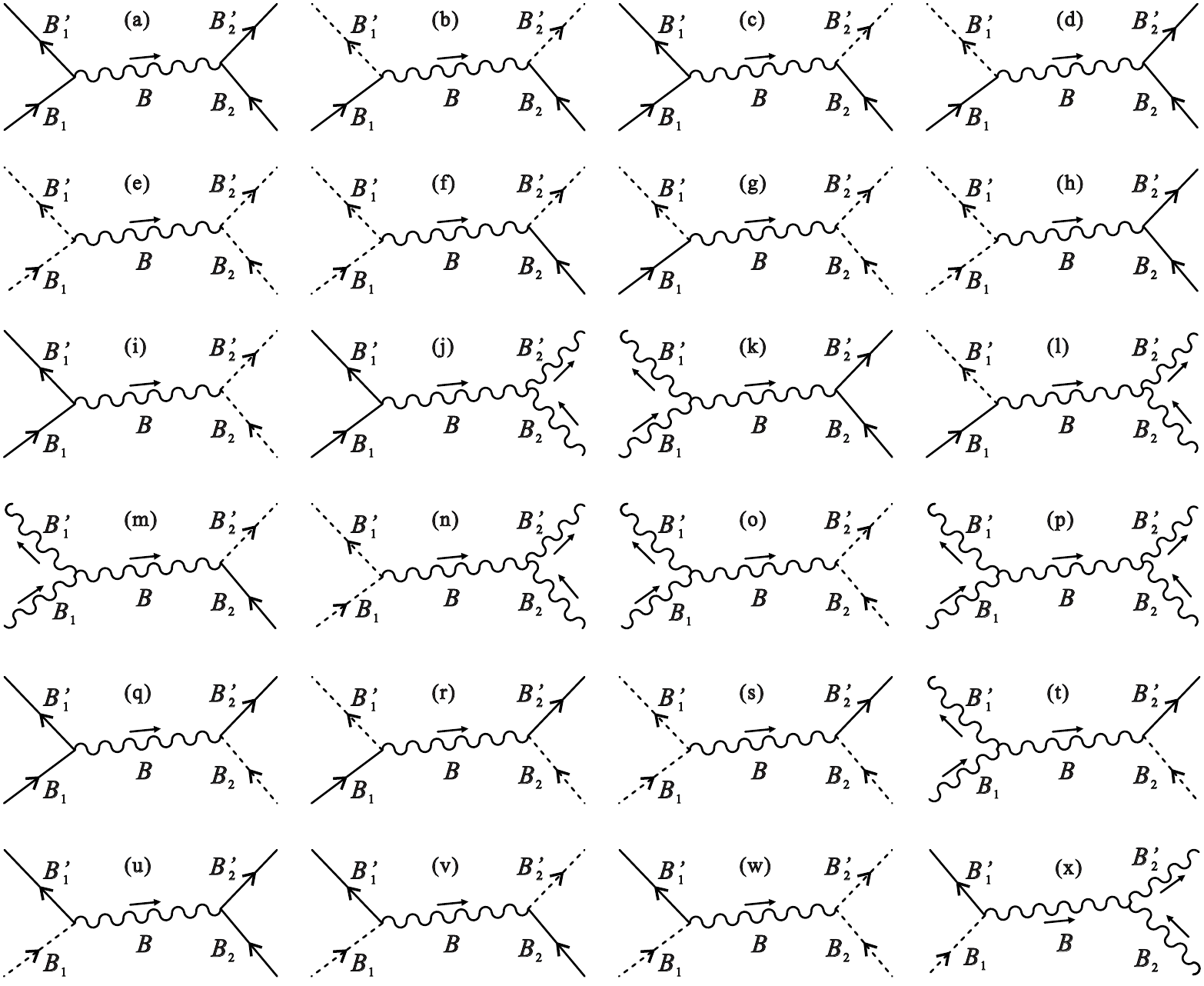}
\caption{Possible right exchange interaction 2-vertices. Time f\/lows
up.} \label{rType2Feynman}
\end{figure}
Fig.~\ref{rType2Feynman} depicts all possible basic 2-vertex diagrams for right exchange interaction, whose left-right mirror images are certainly the basic diagrams for left exchange interaction. These diagrams manifest the invariance of the exchange interaction under C, P, T and their products.

The diagram of an exchange interaction manifests whether it allows symmetric instances. Taking the asymmetry of the f\/irst kind as an example, if a diagram looks formally the same as its left-right mirror image with the arrow over the virtual braid not mirrored, the corresponding exchange interaction allows symmetric instances. It follows that Fig.~\ref{rType2Feynman}(a), (b), (e), (f), (p), and (u)-(x) are such diagrams.

The diagrams in Figs.~\ref{rType1Feynman} and~\ref{rType2Feynman} crystallize the analogy between braids in $\mathfrak{B}^b$ and bosons, as the topological conservation laws permit them to be singly created and destroyed and as exchanges of these excitations give rise to interactions between braids charged under the topological conservation rules.

Multi-vertex and loop braid Feynman diagrams can be constructed out of these basic vertices. As a result, there should exist an ef\/fective f\/ield theory based on these diagrams, in which the probability amplitudes of each diagram can be computed. For this one should f\/igure out the terms evaluating external lines, vertices, and propagators of braids. In a more complete sense, an action of the ef\/fective f\/ields representing braids that can generate these braid Feynman diagrams should be devised. Each ef\/fective f\/ield is a function of the representations of the group elements in the characterizing 8-tuple (and spin network labels if necessary) of a braid; these representations label a line in the corresponding braid Feynman diagram. An easier task is to assign a reasonable probability amplitude of each braid Feynman diagram. In either case, the very f\/irst challenge is to f\/ind an appropriate mathematical language to study the 4-valent scheme analytically. In the next section we will brief\/ly mention three possible formalisms.
\subsection{Discussions and outlook}\label{subSecDisc}
The 4-valent scheme resolves some issues and limitations persisting in the 3-valent approach but also raises new issues. In the f\/irst place, we obtain a $(3+1)$-dimensional evolution of quantum states of spacetime, which has intrinsic dynamics of the braid excitations of these states. Because of the framing and embedding of spin networks, strands of a braid excitation admit twists only in units of 1/3. The twists of a braid are directly related to its electric charge, which naturally, rather than by hand, gives rise to charge quantization and fractional charges such as quark charges. The 4-valent theory also contains another natural selection: braid dynamics magically picks out a group of exactly eight discrete transformations, including the identity, which can be identif\/ied with analogues of C, P, T, and their products.

Some issues the 4-valent scheme raises have been discussed more or less previously. In what follows, we shall analyse some issues that bear on the interpretation of these results. In the last section, however, we shall introduce ideas, future work, and work in progress, which may remove these issues.

\subsubsection*{Stability and locality}
We argued that the stable braids are local excitations of 4-valent braided ribbon networks because they are noiseless subsystems of the networks. Nevertheless, the comparison with topological f\/ield theories that do not bear any local degrees of freedom seems to set the locality of 4-valent braids in doubt. In fact, locality in background independent theories of quantum gravity is delicate, as it is correlated with two other important issues, namely the problem of the concept of spacetime and that of a low energy limit. Moreover, the locality and stability of a braid are also entangled.

A background independent quantum gravity theory usually lacks a metric that directly measures spatial locality. But the graph metric of a spin network may def\/ine the network locality. In this sense, a stable braid is local because it is conf\/ined between two nodes. Unfortunately, the issue in the stability of a braid may damage its locality because Condition \ref{conEvolMove} only protects a~stable braid from being undone but does not prevent the braid's end-nodes from being expanded by $1\rightarrow 4$ moves. In the latter scenario, the three strands of a braid may turn out to be attached to nodes far from each other on the network, causing the braid to be nonlocal. We might strengthen the stability condition by further forbidding the action of a $1\rightarrow 4$ move on either end-node of a stable braid but the price is
the loss of braid decay.

Moreover, the locality discussed above is micro-locality, opposed to which is
macro-locality that is def\/ined in the low energy end of
the theory. Markopoulou and Smolin proposed these two notions of
locality and found that they do not match in
general~\cite{fotini2003}, which is also exemplif\/ied in~\cite{Wan2005}. The quantum spacetime in our context is pre-geometric, as
it is a sum over quantum histories of superposed pre-geometric spin
networks; it is conjectured that continuous spacetime emerges as
a~certain limit of this quantum spacetime. Hence, requiring
the micro-locality def\/ined on each spin network to match the
macro-locality in continuous spacetime makes no sense.

Macro-locality is more relevant to the known physics; however, it is
obtained from micro-locality. This leads back to the problem of low
energy limit, to resolve which Markopoulou et al.\ adapted the
idea of noiseless subsystems with micro-symmetries from Quantum
Information. Therefore, we expect that the symmetry of the braid
excitations will induce emergent symmetries, including time and
space translation invariance, in the low energy ef\/fective
description of the braids.

\subsubsection*{Particle identif\/ication and mass}
The ultimate physical content of the 4-valent scheme is not fully comprehensible at this stage. In the trivalent scheme, \cite{LouNumber} tentatively maps the trivalent braids to SM particles. Whether such a map exists in the 4-valent scheme is as yet obscure. One reason is that although the dynamics of 4-valent braids strongly constrains the def\/ining 8-tuple of a stable braid (in particular the actively interacting braids), the closed form of this constraint is still missing. Consequently, we lack  a means to pick out the 4-valent braids that may be mapped to the SM particles. Nonetheless, we are inclined to consider the prospect that braid excitations are fundamental matter whose low energy ef\/fective theory yields the SM particles.

If the latter is true, the potential instability and non-locality of stable braids may not be an issue because only the low energy ef\/fective counterparts of the braids are physically relevant.

We may also ask how mass arises. There appear to be two possibilities. First, a braid may acquire zero or nonzero mass from some of its intrinsic attributes. Second, mass is not well-def\/ined at the level of spin networks but is emergent in the low energy limit, directly or via certain symmetry breaking. The latter requires working out the ef\/fective theory, which is our future work. As to the former, the number of crossings in a braid can be a candidate for its mass (this is also conjectured in the trivalent scheme~\cite{LouNumber, Bilson-Thompson2006}). Here is the logic. The number of crossings of an actively interacting braid in its unique representation uniquely characterizes the CPT-multiplet the braid belongs to; hence, this number cannot be the charge (already mapped to the braid's ef\/fective twists) or 3-momentum of the braid but probably related to the energy of the braid. Besides, actively interacting braids are equivalent to trivial braids, whereas non-actively interacting ones are not. If we associate a braid's mass to its number of crossings, all actively interacting braids seem massless, consistent with their analogue with (gauge) bosons. and most non-actively interacting braids are massive because they are not fully reducible.

\subsubsection*{Further questions}
\begin{itemize}\itemsep=0pt
\item At the level of spin networks, is there a quantum statistics of braids that can turn the analogy between actively interacting braids and bosons af\/f\/irmative? If true, are there anyonic braid states?
\item Both trivalent and 4-valent schemes need a mechanism to create nontrivial braids on unbraided networks\footnote{Trivial braids $B^{\pm}_0$ in equation~(\ref{eqB0pm}) can be otherwise created and annihilated, in the 4-valent scheme.}. Section~\ref{subsubSecFwork} discusses a possible way out.
\item Our ansatz that spacetime is fundamentally discrete is partly inspired by the LQG area and volume operators with discrete spectra. Nevertheless, whether these area and volume operators are physical is still under debate~\cite{Dittrich2007vol, Rovelli2007comment}. On the one hand, these operators are not gauge invariant. On the other hand, the areas and volumes that we routinely measure are associated to spatial regions determined by matter~\cite{Rovelli1996vol, Rovelli1995AandV} but LQG was devised to be a theory of gravity only. Now that our program of emergent matter shows that matter may be encoded in LQG as emergent braid excitations of spin networks, it may help to settle the debate.

\item Our program of emergent matter is also related to Quantum Graphity, a class of general theories of background independent quantum gravity based on graphs~\cite{Konopka2008, Konopka2006}. \cite{Hamma2008LRb} f\/inds the speed of light\footnote{This is understood as the maximum speed at which information can propagate in a system.} as a Lieb--Robinson bound~\cite{LRbound1972} in certain Quantum Graphity models. As both trivalent and 4-valent braids can propagate, does a Lieb--Robinson bound of braid propagation exist? The 4-valent scheme expects that actively interacting braids saturate the Lieb--Robinson bound of the system but non-actively interacting ones do not, such that they are respectively massless and massive.
\end{itemize}

\subsubsection{Future directions}\label{subsubSecFwork}
We now sketch our plan of reformulating the 4-valent scheme or even our whole program of emergent matter in other frameworks of mathematical physics, such as GFT, Tensor Category Theory, and so on.

{\bf Group f\/ield theories with braids.}
GFTs\footnote{The f\/irst GFT -- the Boulatov model -- originated as a generalization of the Matrix Models of 2D gravity to 3D~\cite{Boulatov1992}. GFTs in 3D and 4D were realised to be generating theories of Spin Foam models~\cite{Freidel2005gft, Oriti2005}. Later, GFTs are suggested to be fundamental formulations of quantum gravity~\cite{Oriti2007}. \cite{OritiThesis} presents an extensive review on the subject.} consider d-dimensional simplicies the fundamental building blocks of $(d+1)$-dimensional spacetime and treat them as f\/ields whose variables are elements in the group def\/ining the simplicies. That is, a GFT is a ``local, covariant quantum f\/ield theory of universes'' in terms of the f\/ields associated with the fundamental building blocks. It would produce a transition amplitude between quantum ``universes'' by summing over the Feynman diagrams of this transition, i.e., summing over all triangulations and topologies as the histories built from the evolution of the fundamental $d$-simplicies. These Feynman diagrams can also be viewed as spin networks and dual to $(d+1)$-simplicies. Group Field Theories encompass most of the other approaches to non-perturbative quantum gravity, such as Loop Quantum Gravity and Spin Foam models, provide a link between them, and go beyond the limitations of them~\cite{Oriti2007}.

We name two viable routes of formulating a GFT of 4-valent braids. Spin networks are purely combinatoric and unembedded in GFTs, so the f\/irst strategy is to enlarge the conf\/iguration space of a~certain $(3+1)$ GFT by adding to its fundamental f\/ield group variables that characterize a~4-valent braid.

Inspired by constructing theories of collective modes in condensed matter physics, our second, simpler strategy is to devise a braid f\/ield as a composite f\/ield of a pair of fundamental group f\/ields whose group variables are identif\/ied in a braided way, and then integrate out the fundamental f\/ields to obtain an ef\/fective theory of the composite f\/ields in certain backgrounds given by the fundamental ones.

Both ways combine spin network labels automatically and are expected to result in a low energy ef\/fective theory of braid excitations in a background spacetime. The former sounds more fundamental and should be able to solve the issue that nontrivial braids cannot be created from spin networks initially free of braids. The latter is what we are currently taking, by which we found it is likely to construct a toy GFT with only certain trivial braids, whose ef\/fective theory is a scalar $\phi^4$ theory.

{\bf Tensor categorical methods.}
Braided tensor categories~\cite{KasselBook} appear to be another elegant and unif\/ied way to resolve many aforementioned issues once and for all. In fact, the connection between LQG and SF Models and Tensor Categories has been recognized for decades~\cite{Crane1991,Kauffman1992,Kauffman2007}. Note that the string-net condensate due to Wen  et al.~\cite{Wen2008tensor1, Wen2008tensor} also illuminates that tensor categories may be the language underlying a unif\/ication of gravity and matter. Braided tensor categories can free our program from embedding by casting both trivalent and 4-valent braids combinatorially~\cite{Kauffman1992}, which is beyond the context of LQG.

A twist of a strand of a braid can be interpreted as characterizing a non-trivial isomorphism from $U(1)$ to $U(1)$. Nonetheless, the concept of twist can be generalized to any vector spaces, which is how it is def\/ined in braided tensor categories. In this manner, we may view spin network labels as if they represent generalized framing of spin networks other than the $U(1)$ framing we have just studied, such that generalized twists can arise, which may of\/fer a unif\/ication of our twists and spin network labels, as well as of internal symmetries and spacetime symmetries.

The end-nodes and external edges of 4-valent braids may exert further constraints on what tensor categories are at our disposal or even motivate new types of tensor categories. Tensor-categorized 4-valent braids and evolution moves may be evaluated by the relevant techniques already def\/ined in theories of tensor categories or new techniques adapted to our case.

\subsubsection{Relation to topological quantum computing}
One should not be surprised to notice that the 4-valent scheme of emergent matter is related to Topological Quantum Computing (TQC). This relation has three facets. Firstly, though seemingly superf\/icial, braids and their algebra are present in both disciplines. A major dif\/ference is that each 4-valent braid have two end-nodes and has only three strands, which is not the case in TQC.

Secondly, as aforementioned, being a concept rooted in Quantum Computing/Information and adapted to models of quantum gravity, noiseless subsystems are a key underlying notion of the program of emergent matter. Furthermore, \cite{fotini2005kribs, fotini2007bi} suggest that background independent quantum gravity is a quantum information processing system. On the other hand, in~\cite{Wen2006gravity,Wen2004light, Wen2005phase} topological quantum phase transitions have proven to give rise to emergent gauge and linearized gravitons.

Thirdly, one of our future directions is to employ tensor categories~-- in particular braided ribbon categories~-- to make an elegant reformulation of the 4-valent scheme, while TQC is also naturally described in the language of tensor categories~\cite{Rowell2008tqc, Wang2006tqc} and related to framed spin networks~\cite{Kauffman2007}.

Therefore, it is interesting to study TQC from the viewpoint of quantum gravity and vice versa, which may shed new light on both disciplines. For example, we may interpret each 4-valent braid as representing a process of quantum computation, with an end-node of the braid as a~fusion rule of anyons or a quantum gate in TQC. We wonder if the interactions of 4-valent braids can be introduced to TQC to study how two quantum processes can join, how one quantum process can split, and when two sequences of quantum processes can be equivalent. Conversely, TQC may be useful in understanding the signif\/icance of the conserved quantities of 4-valent braids.

\section{A unif\/ied formalism}\label{secJon}
Recently in \cite{Hackett2011a} the trivalent nodes were recast into the tetravalent scheme, giving a consistent footing to study which results from each scheme could be transferred over to the other. Here we reproduce the unif\/ied def\/inition of braided ribbon networks of valence $n$ (with $n \geq 3$) as follows:
\begin{itemize}\itemsep=0pt
\item We begin by considering an $n$-valent graph embedded in a compact $3$ dimensional manifold. We construct a $2$-surface from this by replacing each node by a $2$-sphere with $n$ punctures ($1$-sphere boundaries on the $2$-sphere), and each edge by a tube which is then attached to each of the nodes that it connects to by connecting the tube to one of the punctures on the $2$-sphere corresponding to the node.

\item Lastly we add to each tube $n-1$ curves from one puncture to the other and then continue these curves across the sphere in such a way that each of the $n$ tubes connected to a node shares a curve with each of the other tubes.

\item We will freely call the tubes between spheres \textit{edges}, the spheres \textit{nodes} and the curves on the tubes \textit{racing stripes} or less formally \textit{stripes}.

\item We will call a braided ribbon network the equivalence class of smooth deformations of such an embedding that do not involve intersections of the edges or the racing stripes.
\end{itemize}

We immediately face the following consequence: under this def\/inition\footnote{See~\cite{Bianchi2011} for another point of view.} there are only braided ribbon networks of valence $2$, $3$ or $4$ (with valence $2$ being a collection of framed loops). To see this fact we consider a $5$-valent node~-- a $2$-sphere with $5$ punctures, with each puncture connected to each other puncture by a non-intersecting curve. Taking each puncture as a node, and the curves as edges, we then get that these objects would constitute the complete graph on $5$ nodes and as they lie in the surface of a $2$-sphere, such a graph would have to be planar. This is impossible by Kuratowski's theorem~\cite{kk}: the complete graph on $5$ nodes is non-planar. Likewise, we have for any higher valence $n$ that the graph that would be constructed would have the complete graph on $5$ nodes as a subgraph, and so they too can not be planar. If the reader desires an intuition for this, it may be instructive to recall that these statements follow from the four colour theorem~-- the existence of such a node would imply the existence of a map requiring f\/ive (or more) colours.
\begin{figure}[!h]
\centering
\includegraphics[scale=0.4]{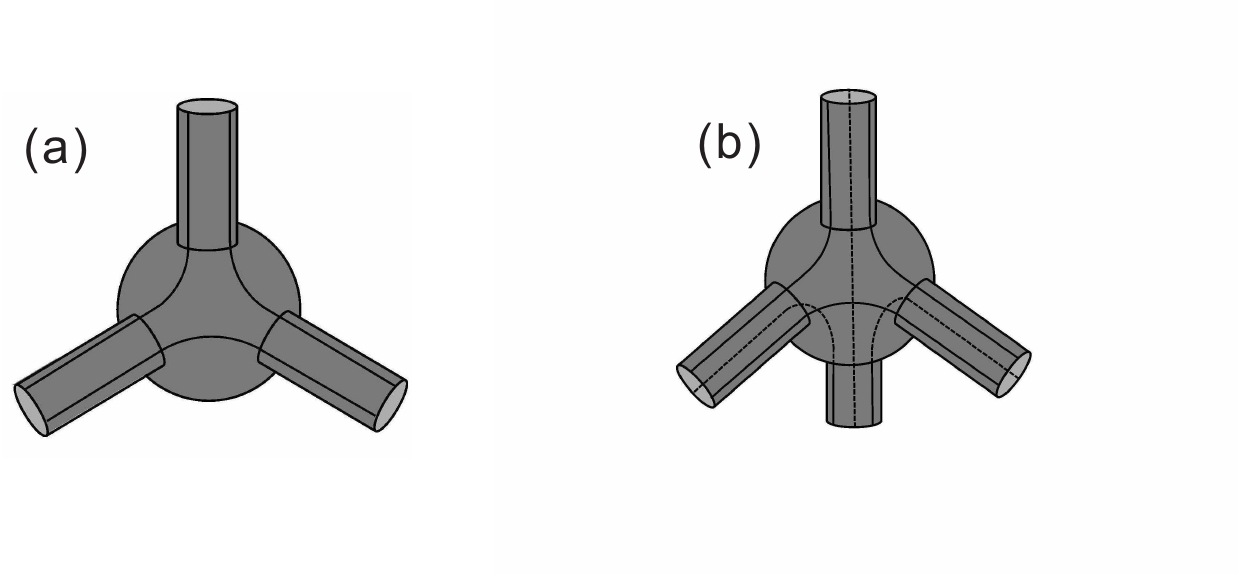}
  \caption{(a) Trivalent node. (b) Four-valent node.}
  \label{13}
\end{figure}

We can also introduce a modif\/ication to the framework that allows for higher valence vertices. To do this we f\/irst make a few def\/initions.

\begin{definition}
We def\/ine the \textbf{natural valence} of a braided ribbon network to be the number of racing stripes on each edge.
\end{definition}

\begin{definition}
We say that a node is \textbf{natural} if each of the tubes which intersect share a~racing stripe with each of the other tubes.
Otherwise we will say that a node is \textbf{composite}.
\end{definition}

\begin{figure}[!h]
  \centering
    \includegraphics[scale=0.4]{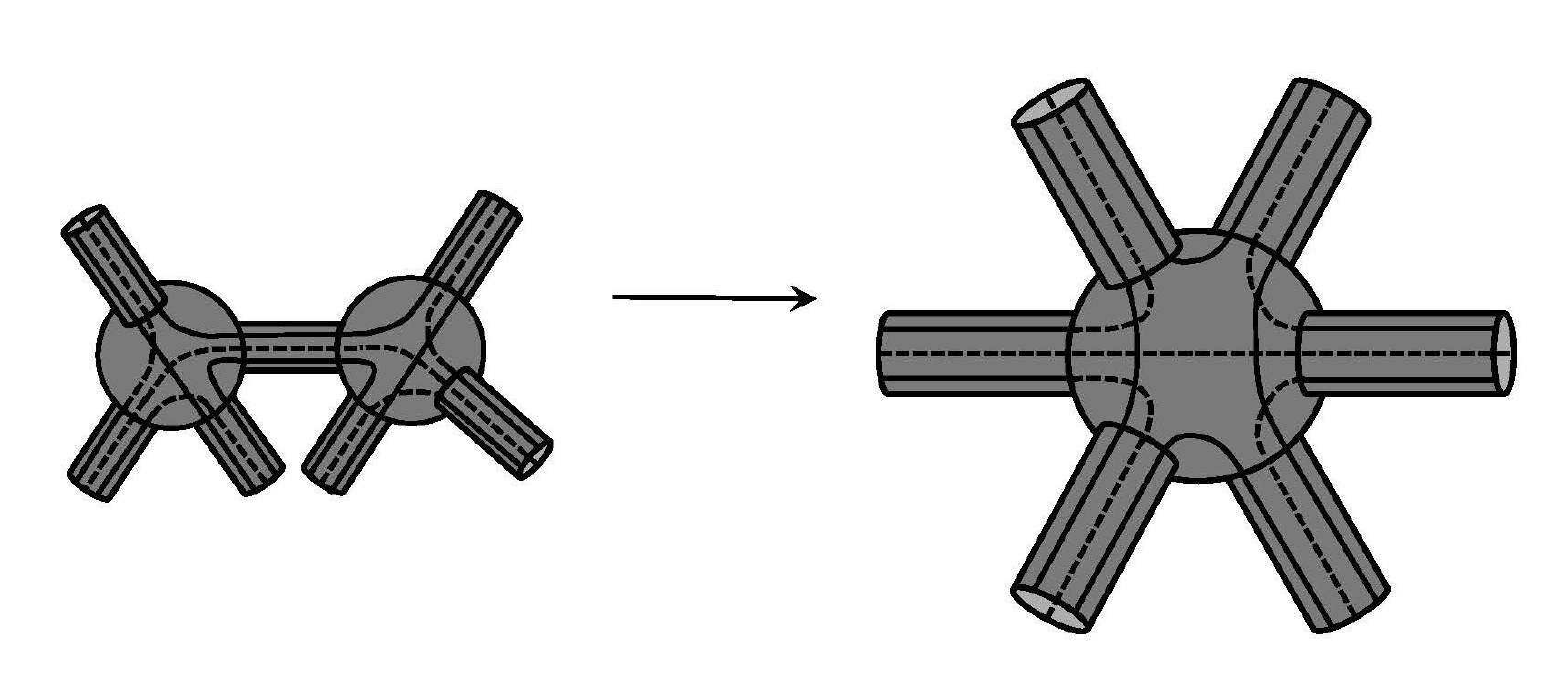}
    \includegraphics[scale=0.4]{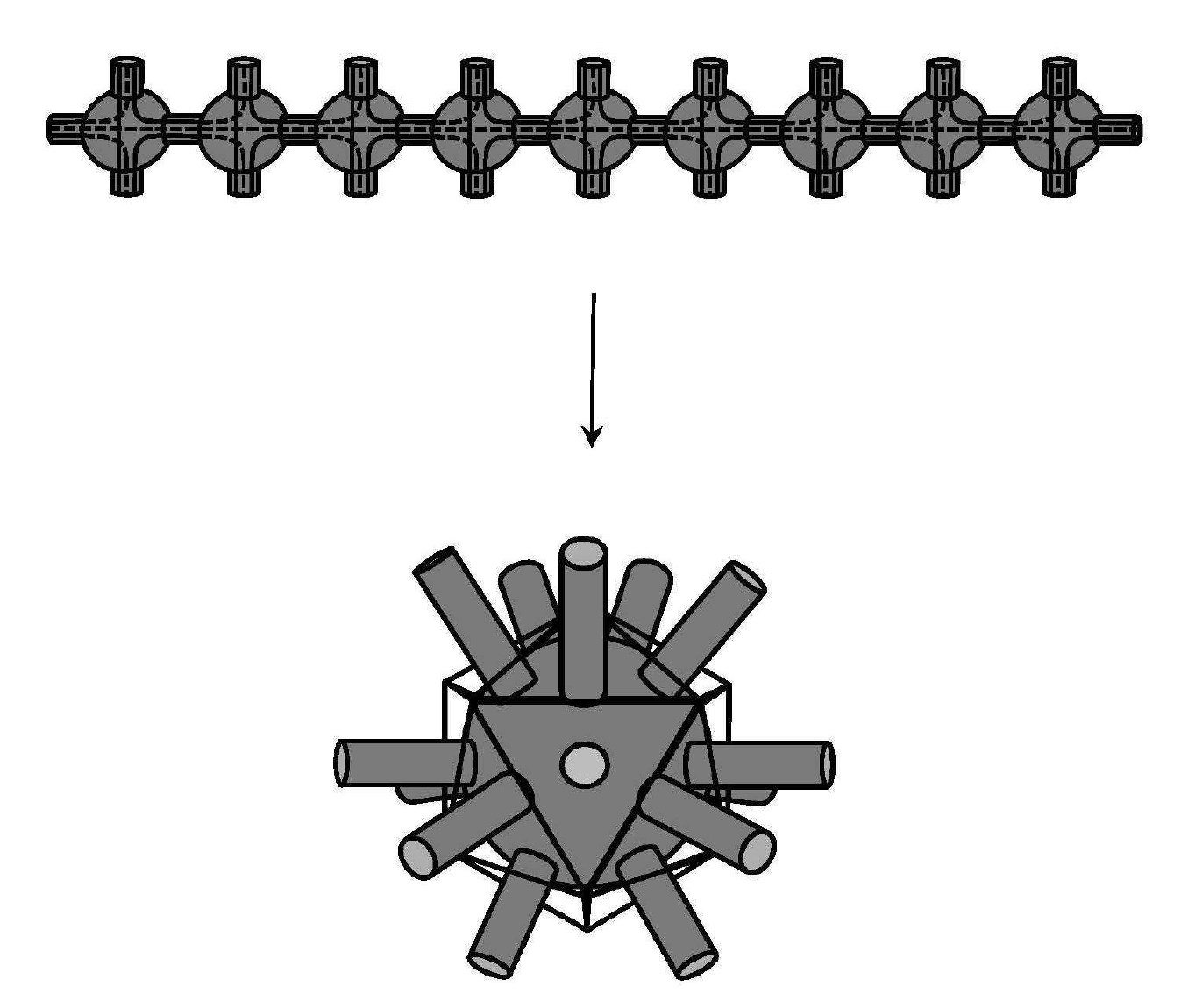}
\caption{Forming composite nodes.}\label{composite}
\end{figure}

We can then def\/ine a $n$-valent BRN with natural valence $m$ (here $n$ can take values of $n = km-2(k-1)$ for any integer $k$) as a braided ribbon network where each of the nodes has $n$ tubes which intersect it but where each of the tubes has $m-1$ racing stripes. Likewise we can def\/ine a multi-valent BRN with natural valence $m$ in a similar manner but without f\/ixing the value of~$k$ for all nodes. We then construct composite nodes by connecting natural nodes in series by simple edges and shortening the edges which connect them internally until all of these nodes combine into a single sphere with the appropriate number of punctures (see Fig.~\ref{composite}). As these combined nodes are simply glued they are then dual to gluings of simplices which when grouped together would be equivalent to a polygon (for a natural valence of~3) or a polyhedron with triangular faces(for natural valence of~4).

\subsection{Relating to the ribbon pictures}\label{sec:relatingtoribbon}

\begin{figure}[!h]
  \centering
    \includegraphics[scale=0.4]{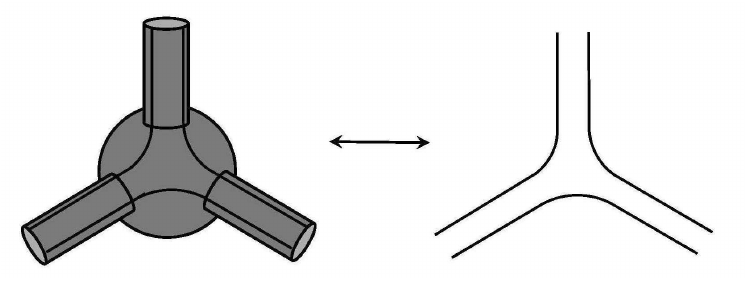}
\caption{From BRN to trinions.} \label{15}
\end{figure}
We can construct from the previous form of trivalent braided ribbon graphs a braided ribbon network as we've now def\/ined them as follows: for each node of the network we consider a~closed ball in the embedding space which has the node on its surface but which has an empty intersection with the rest of the ribbon graph. These spheres then def\/ine the nodes of the braided ribbon network. The edges of the braided ribbon networks are then def\/ined by similarly constructing tubes between these spheres so that the boundaries of the edges of the ribbon graph coincide with the boundary of the tubes. The boundaries of the surface of the ribbon graph then become the racing stripes of the braided ribbon network.

Likewise we can construct a traditional braided ribbon network from a 3-valent braided ribbon network by making the following observation: at each node the racing stripes divide the sphere into two parts, likewise along each edge the tube is divided in two by the racing stripes. Provided each closed path through the network encounters and even number of twists through $\pm\pi$ the network will split along the racing stripes into two orientable surfaces. We can consistently choose one side or the other and identify this as the surface of a traditional braided ribbon network (alternatively we can think of `squishing' the two halves into a single surface, in a way def\/ining one side to be the `front' and the other the `back').

\subsection{Applications of the unif\/ied formalism}
In \cite{Hackett2011b} and \cite{Hackett2011a} this formalism was used to demonstrate several general results for braided ribbon networks and embedded spin networks. We shall not reproduce these results here, but instead direct the reader to those papers for demonstrations of:
\begin{itemize}\itemsep=0pt
\item The generalization of the reduced link to the unif\/ied formalism, and hence to $4$-valent BRNs.
\item The demonstration of the conservation of the $4$-valent reduced link.
\item The Construction of maps between BRNs and Spin Networks.
\item The demonstration that the reduced link is a conserved quantity for Spin Networks.
\end{itemize}

These results give us a new use for braided ribbon networks: they have become an ef\/fective tool for understanding the information in the embedding of Spin Networks. They also demonstrate that a great deal of the structures that we study in BRNs also exist and are conserved in embedded spin networks.

\subsection{Correspondence between the trivalent and tetravalent cases}\label{sec:3v4vcorrespondence}
The natural formulation of framed tetravalent networks, as mentioned in Section~\ref{subSecGraphNotation}, is as tubular links between spherical nodes. It is in fact quite easy to see that such a network can be matched up to a framed trivalent (ribbon) network, by simply ``slicing" a tubular link down opposite sides, as discussed in Section~\ref{sec:relatingtoribbon}.

Given any framed trivalent network, we can always combine adjacent nodes to create composite tetravalent nodes. Likewise, the tetravalent nodes of the ribbon networks obtained by the splitting process described above can be decomposed into pairs of trivalent nodes. This allows us to switch between braids in the framed trivalent and tetravalent cases.

\begin{figure}[!h]
\centering
    \includegraphics[scale=0.3]{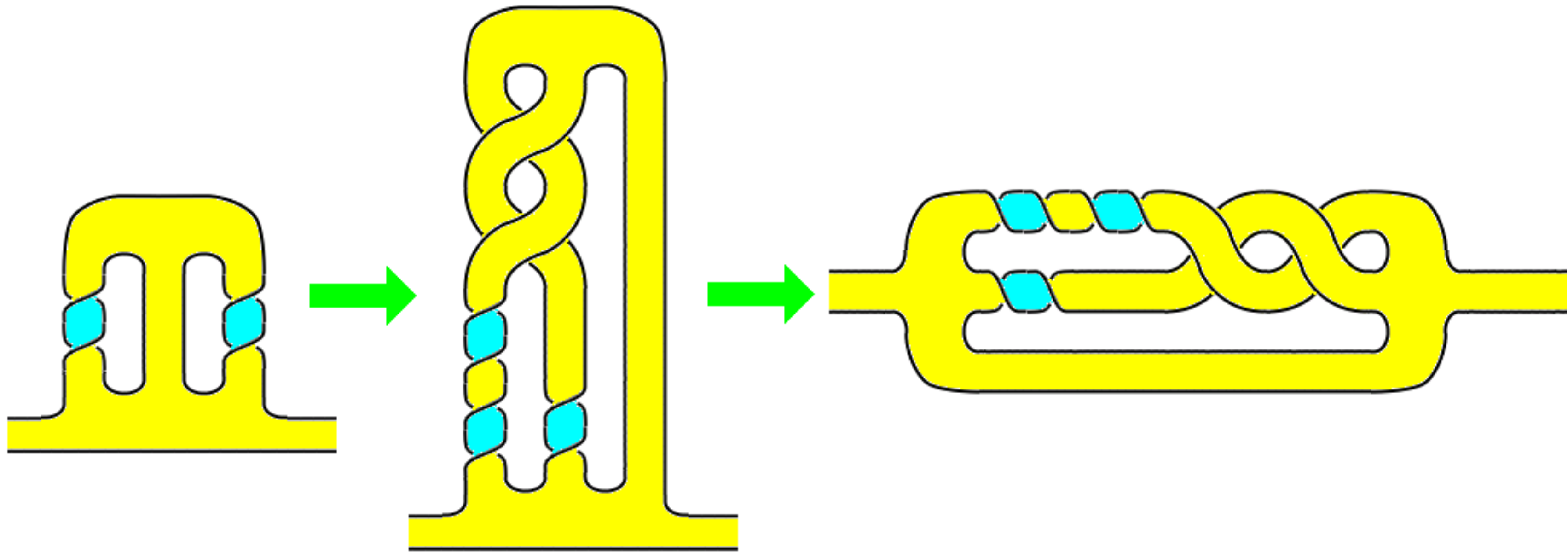}
\caption{Forming composite nodes allows us to convert between trivalent and tetravalent style braids.}\label{fig:tri-to-tetra-valent}
\end{figure}

Suppose we begin with a trivalent framed braid. We are always able to reduce this braid to its pure twist form, as noted above. Once in this form, in which all crossings have been removed, it is always possible to rotate the node at the top of the braid in such a manner that all the twisting on one strand (say, the rightmost strand) is removed, and extra twists and crossings are induced on the other two strands. We thereby arrive at a braid on three strands in which a single strand does not carry any twisting or crossing. The node at the bottom of this strand may then be freely combined with the node at the top of the braid to form a single tetravalent node. Likewise the nodes at the bottom of the two twisted strands may be combined to form a single tetravalent node. This process is illustrated in Fig.~\ref{fig:tri-to-tetra-valent}. By this process we obtain a braid located between two tetravalent nodes, just as occur in the framed tetravalent case (Section~\ref{subSecGraphNotation}). The braid obtained is, of course, embedded in a ribbon network, but it can always be used to reconstruct a tube-and-sphere framed tetravalent BRN.

The signif\/icance of this construction is that it allows us to associate framed tetravalent networks with structures occurring in the helon model (and hence with SM fermions), and allows the structures in the helon model to interact via the results on framed tetravalent networks~\cite{WanLee2007}. We thereby obtain a model which allows us to reproduce both kinematic and dynamical aspects of the Standard Model.

\section{Conclusions}\label{sec:conclusions}

While the idea of matter emerging from spacetime as topological substructures is an old one, it is only recently that our understanding of the subatomic structure of matter has made models of such emergent matter viable. In this review article we have discussed two parallel approaches, the trivalent and tetravalent scheme, which grew out of the suggestion that the most basic level of substructure within matter may be modelled by braided ribbons. The tetravalent scheme has proven to embody a rich dynamical theory of braid interactions and propagation ruled by topological conservation laws, but has until now not been able to construct a direct mapping to the particle states of the SM, instead producing a seemingly inf\/inite range of equivalence classes of braid states that fall into two types respectively analogous to bosons and fermions. The trivalent scheme has been unable to model interactions, but has been quite successful at taming the profusion of braid states present by constructing equivalence classes of braids, each equivalence class being mapped to a single type of particle. The unif\/ication of trivalent and tetravalent approaches we suggest here promises to allow the development of a fully dynamical theory of interacting particles, to restrict the range of particle states existing within the theory, and to provide a Rosetta stone that allows trivalent braids, tetravalent braids, and the particles of the SM to be equated in a satisfying manner. If successful, this will be a compelling theory of quantum spacetime and emergent matter.

\subsection*{Acknowledgements}
SBT is grateful to the Ramsay family for their support through the Ramsay Postdoctoral Fellowship. JH is grateful to his Thesis Advisor Lee Smolin for his discussion and critical comments. YW is in debt to his Supervisor Mikio Nakahara for his constant support and generosity. YW is also supported by ``Open Research Center'' Project for Private Universities: matching fund subsidy from MEXT, Japan.

\pdfbookmark[1]{References}{ref}
\LastPageEnding

\end{document}